\documentclass[usegraphicx,onecolumn,usenatbib]{mn2e}

\usepackage{graphicx}	
\usepackage{amsmath}	
\usepackage{amssymb}	
\usepackage{bm}		
\usepackage{pdflscape}	
\usepackage{newtxtext,newtxmath}

\usepackage[T1]{fontenc}
\usepackage{ae,aecompl}
\usepackage{amsmath,amssymb}
\usepackage{graphicx}
\usepackage{latexsym,color}
\usepackage{hyperref}
\usepackage{cleveref}
\usepackage[utf8]{inputenc}
\usepackage{mathtools,cuted}
\usepackage{natbib}

\usepackage{graphicx}	
\usepackage{amsmath}	
\usepackage{amssymb}	

\title[Limiting effects in misaligned tori clusters]{Limiting effects in clusters  of  misaligned toroids  orbiting   static SMBHs}

\author[D. Pugliese et al.]{
D. Pugliese,\thanks{E-mail: d.pugliese.physics@gmail.com}
Z. Stuchl\'{\i}k
\\
Research Centre of Theoretical Physics and Astrophysics,
Institute of Physics,
  Silesian University in Opava,\\
 Bezru\v{c}ovo n\'{a}m\v{e}st\'{i} 13, CZ-74601 Opava, Czech Republic
}

\date{Accepted XXX. Received YYY; in original form ZZZ}

\pubyear{2015}

\begin{document}
\label{firstpage}
\pagerange{\pageref{firstpage}--\pageref{lastpage}}
\maketitle
\count\footins = 1000
\begin{abstract}
We consider  agglomerates of  misaligned,  pressure supported tori orbiting a Schwarzschild black hole. A leading function  is introduced, regulating   the toroids distribution   around  the central static attractor--it enables to model the  misaligned tori aggregate as  a single   orbiting (macro-)configuration. We first   analyze  the leading function for  purely hydrodynamical perfect fluid   toroids.   Later, the function is modified for  presence of a toroidal magnetic field.
 We study the  constraints on the  tori collision  emergence  and  the instability  of the  agglomerates of misaligned  tori with  general relative inclination angles.
    We  discuss  the possibility that the twin peak high-frequency  quasi-periodic oscillations (HF-QPOs) could be  related to the agglomerate inner ringed structure.  The   discrete  geometry of the system is related  to HF-QPOs  considering several oscillation geodesic models associated  to the toroids inner edges.  We also study possible effect of the tori geometrical thickness on the oscillatory phenomena. \end{abstract}




\def\be{\begin{equation}}
\def\ee{\end{equation}}
\def\bea{\begin{eqnarray}}
\newcommand{\cc}{\mathrm{C}}
\newcommand{\jj}{\mathrm{J}}
\def\eea{\end{eqnarray}}
\newcommand{\tb}[1]{\textbf{{#1}}}
\newcommand{\ttb}[1]{\textbf{#1}}
\newcommand{\rtb}[1]{\textcolor[rgb]{1.00,0.00,0.00}{\tb{#1}}}
\newcommand{\btb}[1]{\textcolor[rgb]{0.00,0.00,1.00}{\tb{#1}}}
\newcommand{\otb}[1]{\textcolor[rgb]{1.00,0.47,0.22}{\tb{#1}}}
\newcommand{\ptb}[1]{\textcolor[rgb]{0.57,0.14,0.58}{\tb{#1}}}
\newcommand{\gtb}[1]{\textcolor[rgb]{0.00,0.47,0.24}{\tb{#1}}}
\newcommand{\pp}{\textbf{()}}
\newcommand{\non}[1]{{\LARGE{\not}}{#1}}
\newcommand{\mso}{\mathrm{mso}}
\newcommand{\Qa}{\mathcal{Q}}
\newcommand{\mbo}{\mathrm{mbo}}
\newcommand{\il}{~}
\newcommand{\rc}{\rho_{\ti{C}}}
\newcommand{\dd}{\mathcal{D}}
\newcommand{\Sie}{\mathcal{S}}
\newcommand{\Sa}{\mathcal{S}}
\newcommand{\mnras}{MNRAS}
\newcommand{\aap}{A\&A}
\newcommand{\prd}{Phys. Rev. D}
\newcommand{\actaa}{Acta Astron.}
\newcommand{\Fa}{\mathcal{F}}
\newcommand{\Mie}{\mathcal{M}}

    \newcommand{\oo}{\mathrm{O}}

\section{Introduction}\label{Sec:wolv}
In various phases of the  black hole (\textbf{BH}) accretion in active galactic nuclei (\textbf{AGN}), the angular momentum orientation of the infalling material  related to various accretion periods,  can be  expected to be  misaligned, leading to aggregates of misaligned toroidal structures.   These and similar processes open up the prospect of investigation into the possibility that during  more or less  long periods of the attractor life there may be different orbiting  toroidal structures, demonstrating substantially  different inclinations of their symmetry planes relative to a distant observer. Misaligned
   tori could be created even  around a \textbf{BH} accreting matter in a binary system, being raised from a  warped accretion disk. This scenario is now widely focused  in several  studies and simulations evidencing   how such complex orbiting structures are essentially regulated  by initial data of their formation.

Above all  these  toroidal structures  are governed by the  geometry of their   attractor. Particularly  essential    factor governing  the tori formation and misalignment is the presence or absence of  the central  \textbf{BH} spin and its magnitude. Purely hydrodynamic (HD) models of these toroidal structures  can provide  effective descriptions of  limiting effects associated with these systems and   indicate  observable situations  constraining  their formation and stability.
  The interest in these structures is  then manifold, they can throw light on the processes of accretion  disk formation and  tori instability associated with the accretion phases.  Furthermore,  such complex configurations are of  clear  observational interest due to existence of  different phenomena possibly caused by their non-homogeneous inner structure. A multi-disk model can be  also a  frame  for the interpretation of the mass accretion rates of \textbf{SMBHs} in \textbf{AGNs},  and for the evolution of the attractor spin and  jet emission.
(We note that the Kerr \textbf{BH} warped torus can evolve  together with its attractor changing its mass and  spin--see \cite{[BP75]}--
 see also \cite{Martin:2014wja,2012ApJ...757L..24N,2012MNRAS.422.2547N,
 2015MNRAS.448.1526N,2006MNRAS.368.1196L,Feiler,Aly:2015vqa}).

In this article  we consider  some of these limiting effects for aggregates of   misaligned (tilted)  tori orbiting a central spherically symmetric, Schwarzschild  \textbf{BH}.
 In  \cite{mis(a)} agglomerates  of  misaligned  tori were framed within an adapted modified  ringed accretion disks (\textbf{RADs}) model.   These ringed structures were introduced in \cite{pugtot,ringed}  as aggregates of axisymmetric accretion configurations, \emph{coplanar} and centered on the \emph{ equatorial plane} of the central  Kerr super massive \textbf{(SM)BH}. From now on we will refer to  the original equatorial  model considered in \citet{ringed,open,dsystem,Multy,long,letter,proto-jet}  as \emph{equatorial-RAD} or \textbf{eRAD}  to distinguish it from the misaligned case, referred  here for now on  as \textbf{RADs}. The \textbf{RAD} and \textbf{eRAD} models are essentially "constraining-models"   from    methodological and technical viewpoints. One of the main goals of these models was  to provide constraints  as initial data for  dynamic situations, for example in simulations of complex structures  such as GRMHD  (general relativistic magneto-hydrodynamic) supported tori orbiting a Kerr \textbf{BH}.
A fundamental characteristic  in these approaches is that the strong gravity of the \textbf{SMBH} attractor has a dominant role in determining  the morphology and   stability  of the aggregate toroidal components--\cite{open,dsystem}.
An essential  methodological aspect of the   \textbf{eRAD}-approach was the adoption of a  leading function capable of describing the distribution of toroids in the \textbf{eRAD}, and therefore its inner ringed structure--\cite{ringed,mis(a)}.
Here we use the leading function  for the  \textbf{RAD} model  of the misaligned structures in the case of  HD toroids--we elaborate in addition  an energy function linked to  the \textbf{RAD}  tori densities and  energetics. (The leading function can be  modified  to  consider the contribution of a  toroidal magnetic field  for some of the  aggregate toroids).

A further relevant  aspect of the \textbf{eRAD} investigation  consisted in the discussion of  significant observational tracks of a ringed structure which we expect to appear particularly through optical effects \citep{KS10,S11etal,Schee:2008fc,Schee:2013bya}--see also discussion in \citep{Marchesi,Gilli:2006zi,Marchesi:2017did,Masini:2016yhl,DeGraf:2014hna,Storchi-Bergmann}. Here we  return to this aspect showing  typical structures of \textbf{RAD} tori that should be evident as signature of their inner discrete inhomogeneous structure.

In these first  adaptations of the  \textbf{eRAD} into a \textbf{RAD} model, the advantage of considering    the spherically symmetric case is essentially methodological. Then for the description of the situation for a rotating and axially symmetric Kerr \textbf{BH}, a perturbative approach   can be considered having the limiting case of central Schwarzschild \textbf{BH}.  In the special case of a static, Schwarzschild black hole  each plane  crossing  the center  of the   attractor can serve as central plane of the toroid; only one of them  can be related as the  possible symmetry plane for a given orbiting toroid.
A \textbf{RAD} aggregate on static spacetime has a number of similarities  with the \textbf{eRAD} on a Kerr spacetime,  specifically with the case of  the "$\ell$corotanting"  tori sequences   which are  sequences of orbiting  coplanar tori orbiting on the  equatorial plane of a Kerr   \textbf{BH}   being   all corotating or counter-rotating with respect to the central Kerr attractor.
  Nevertheless, as discussed in \cite{mis(a)},  for   very slowly rotating attractors,   tori misalignment allows to reconsider in some extent the possibility of the presence of multi accreting  tori on different planes, enlightening interesting situations and phenomenology   which were considered  not possible for the \textbf{eRAD}.
  Geometry of the  \textbf{RAD} accreting or equilibrium  tori, stability and collision emergence, were studied in  \cite{mis(a)} where evaluations  of  quantities  related  to tori energetics such as the  mass-flux,  the  enthalpy-flux (evaluating also the temperature parameter),
and  the flux thickness, in dependence on the model parameters were provided for
polytropic fluids.
Some notes
on  the \textbf{RAD} models including  proto-jets, which are open and cusped solutions  associated to geometrically thick tori were also reported.
In the present  investigation we  consider also  the possibility of  oscillating tori causing  quasi-periodic oscillations (QPOs) oscillations emission.

{We focus in particular on  the  fluid specific  angular momentum  $\ell(r)$ in the HD-torus, as "leading-\textbf{RAD} function", i.e.  as a "reference" function constraining the location of the maximum and minimum points  of the hydrostatic pressure and density in the \textbf{RAD}. Therefore  the function  represents   the distribution of  tori  centers and inner and outer edges of the toroidal  aggregate components.}
{The idea behind the elaboration  of a leading function is to represent the possible tori "distribution" of the  aggregate  in the \textbf{RAD}.  In the case  focused  in this work, considered as reference  for more complex situations, the distribution  is described by the specific angular  momentum  $\ell(r)$, this function   of  the radius  $r$ (radial distance from  the central Schwarzschild black hole),  has  mainly a geometrical and centrifugal origin. The analysis via  the leading function frame has to be developed along three main steps.}

{Firstly, the effective  existence of such  leading  function  has to be established-- it has to  assess the possibility of representing the macro-structure through one function which can provide effective constraints to  the \textbf{RAD} tori. In this investigation   we have discussed this issue  through the analysis  of the force balance equation (the relativistic Euler equation) of the hydrodynamical model, where the existence of a leading  function is  guaranteed by an integral condition on an  ordinary differential equation due  to the application of the  von Zeipel theorem.}

 {A second step of the investigation concerns the identification   of a proper leading function which in general will depend on  the specificity of the accretion models. In Sec.\il(\ref{Sec:magne}) we  consider a deviation from the hydrodynamic case, considered as a (geometrical) reference case,  including the contribution of  a toroidal magnetic field in the force-balance equation.}

 {Thirdly,    the constraints  obtained in the leading-function approach  are  applied   in  different models which can   consider  this setting as  reference.
Indeed, together with the  conceptual relevance of constraining  the  orbiting tori system with \emph{one}  distribution leading function,
this approach has the  advantage  to   provide general constraints  for the macro-structure which, in some extent, avoids the explicit solution of the GRHD equations or, eventually, the study of  a series of tori characteristic effective potentials. In this last case the effective potentials, recovered from some integrability conditions on the Euler equations, would require a  fine tuning of some model parameters related to each toroidal component of the \textbf{RAD}  and  the  management of the particular boundary conditions adapted to the specific case in examination  arising  from the  ringed structure of  the misaligned orbiting configurations.  This especially holds when the \textbf{RAD} tori are considered as set of initial data in complex models which have to be  treated through numerical integration.
{(We report, with regard to this aspect of the applicability of this method that these  toroidal analytic models are commonly  used as  starting condition for numerical studies of black hole accretion. In these studies, the   simulations of accretion flows  largely verify
the   agreement with the
model predictions.)}}

{Since we consider geometrically thick tori as \textbf{RAD} constituents it is convenient  to describe  here  the accreting toroids  in comparison with different models. There are several  accreting disks models which differently consider   the   variety of
   processes of very diversified  nature  characterizing the  accretion disk physics.
It turns useful to distinguish  these  models according to the following three  aspects. \textbf{(1)} The disk  geometry is a first significant aspect of disk  physics which, distinguishing the   geometrically thin from  the geometrically thick disks (tori), is   essentially defined by the disk vertical thickness  (on the disk symmetry plane). \textbf{(2)} A second characterizing element is    the matter accretion rate of the  accretion  disk   (correlated to sub or super-Eddington luminosity). \textbf{(3)} Third featuring   element  for the accretion disk is its  optical depth (for transparent or opaque disks)--see \citep{abrafra}.
Typically,  thick disks, for example the Polish doughnuts  tori (P-D), have  very high, super-Eddington, accretion rates, and  high optical depth, while the ion tori,   the ADAF (Advection-Dominated Accretion Flow) disks have low optical depth and relatively low accretion rates (i.e. sub-Eddington).
Importantly for our analysis, in thin disks the dissipative viscosity processes, which are usually framed  with  the  local magnetic fields in the magnetorotational instability (MRI), are relevant for accretion.}

{In the toroidal disks, as those considered here, pressure gradients are crucial  for the accretion mechanism as well as the disk morphology.
Given the relevance of the issue, we studied in section\il(\ref{Sec:epi}) the tori geometrical thickness  in  \textbf{RAD} frame (for geometrically thick disks including viscose effects see also  \cite{expect}).
(Recently astronomers of   observatory Karl G. Jansky Very Large Array (\textbf{VLA})    provided a picture of  the dusty, thick torus and associated emitted   jets of material ejected by the disk orbiting the \textbf{SMBH}  at the core \textbf{Cygnus A} \citep{Carilli2019}).
Tori in this analysis are   therefore a case of radiation pressure supported thick disks with Super-Eddington luminosity,
generally having very small accretion efficiency  and consequentially highly super-Eddington accretion rates,
often with strong outflows, and  advection cooled (for some $q \approx 1$, i.e.  energy flux per advection$\approx$energy flux per radiation).  In the  case considered here the accreting  flow "starts" across a "Lagrange  point" (Roche lobe overflow, due to Paczy\'nski accretion mechanics \cite{Pac-Wii,cc}). This is  the  cusp of the orbiting toroidal surface, which is an  important aspect of thick disks  since its presence also stabilizes the tori against several instabilities (thermal and viscous local instabilities, and globally   against the Papaloizou-Pringle instability-PPI and it could be possibly connected to QPOs emission.)}

{This  fully general
relativistic model of an opaque,  pressure supported  and super-Eddington torus,
 traces back to the Boyer theory of     the equilibrium and rigidity in general relativity, i.e. the  analytic theory of equilibrium configurations of   rotating perfect fluids \citep{Boy:1965:PCPS:}.
Within the so  called ``Boyer's condition'', we   can  determine    the boundary of  stationary, barotropic, perfect fluid body as  the
surfaces of constant pressure (eventually also   equipotential surfaces $\partial_r V_{eff}=0$). This   occurs in many models, essentially thanks to  the condition $\Omega=\Omega(\ell)$ on the fluid relativistic frequency $\Omega$   that has   to be a     function of  $\ell$ (fluid specific angular momentum), a result known as
von Zeipel condition \citep{Zanotti:2014haa,Koz-Jar-Abr:1978:ASTRA:,M.A.Abramowicz,Chakrabarti,Chakrabarti0}.
The advantage of this model turns  to be both conceptual and technical: from a technical view-point, essential   features of the disk  morphology  like the thickness, the elongation   on its symmetric (equatorial) plane, the distance from the attractor are predominantly regulated  by the geometric properties of spacetime via  the pressure gradients in  the relativistic Euler equation, reducible to an ordinary differential equation (ODE), often integrable with the introduction of an effective potential. In this context also the torus inner (outer) edge,  in the different torus  topological phases related to the stable phases of the disk (emergency of the Roche lobe and  the cusp formation) are well defined and constrained.
}

{
The effective potential function, $V_{eff}(r;a,\ell)$, expressible here for each \textbf{RAD} component in the purely hydrodynamic model (without the presence of additional  factors such as magnetic fields) contains  two  essential features. \textbf{1.} It encodes  a  geometrical factor   related to the properties of the  spacetime background. This property  constitutes  essentially the conceptual advantage of having such a model as a reference model, even for decidedly more complex systems. In fact it allows to extrapolate geometrical basic information, when the "curvature" effects due to the very strong gravity of the central \textbf{SMBH} attractor are relevant, for example where these models are used as initial data of GRMHD simulations. \textbf{2.} The second factor expressed in  the effective potential $V_{eff}(r;a,\ell)$, which we  can consider as a dynamical factor, is related to the orbiting matter, and represents the centrifugal component of the forces balance in the toroid. This is  expressed here by the fluid specific angular momentum $\ell$ on which we will dwell long in this analysis. The function $\ell(r)$ is generally  assumed,    in many applications of these toroidal models, constant in the torus (see  also \cite{Lei:2008ui} for a more general discussion on the functional form of the specific angular momentum). In this model the entropy is   constant along the flow  (also in the magnetized case) and  the rotation law $\ell=\ell(\Omega)$ is independent of the equation of state \citep{Lei:2008ui,Abramowicz:2008bk}. In these tori
in fact  the functional form of the angular
momentum and entropy distribution during the evolution of dynamical processes, could be considered as dependent only on the initial conditions of the system and on
the details of the dissipative processes--for the relevance of this assumption see \cite{abrafra}.   However, in the present paper the specific angular momentum $\ell(r)$ assumes a broader meaning  by adapting it to the macro-structure, and we also discuss  the fundamental of this assumption.  We use this function as leading function for the \textbf{RAD}, representing the distribution of the  pressure   gradient points  in the \textbf{RAD}. Thus,  we assume in the hydrodynamical model, the radial profile of the  Keplerian fluid  specific  angular momentum  $\ell(r)=\ell_K(r)$ as the leading function.
 }

{
Finally, regarding the possible associated phenomenology,
  a composite model of accretion tori obviously opens a scenario to numerous phenomena derived from the ringed inner structure and particularly related to  instabilities, tori oscillations and their possible traces due to the  QPOs emission.
The  accretion disks (tori) are characterized by natural oscillation modes depending  mainly on the disk geometrical thickness and  the location of the inner edges. These  modes arise  consequently to  disks internal processes  whose traces would appear in the typical   light curves of the disk which is the mechanism at  base of  the QPOs models from accretion disks--\cite{2013A&A...552A..10S}.}

{
 Thick tori are dynamically unstable for the  non-axisymmetric oscillation modes, Papaloizou and Pringle   instability (PPI), affecting  the non accreting torus, particularly for the  $\ell=$constant  tori  considered here.
However, in these   tori   the proper accretion process
across  the cusps of the closed configurations (modulated by global  oscillations)  regulates the accretion rate (due to  the mass loss)
 stabilizing the  torus for  (local) thermal and viscous instabilities  and  globally
against the PPI--\cite{Blaes1987,Abramowicz:2008bk,Pac-Wii,cc,Koz-Jar-Abr:1978:ASTRA:}.
Thus, thick torus can have global
instabilities although the  flow can be locally stable. Eventually,  it has been shown that  such tori turn to be  (marginally) stable for  local axisymmetric
perturbations and  unstable to  non-axisymmetric modes. In this respect the PPI  is the tori characteristic global instability,  mainly   regulated by the  boundary
conditions  which  are extensively considered here in the aggregate model  of misaligned tori. These modes are stabilized  (suppressed)  by the  overflows across the cusp (for a discussion on the  PPI in the  presence of  a toroidal magnetic field see for example \citep{Bugli} and discussion in \cite{mnras})
Tori oscillations and perturbations  are strongly  dependent on the geometrical torus thickness (i.e. the  "vertical" direction on its symmetric plane). Nevertheless,
 so far a complete spectrum of modes  for dynamical oscillations of geometrically thick tori is  still an open (technical) problem.  For this reason,  in this paper we  extensively consider the evaluation of the torus geometric thickness and the related $\beta$ parameter  to guarantee  a  validity regime of the approximations considered in the analysis.
 In the  hydrodynamic models of thick disks incompressible and   axisymmetric  modes of  global oscillations are associated with typical characteristic frequencies. We consider as the frequencies relevant in the thick torus, the  Keplerian frequency and the two epicyclic (geodesic) frequencies (radial and vertical);  their applicability is discussed in \cite{2013A&A...552A..10S}.
 Other modes  (surface gravity,  acoustic and  internal inertial modes) can be  studied in the so called relativistic Papaloizou-Pringle equation.
  To complete the discussion on the tori instabilities we mention the  Runaway   instability  (RI) affecting thick tori and their \textbf{BH} attractors. In this article we have considered a frozen-background spacetime (the black hole mass and zero spin do not change following matter accretion).  Runaway instability follows the large accretion rates  typical of these tori, the \textbf{BH} mass  increases changing  the spacetime properties and  in turn  affecting   the orbiting accreting   disk, the inner edge moves inwards,  this can leads to a  a stable situation, or  the cusp moves
 inside the disk inducing an increases of mass transfer.
  In \cite{dsystem} we have considered the possibility of Runaway-Runaway instability (RRI) when the Runaway instability affecting the  \textbf{BH} and the inner accreting  torus of the agglomeration  is accompanied by the consequences of the background modification on the outer tori of the \textbf{RAD} which can collide, accrete or stabilize depending on  initial conditions.}

\textbf{Article overview:}
{To make more clear and simple the
description and explanation of the \textbf{RAD} model, we provided Table (\ref{Table:pol-cy})
with a list of the main symbols and relevant notation used throughout this article.
  The article is structured in two parts: in first part,
 section (\ref{Sec:Misal}), we develop the model  introducing its  essential quantities and discussion of its  main  aspects that will be  used in the second part.
In details,
we  introduce  the agglomerate of  misaligned perfect fluid tori orbiting a central Schwarzschild black hole including a description of the  thick disk  model.
 In this analysis we  use also  results of \cite{mis(a)}, reported  for convenience in
Appendix\il(\ref{Sec:doc-ready}), which  particularly  contains a  review  of the main characteristics of misaligned (accreting) tori morphology.
 The second part  of the article, section\il(\ref{Sec:epi}), develops   application of the  results of our analysis to oscillation models  often considered in the modeling of high-frequency QPOs.
Specifically, we  discuss  the possibility that the twin peak high-frequency  QPOs could be  related to the \textbf{RAD} inner structure,  relating  the \textbf{RAD} discrete  geometry  to the QPOs emission.
{In Sec.\il(\ref{Sec:pen-comme}) we include some comments on the \textbf{RAD} structures and  a brief discussion on  the     outcomes  of  our analysis. }
Finally, as sideline of this analysis,   in  Sec.\il(\ref{Sec:magne}) we  discuss  the case when the
 leading \textbf{RAD} function, defining the distribution of the  tori in the \textbf{RAD}, has been modified  to an alternative definition, considering the case of  \textbf{RADs} where some of their components  can be  magnetized tori with the toroidal magnetic field introduced in \cite{Komissarov:2006nz}, in the approach considered in \cite{epl,mnras}.
Concluding remarks follow in section (\ref{Sec:conclu}).}

{
\begin{table*}
\centering
\resizebox{.92\textwidth}{!}{%
\begin{tabular}{lll}
 \hline \hline
 $\ell_K(r)$& \textbf{RAD} rotational law-- \textbf{RAD} specific  angular momentum  distribution &
equation\il(\ref{Eq:lqkp})
\\
 $K(r)$ & \textbf{RAD} energy function--distributions of \textbf{RAD} maximum and minimum density/pressure points
  &
equation\il(\ref{Eq:Kdir})
\\
$r_{\gamma}=3M$& Marginally circular orbit  (photon  orbit) & Sec.\il(\ref{Sec:Misal})
\\
$r_{mbo}=4M$& Marginally bounded circular orbits& Sec.\il(\ref{Sec:Misal})
 \\
$r_{mso}=6M$& Marginally stable  circular orbit (ISCO)& Sec.\il(\ref{Sec:Misal})
 \\
$r_{cent}$& Torus center, maximum density and pressure point in a torus & equations\il(\ref{Eq:schaubl})
(\ref{Eq:rcentro})
\\
 $r_{crit}=r_{cusp}=\{r_{\times},r_j\}$&  Effective potential maximum&   equation\il(\ref{Eq:schaubl})
\\
$r_{\times}$ & (Accreting) Torus cusp (minimum density and pressure point) & equations\il(\ref{Eq:schaubl},\ref{Eq:r-inner-A1})
\\
$r_{j}$&  Proto-jet (open) configuration cusp &  equation\il(\ref{Eq:schaubl})
\\
$r_p^{\ell}(r)$&  Solution of  $\ell_K(r)=\ell_K(r^{\ell}_p)$ relates $r_{crit}$ and $r_{cent}$& equation\il(\ref{Eq:schaubl})
 \\
$r_{mbo}^b\approx10.4721M$& Solution of $\ell_K(r_{mbo})=\ell(r_{mbo}^b)$& equation\il(\ref{Eq:schaubl})
\\
 $r_{\gamma}^{b}=22.3923M$& Solution of $\ell_K(r)=\ell_{\gamma}$  & equation\il(\ref{Eq:schaubl})
 \\
 $(r_{in},r_{out})$& Torus inner and outer edges & equations\il(\ref{Eq:outer-inner-l-A1},\ref{Eq:over-top},\ref{Eq:over-top1})
\\
 $r_p(r)$&  Solution of $K(r)=K(r_p)$,
  relates tori  ($T_1$, $T_2$) with  $K_{cent}(T_1)=K_{crit}(T_2)$& equation\il(\ref{Eq:Krrp})
\\
$r_{mbo}^k\approx4.61803M$& Solution of  $K(r_{mbo}^k)=K(r_{mbo}^b)$ & equation\il(\ref{Eq:Krrp})
\\
$r_{\gamma}^k\approx4.21748M$& Solution of $K(r_{\gamma}^k)=K(r_{\gamma}^b)$& equation\il(\ref{Eq:Krrp})
\\
 $\lambda\equiv r_{out}-r_{in}$& Torus elongation on its symmetry plane & equation\il(\ref{Eq:elong-l-A1})
\\
 $\bar{\lambda}=r_{in}^o-r_{out}^i$&  Inner $T^i$ and outer $T^o$ tori spacing in the \textbf{RAD}& appendix\il(\ref{Sec:doc-ready})
 \\
$ \Sa$& Torus geometrical  thickness  and  $\Sa_{\times}=2h_{\times}/\lambda$ geometrical thickness of accreting torus& appendix\il(\ref{Sec:doc-ready})
\\
$h_{\times}$ & Height of the accreting toroidal surface  an inner $T^i$ and outer $T^o$ & equations\il(\ref{Eq:xit-graci},\ref{Eq:max-prob-A1})
\\
$r_{max}\equiv (x_{\max},y_{\max})$&  Torus  geometric maximum& equations\il(\ref{Eqs:rssrcitt},\ref{Eq:r-quest-v}).
\\
\hline\hline
\end{tabular}}
\caption{{Lookup table with the main symbols and relevant notation  used throughout the article. Links to associated sections, definitions and figures are also listed.  In general we  adopt notation $\mathcal{Q}_{\bullet}\equiv \mathcal{Q}(r_{\bullet})$ for any quantity $\mathcal{Q}$ evaluated  on  a general radius $r_{\bullet}$.}}
\label{Table:pol-cy}
\end{table*}}
%
\section{Model of  aggregates of orbiting geometrically thick   tori}\label{Sec:Misal}
We start by considering  misaligned  one-specie perfect   fluid  tori orbiting a central Schwarzschild \textbf{BH}. All the relevant fluid  quantities $\Qa$ satisfy the conditions $\partial_i \Qa=0$ where $i\in\{t, \phi\}$, in the spherical  standard Schwarzschild  coordinates  $\{t,r,\theta,\phi\}$. In the \textbf{RAD} model construction   discussed in  \cite{mis(a)}, we
 considered the tori boundary conditions limiting   the aggregate \textbf{RAD} inner structure    depending  on  the   tori  with tilted relative inclination angle   $\theta_{ij}$.
 {Below we include a  general discussion of the main features  of geometrically thick tori considered here as \textbf{RAD} components which are  relevant to  the \textbf{RAD} framework. We mention, as a general reference for   these well-known disk  models, the  general  review \cite{abrafra}  and with regards  to their role in the \textbf{eRAD} frame we refer to  \cite{ringed}. This section closes with the  introduction of  the leading \textbf{RAD} function $\ell_K(r)$ and the energy function $K(r)$ and discussing the derived constraints for the \textbf{RAD}.}

\medskip

\textbf{{Main properties of geometrically thick tori}}

\medskip

\begin{figure*}
\includegraphics[width=9cm,angle=90]{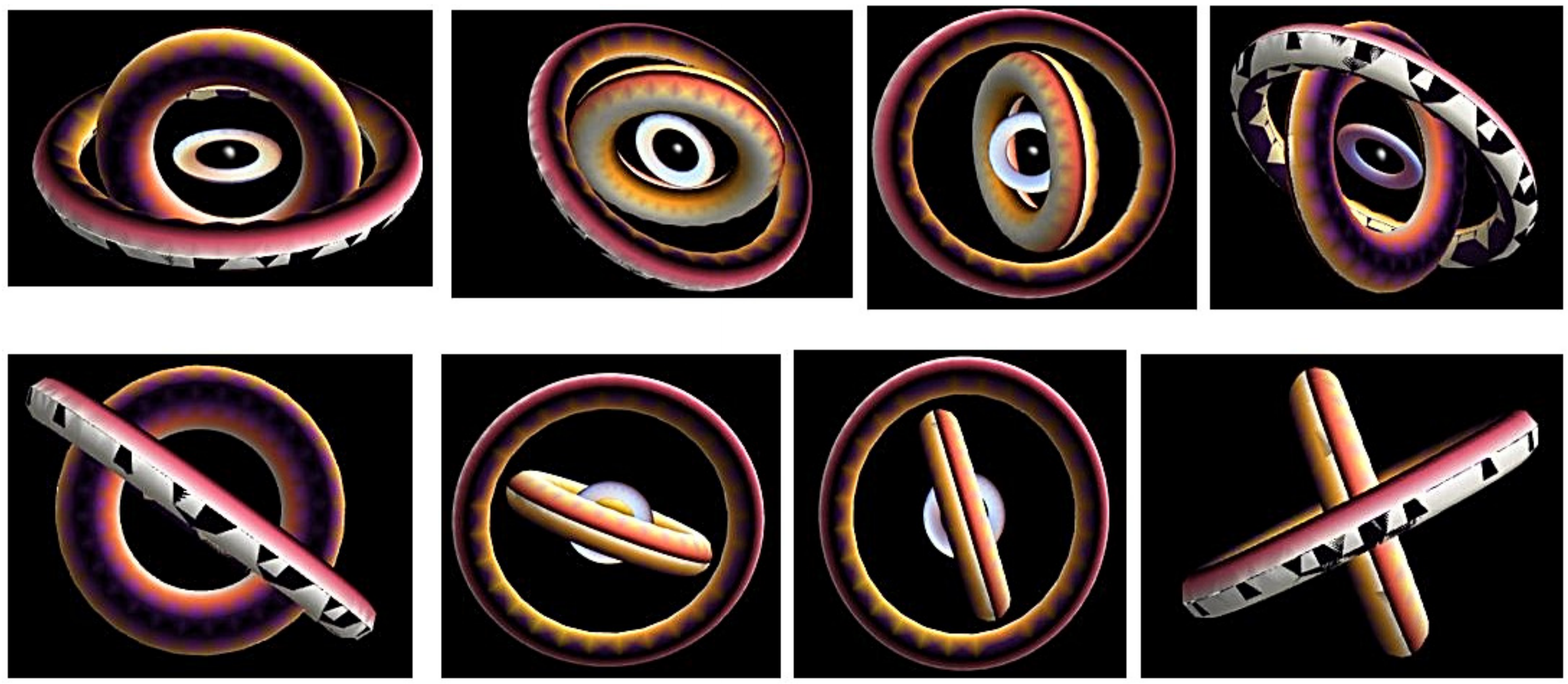}
\caption{Density profiles of a \textbf{RAD} orbiting misaligned tori obtained from the  3D {HD} integration of the perfect fluid Euler equation (\ref{Eq:Eulerif0}).  Black region is the central Schwarzschild \textbf{BH}.  The \textbf{RAD} is of the order $n=3$ (number of orbiting disks).    Different viewing angles are shown, where the central black hole (outer horizon) is embedded into the orbiting \textbf{RAD},  the  ringed  structure  also be recognized   by  the observation. Tori parameters $\ell$, being the fluid specific angular momenta and $K$, a parameter related to matter density and tori energetics are  as follows:   $\mathbf{T_1}-(\ell _1=3.75, K_1=0.95)$,  $\mathbf{T_2}-(\ell_2=4.7958, K_2=0.976)$,  $(\mathbf{T_3}-\ell _3=6, K_3=0.98487)$, where $\mathbf{T_1}<\mathbf{T_2}<\mathbf{T_3}$, indicates the tori closest to the central attractor. Torus $\mathbf{T_1}$  is at distance  $r_{in,1}=5.75538M$ from the central \textbf{BH} (location of the inner edge of the orbiting torus),   and elongation on the equatorial plane $\lambda_1=4.65846M$,  $\mathbf{T_2}$ is at distance  $r_{in,2}=14.168M$ from the central \textbf{BH}, with elongation  $\lambda_2\approx 11.2734M$. $\mathbf{T_3}$  is at distant $r_{in,3}=27.867M$  from the central Schwarzschild \textbf{BH} with elongation $\lambda_3= 8.57362M$. In Figures\il(\ref{Fig:fondam}),  schemes of \textbf{RAD} of order  $n=5$ of quiescent and non-interacting  tori are represented.
\label{Fig:solidy} }
\end{figure*}

 These tori are regulated by
 the
  Euler equation\footnote{The continuity equation,
  $u^a\nabla_a\rho+(p+\rho)\nabla^au_a=0\, $  is identically satisfied because of the symmetries. {The choice of a perfect fluid stress-energy tensor
 is closely correlated with the typical relation between the (fictitious) time-scales of the main physical processes  assumed relevant for  geometrically thick disks, which in turns is  linked to the forces balance inside the disk.  The disk physical processes are generally  conveniently  considered having  three main origins: (1) a dynamical origin, (2) a thermal and (3) a  viscous origin. Specifically, the  relation  between the related timescales assumed for these  tori  is
$\tau_{dyn}\ll\tau_{therm}<\tau_{visc}$. The dynamic part is  represented by  the time reached  by the pressure forces to balance the centrifugal  and gravitation component. The thermal time scale  concerns the entropy redistribution, dissipative heating  and the cooling processes.
The part most directly related to our choice of the rotation law (i.e $\ell(r)$ interpreted for each torus, and here considered as \textbf{RAD} leading function) is the time scale  involved when the angular momentum changes because of
 torque and  dissipative effects--time scale of the viscous effects.}}:
\bea
&&
\label{Eq:Eulerif0}
(p+\rho)u^a\nabla_au^c+ \ h^{bc}\nabla_b p=0
\eea
 on the  symmetry  plane, at an inclination angle  $\theta$ of each  toroid of the orbiting agglomeration, where
$\rho$ and $p$ are  the total energy density and
pressure, respectively, as measured by an observer moving with the fluid with fluid
four-velocity $u$ \footnote{The fluid  four-velocity  satisfy $u^a u_a=-1$. We adopt the
geometrical  units $c=1=G$ and  the $(-,+,+,+)$ signature.  The radius $r$ has unit of
mass $[M]$, and the angular momentum  units of $[M]^2$, the velocities  $[u^t]=[u^r]=1$
and $[u^{\varphi}]=[u^{\vartheta}]=[M]^{-1}$ with $[u^{\varphi}/u^{t}]=[M]^{-1}$ and
$[u_{\varphi}/u_{t}]=[M]$. For the seek of convenience, we always consider the
dimensionless  energy and effective potential $[V_{eff}]=1$ and an angular momentum per
unit of mass $[L]/[M]=[M]$.},
$h_{ab}=g_{ab}+ u_a u_b$ is the projection tensor and $g_{a b}$
the Schwarzschild metric tensor (where $\nabla_a g_{bc}=0$),  $M$ is the \textbf{BH} mass.
 In Figures\il(\ref{Fig:solidy}) we show a solution of  the Euler equations with appropriate boundary conditions for misaligned barotropic tori, governed by
the  integral of Eq.\il(\ref{Eq:Eulerif0}) that states the form of definition of an effective potential due to relation
\bea\label{Eq:c-i-va}
&&
\int\frac{dp}{\rho+p}=-\ln V_{{eff}}=-\ln\left[\sqrt{\frac{(r-r_+) r^2}{r^3-\ell^2  (r-r_+)}}\right],
\eea
%
where $r_+=2M$ is the \textbf{BH} horizon, and $\ell(r)$ denotes  radial profile of the specific angular momentum of the orbiting fluids in the symmetry plane, here and in the following we shall use dimensionless quantities if not otherwise specified --\cite{abrafra,Font,Pugliese:2012ub,epl,Salny,
2005PhRvD..71b4037S}.
The model we adopt here for each  component of the  \textbf{RAD} aggregate provides  the
 the (rigid) boundary of a stationary, barotropic, perfect fluid toroid constituted by the
equipotential  and equipressure surface given by the conditions  $V_{eff}(\ell,\theta)=K=$\mbox{constant}.
 The main features of the equipotential
surfaces for a generic rotation law $\Omega=\Omega(\ell)$
 are described here  by the
equipotential surface for  uniform distribution of the
angular momentum density $\ell\equiv L/E$ (also the fluid \emph{specific}  angular momentum where
 $E$ and $L$   are the particle energy and angular
momentum per unit of mass as seen by infinity).
The choice $\ell=$constant for each torus is a well known assumption, widely used in several contexts where geometrically thick tori are considered--\cite{abrafra}.  Conditions
$\ell=\mbox{constant}$, and   $\Omega=\mbox{constant}$ for the  fluid relativistic angular velocity $\Omega$,  define the surfaces  known as {von Zeipel's cylinders}. From the series of results related to  the von Zeipel theorem it follows that   the equipotential surfaces of the marginally stable
configurations orbiting in a Schwarzschild spacetime correspond to constant  definition of  $\ell$.
In the static spacetimes, the family of von
Zeipel's surfaces  depend only on the background
spacetime,  therefore  they do  not depend on  the  particular fluid rotation law ($\Omega=\Omega(\ell)$). If the fluid is barotropic, as we are considering here, then    von Zeipel's theorem guarantees that
the surfaces $\Omega=\mbox{constant}$ coincide with the surfaces  $\ell=\mbox{constant}$.
Solutions of equation\il\ref{Eq:Eulerif0} with appropriate boundary conditions\footnote{{
We note that the von Zeipel condition guarantees an integrability condition   on the Euler ODE, and it is therefore essential in many models where the reduction of ODE  considered here (effective potential approach, leading function) is to be extended to  more complex objects, for example in the extension to poloidal magnetic field  a la  Komissarov which is  now present in integral form only as a  toroidal  magnetic field \cite{Komissarov:2006nz,Zanotti:2014haa}.
The von Zeipel conditions ($\ell(\Omega)$)  is closely related to the barotropic EoS of the fluid.
In other flows, where dissipative effects must be considered, $\ell(\Omega)$  depends on these in a way that is still unclear and often given  on assumptions on various parameters, usually related to viscosity factors to be taken ad hoc, for example in the so called alpha-prescription in MRI or GRMRI models \cite{Balbus2011}. Compared to these assumptions, the Paczy\'nski prescription has the notable advantage of having a narrow geometric sense, which in fact makes it the bottom boundary condition on practically  any accreting disk, with the respect to the quasi-spherical Bondi accretion \cite{abrafra}. On the function  $\ell_K $ generalization  (that  can have also  different regimes in   regions  discriminated by $r_{mso}$) see \cite{Lei:2008ui}.}} lead to  four  classes of configurations  corresponding  to  {closed} and  {open} surfaces, and surfaces  with or without
a cusp, i.e. self-crossing open or closed configurations. The
closed, not cusped,   surfaces are  associated to  stationary equilibrium (quiescent)  toroidal configurations.
 For  the  cusped and   closed equipotential surfaces,   the accretion onto the central black hole can occur through the
cusp of the equipotential surface: the torus  surface exceeds
the critical equipotential surface (having a cusp), leading to a mechanical
non-equilibrium process where  matter inflows  into the central  black hole (a violation of the hydrostatic equilibrium known as Paczy\'nski mechanism) \cite{abrafra}.
Therefore, in this
accretion model we shortly indicate the  cusp of the  self-crossed closed toroidal surface as  the "inner edge of accreting torus".
Finally, the open equipotential surfaces, which we do not consider explicitly here, have been associated to  the formation of  proto-jets \cite{open,proto-jet}.

\medskip

\textbf{Tori agglomerate in the RAD framework}

\medskip

We describe the aggregate of misaligned  tori  adopting the \textbf{RAD} framework developed in
 \cite{ringed,open,dsystem} in the case of an   agglomerate composed by tori orbiting on the equatorial plane of a Kerr attractor (the  \textbf{eRAD}).
As  in \cite{mis(a)}, it is convenient to introduce a
 \textbf{RAD} \emph{rotational law }$\ell(r)$ as the distribution of specific angular momentum of  its toroidal components. This function plays the part of  leading \textbf{RAD} function proving also the  toroids \textbf{RAD} distribution in its  ringed structure. Therefore
 $\ell(r)$ is intended to be  the absolute magnitude of the fluid specific angular momentum distribution  of each  toroid of the aggregate  orbiting
at distance $r$ from the central static \textbf{BH} on its general  symmetry  plane (while we stress the specific angular momentum is constant inside each toroid).
The  Keplerian, geodesic distribution of the  specific angular momentum   $\ell_K(r)$ then  parameterizes each torus in the \textbf{RAD}, due to  \textbf{RAD rotational law} 
\bea\label{Eq:lqkp}
\ell_K(r)\equiv {\frac{r^{3/2}}{(r-r_+)}}.
\eea
We introduce also the \textbf{RAD energy function} parameterizing each torus of the agglomerate with  further $K$-parameter governing extension of the torus \cite{mis(a)}
\bea\label{Eq:Kdir}
K(r)\equiv\frac{(r-r_+)}{\sqrt{(r-r_{\gamma}) r}},
\eea
$r_{\gamma}=3M$ is the location of the last photon circular orbit.

\medskip

\textbf{Origin of $\ell(r)$ and $K(r)$}

\medskip

{The specific angular momentum adopted in  Eq.\il(\ref{Eq:lqkp}) and energy function $K(r)$ of Eq.\il(\ref{Eq:Kdir}), have a clear  geometric origin related to the symmetries of the Schwarzschild  background.
 It is convenient to explicit    the following   quantities as follows
\bea\label{Eq:pro-conserva}
&&
E \equiv -g_{ab}\xi_{t}^{a} \mathrm{p}^{b}=-g_{tt} \mathrm{p}^t ,\quad L \equiv
g_{ab}\xi_{\phi}^{a}\mathrm{p}^{b}=g_{\phi \phi}\mathrm{p}^{\phi}, 
\eea
\cite{pugtot,Pugliese:2012ub}, where the metric components are
 $g_{tt}=-e^{\nu(r)}$ and  $g_{\phi\phi}=+r^2\sin^2\theta $, written in standard spherical coordinates
$(t,r,\theta,\phi)$, where  $ e^{\nu(r)}\equiv\left(1-2M/r\right)$.
Expressed  in terms  of  the four  momentum  components and   $(E, L)$ are   constants of motion for test particle geodesics with    four-velocity $u^a$,  related to the  Schwarzschild geometry Killing vectors $\xi_t$ and $\xi_\phi$.
 We can  define the
 quantity $ V_{eff}$, considering the normalization condition on the fluid four-velocity (assuming  $\mathrm{p}^{\theta}=0$  for the circular configurations), we introduce also  the  relativistic angular frequency  $\Omega$ and the fluid specific angular momentum $\ell$:
\bea\label{Eq:poll-delh}
&&u^r \equiv\sqrt{E^2-V_{eff}^2},
\quad\Omega \equiv\frac{u^{\phi} }{u^t}=-\frac{g_{tt} L}{{E} g_{\phi \phi}}= -\frac{g_{tt} \ell}{g_{\phi\phi}},\quad
\ell\equiv\frac{L}{{E}}=-\frac{g_{\phi\phi} \Omega }{g_{tt} },
\eea
and considering $u^r=0$, we obtain $V_{eff}=E$  and
\bea\label{Eq:pa-bal}
  L_K=\pm\sqrt{\frac{(\sin\theta)^2  r^2}{(r-3)}} \quad  \ell_K=\sqrt{\frac{(\sin\theta)^2 r^3}{ (r-2)^2}},\quad V_{eff}=\sqrt{\frac{-g_{tt} g_{\phi \phi}}{g_{\phi \phi} +\ell^2g_{tt}}},
\eea
where
 $L=L_{{K}}$  and  $\ell=\ell_K$  are the test particles and fluids angular momentum
  respectively, correspondent to  the  extremes for the $V_{eff}$ $(\partial_r V_{eff}=0)$ when expressed in terms of the particle momentum $L$ or the fluid specific momentum $\ell$ respectively. Energy function $K(r)$ of Eq.\il(\ref{Eq:Kdir}) can be found  considering $V_{eff}$ of  Eq.\il(\ref{Eq:pa-bal}) evaluated in $\ell=\ell_K(r)$.}

{We stress  that the function $\ell(r)$,
adopted  for the most part here as the \textbf{RAD} leading function, has a meaning for the single toroidal \textbf{RAD} component  as discussed in the previous paragraph.
Here, we reinterpret $\ell(r)\equiv\ell_K(r)$  as a tori distribution in the \textbf{RAD}, this  setting has been detailed  discussed  in \cite{ringed,dsystem,open,letter} for the \textbf{eRAD}. It follows that, because of its geometric and centrifugal origins,  function $\ell(r)$   is naturally   considered  as a possible reference distribution. (It is  then  the  boundary level for the   "Bondi condition"\footnote{
{Thick accretion
disks are characterized by a significant contribution of the centrifugal force represented by  "high" angular momentum of matter,  that is  superior or equal the Keplerian $\ell_K$ angular momentum.
The slow rotation cases are often referred to as "Bondi flows" in assonance with  the spherically-symmetric (non-rotating) accretion  \citep{Bondi}.
Angular momentum distribution $\ell_K(r;a)$ turns thus to be a natural reference function of the accretion model. More precisely, in the  reference accretion flow  known as  the   (quasi)--spherical "Bondi" accretion,  the angular momentum is not so relevant in the dynamical forces compared with others ingredients of the forces balance  and the (specific) angular momentum  in the disk  turns    smaller
than the Keplerian one. This frame is generally expanded by  assuming the  condition that an accretion  disk must have  an extended region
   where matter has a  large (fast) enough  centrifugal component ($\ell\geq \ell_K$)--\citep{abrafra}.
"Bondi accretion" in this sense can be associated to  small accretion rates (for example in the limit free fall accretion disks--slow rotation),
  in this sense fast rotation disks are for example thick disks--\cite{Bondi} see also \cite{lowcgq}.
  Thus in these  thick  tori, where the  strong gravitational field is assumed  dominant with respect to the   dissipative forces,  and a    perfect fluid energy-momentum tensor is assumed,    the specific angular momentum $\ell_K$ turns to be an upper boundary conditions with the respect to the  "Bondi condition".  These tori represent therefore a reference model  for more complex systems.}} on the fluid angular momentum in an extended region  of  any ("fast" rotating) accretion disk). Similar argument applies   for the energy function $K(r)$. Each value of the function (\ref{Eq:Kdir}) has, in fact, a role for a toroidal \textbf{RAD} disk, being directly related to the maxima and minima of the \textbf{RAD} tori effective potential $V_{eff}(a,\ell;r)$ and therefore to the  notable points of the tori topology.

\medskip

{\textbf{Critical density and pressure points in the tori}}

\medskip

According to Eq.\il(\ref{Eq:Eulerif0}), the maxima of the effective potential correspond to the minima of the HD pressure (and density because of the barotropic EoS), viceversa the minima of the potential correspond to the maxima of the HD pressure and density (the tori centers). Specifically:
\begin{description}
\item[
\textbf{(a)}] The constant values  $K(r)=K\in]K_{\mso},1[$--are associated to
\begin{description}
\item[\textbf{i.}] closed configurations, that is  to quiescent tori, if the $K< \max V_{eff}(r)<1$ i.e. $K$,  constant for each torus, is mower then the torus effective potential maximum.
\item[\textbf{ii.}] cusped tori if the $K\equiv \max V_{eff}(r)<1$ i.e. $K$,  constant for each torus, corresponds to the maximum point  of the torus effective potential. Here $K_{\mso}\equiv K({r_{\mso}})$ is the minimum point of the energy function $K(r)$--see Table\il(\ref{Table:pol-cy}). This maximum point  will be located on the cusp  $r_{cusp}\equiv r_{\times}$ of the torus Roche lobe. The  torus center, $r_{cent}$,  is a minimum of the effective potential, and  corresponds to  $K\equiv \min V_{eff}(r)<1$.
\end{description}
\item[
\textbf{(b)}] The  condition $K(r)=K>1$ does not correspond to the minima of the potential, but it  can correspond to  the maximum points. In this last  case $K\equiv \max V_{eff}(r)>1$ is associated  to the critical point,  $r_j$,  of the open surfaces cusp, associated with proto-jets. This radius is located  on
$r_j\in ] r_{\gamma},r_{mbo}[$, where   the boundary radii  of this range  are defined in  Table\il(\ref{Table:pol-cy}). These configurations have very large specific angular momentum $\ell\in]\ell_{\mbo},\ell_{\gamma}[$.
\item[\textbf{(c)}]
For $\ell>\ell_{\gamma}$, the  potential has no maximum points i.e.  the torus  cusp is suppressed by the centrifugal force of the extremely fast rotating fluid, and only quiescent toroidal configurations are possible. Typically, however, these toroids are located quite far from the central attractor i.e. $r_{cent}>r_{\gamma}^b$.\end{description} (Quantities introduced here are given in Table\il(\ref{Table:pol-cy}) and discussed below). The effective potential is widely used in literature for the description of these configurations, however we mention for   analogy with notation and conventions on signs used here
for example
\cite{ringed,Multy,letter}.}

{We should note that, although the energy function  $K(r)$  is  related to the fluid effective potentials in the \textbf{RAD} $V_{eff}(a,\ell;r)$, which is a function of the radius $r$ and depends on the parameters $(a,\ell)$,  function $K(r)$   does \emph{not} depend on  $\ell$, but  it provides,  point by point $r$,  the $K$-parameter values  correspondent to the extremes of the effective  potential for \emph{each} possible toroid in the  \textbf{RAD} agglomeration that is, as explained below, it relates configurations with different  $\ell$. The energy  function $K(r)$ can be recovered directly from the effective potential $V_{eff}$ of Eq.\il(\ref{Eq:c-i-va}) evaluated on the curve $\ell_K(r)$ in Eq.\il(\ref{Eq:lqkp}). }
Function $K(r)$ provides  therefore the \emph{distribution} of the effective  potential values corresponding to the maximum and minimum density (and HD pressure) points.
The dependence from  $\ell^2$, of the tori  characteristics,  including  morphological and stability properties introduced in Appendix\il(\ref{Sec:doc-ready})),
 is expression of the
spacetime  spherical symmetry\footnote{
From the point of view of  \textbf{RAD} model considered  in \cite{mis(a)}, independently of the relative direction of rotation and the angle of inclination of the toroids, in the static spacetime all the toroids  have characteristics similar to the $\ell$corotating pairs composing an \textbf{eRAD}  orbiting a Kerr central \textbf{SMBH} as discussed in  \cite{dsystem}.}.
For fixed  $\ell=$constant of a torus, the solution of $\ell=\ell_K(r)$ gives the torus center  $r_{cent}$ which  is also the maximum density and HD pressure points; \textbf{RAD} tori  have also equal minimum  pressure  (and  density)-points,  but not in general  equal geometric center which  depends also on  $K$. The  function $K(r)$ of Eq.\il(\ref{Eq:lqkp}) is related instead to  an independent tori parameter $K$ which  regulates the torus elongation $\lambda$ on {its} symmetry plane and  the emergence of hydro-dynamical instability. Furthermore, this is also associated  to the torus density,  the torus thickness and other characteristics related to the tori energetics, as cusp luminosity and accretion rates \cite{mis(a)}.

{Below we  use the leading function $\ell_K(r)$ and the energy function $K(r)$ of the \textbf{RAD}, constraining  its inner structure. The idea is to consider limiting points representing constraints on the toroidal components and therefore constraints  on the \textbf{RAD} as the entire system (inner structure, inner and outer boundaries, limits on instabilities, as accretion or tori collision--\cite{ringed}). Since $\ell_K(r)$ and $K(r)$ correlate different points,  we study them by introducing some functions (radii) obtained from the conditions on the  leading function. In brief,  these   represent the boundary conditions for the existence and location of   the  inner and  outer  edges, clearly regulated by the  Euler equations, or maxima and minima of the pressure (and  density in barotropic  models like tori considered here where the pressure is the hydrostatic pressure).
}

The phases of the torus evolution towards accretion  are supposed to be associated  generally to a decrease of the momentum magnitude $\ell$ and  possibly an increase  of the  $K$-parameter \citep{pugtot}.
%
Moreover, the study of $\ell(r)$ and $K(r)$ functions is also important to set constrains on the tori  collision.
Equal  $\ell$ tori, not possible in the \textbf{eRAD} frame, might be possible configurations in the \textbf{RAD} because of the  different tori   inclination angles. They  present a doubled collisional region, minimized in case of torus maximum relative inclination--$\theta_{ij}=\pm\pi/2$--coinciding with the toroidal section with $\min(K(\ell))$ and having equal maximum density points.
{We introduce therefore the two functions $r_p^{\ell}(r)$ and $r_p(r)$.}

\medskip

\textbf{{The function $r_p^{\ell}(r)$ (from the leading function $\ell_K(r)$)}}

\medskip

Condition  $\ell(r)=$constant identifies the center $r_{cent}>r_{mso}$  (maximum pressure points), where $r_{mso}$ denotes radius of marginally stable orbits, and eventually, the  instability point $r_{\times}<r_{mso}<r_{cent}$ of an cusped  torus with cusp $r_{\times}$. We can relate these radii with a function $r_p^{\ell}(r)$
\bea&&\label{Eq:schaubl}
r_p^{\ell}(r)\equiv \frac{2r}{(r-r_+)^2} \left[\sqrt{(2 r-r_{\gamma})}+r-r_{\gamma}^b\right]:
\quad
\ell_K(r)=\ell_K(r^{\ell}_p),\quad \mbox{where}\quad  r\in[r_+,r_{\gamma}^b], \quad \mbox{and}\quad r_{\gamma}^b=M
 \eea
Function  $r_p^{\ell}$ relates radii  $r_{cent}$ or {$r_{crit}(=r_{cusp})$}  for the configurations with equal $\ell$,  for closed cusped tori  the critical point $r_{crit}$ is the  accretion point $r_{\times}$ as function of the other radius of the couple. Note that the evaluation of $K(r=r_p^{\ell})$ from equation\il\ref{Eq:Kdir} provides the $K$-parameter value of the maximum pressure $r_{cent}$ or minimum  ($r_{crit}$) (if this exists), turning  in a  \textbf{RAD} parametrization  in terms of disks pressure gradients, {where the disk center is  expressed in terms of the instability point and vice versa. }
\medskip

{The location of the accretion tori edges and  center.}

\medskip

\begin{figure}
\includegraphics[width=5.3cm,angle=90]{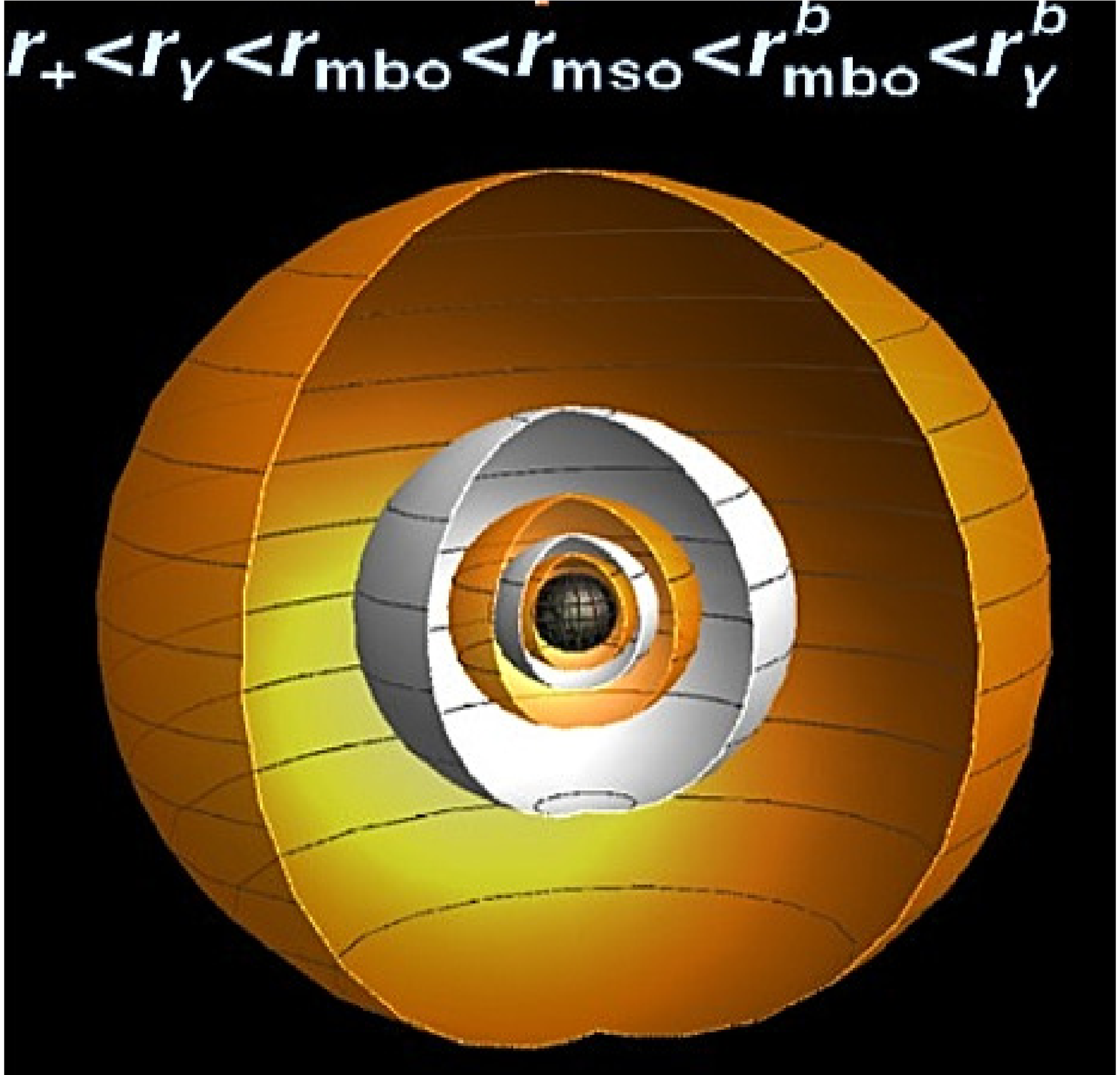}
\vspace{0.3cm}
\includegraphics[width=6.4cm]{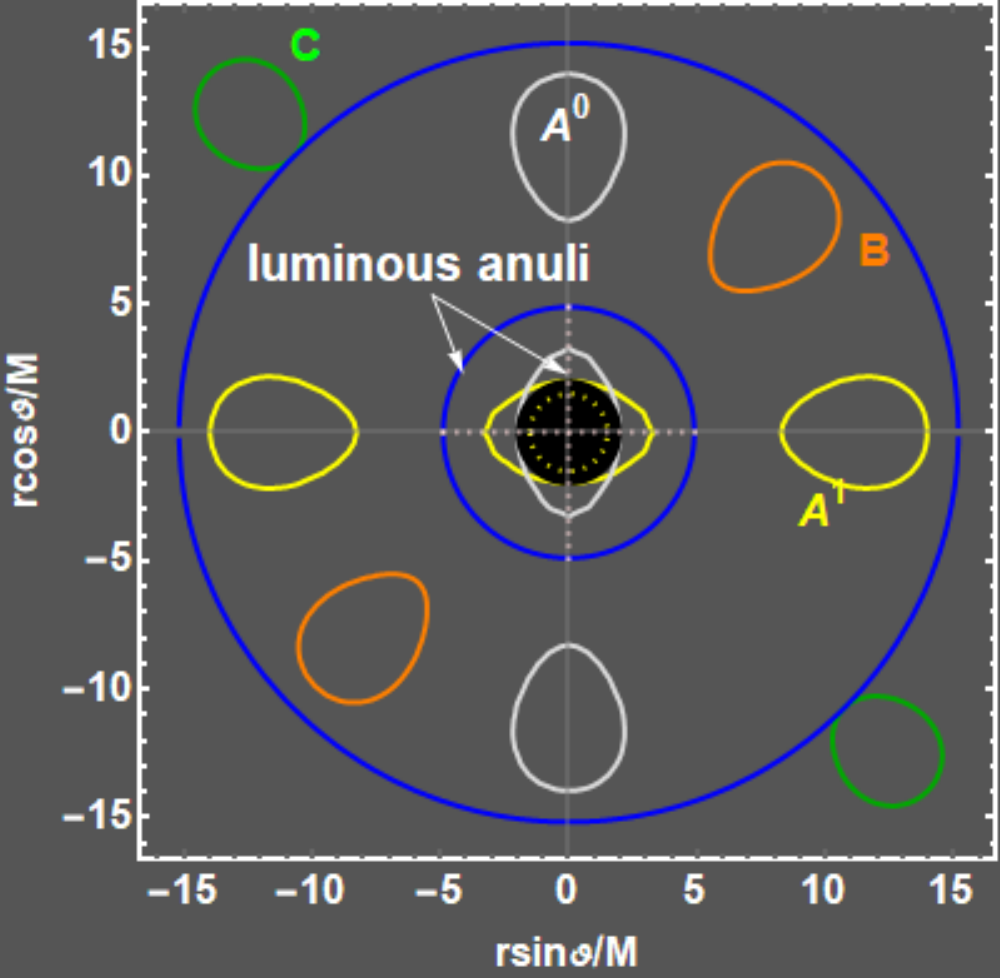}
\caption{Left panel: Stability spheres for the Schwarzschild spacetime. Black sphere is the central $\textbf{BH}$.
Cusps of accreting tori are in $\left]r_{mbo},r_{mso}\right[=\left]4M,6M\right[ $ with tori  center (maximum)
in $\left[r_{mso},r_{mbo}^b\right[ $.
Cusps of open cusped proto-jets configurations are in $ \left]r_{\gamma},r_{mbo}\right[=\left]3M,4M\right[ $, the center in $\left[r_{mbo}^b,r_{\gamma}^b\right[$. Configurations in $r>r_{\gamma}^b$ are quiescent--see discussion after equations\il\ref{Eq:schaubl}. The spheres are also the regions of existence of the accreting globules conjectured in \protect\cite{mis(a)}
   embedding the static \textbf{BH}  in the set of orbiting \textbf{RAD} tori with the \textbf{BH}  horizon  ``covered'' to an observer at   infinity, with  the \textbf{RAD}--see Figure \ref{Fig:solidy}    Right panel: Equi-pressure tori surfaces, obtained from the integration of Euler equation (\ref{Eq:Eulerif0}), with appropriate boundary conditions, for specific angular momentum $\ell=\sqrt{14}$ and $K=0.950631$ for $\mathbf{A^1}$-model at $\theta=0$, and  $\mathbf{A^0}$-model with $\theta=\pi/2$;    $\mathbf{B}$-model: $\theta=\pi/4$, $\ell=4$ and  $K=0.96$; $\mathbf{C}$-model: $\theta=-\pi/4$, $\ell=\sqrt{22}$, $K=0.973$. Configurations $(\mathbf{A}^0,\mathbf{A}^1)$ are orthogonal (crossing) accreting tori, $\mathbf{B}$ crosses with $(\mathbf{A}^0,\mathbf{A}^1)$. The tori couple $(\mathbf{A}^0,\mathbf{A}^1)$ correspond to the same point of the curves $\ell(r)$ and $K(r)$ of equations\il\ref{Eq:lqkp} and (\ref{Eq:Kdir}) respectively. Boundary spheres for the inner edge  are also shown.  Luminous anuli, corresponding to the apparent inner edge of accreting tori are also show from different  prospects. Central black region is the Schwarzschild \textbf{BH} . \label{Fig:SIGNS}}
\end{figure}

(Here we adopt notation $\mathcal{Q}_{\bullet}\equiv \mathcal{Q}(r_{\bullet})$ for any quantity $\mathcal{Q}$ and for a general radius $r_{\bullet}$, in particular $\mathcal{Q}_{\times}$ refers generally to quantities evaluated at the cusp $r_{\times}$ or  related to cusped tori.There is $r_{\gamma}=3M$, the marginal circular orbit (photon circular orbit), $r_{mbo}=4M$ (marginally bounded orbit), $r_{mso}=6M$ (marginally stable circular orbit).)
\begin{description}
\item[--]
 The proto-jet  (open cusped configurations) cusp is at $r_{crit}=r_{j}\in]r_{\gamma},r_{mbo}[$ with specific angular momentum $\ell\in [\ell_{mbo},\ell_{\gamma}]$;
  \item[--] The torus  cusp is at $r_{crit}=r_{\times}\in]r_{mbo},r_{mso}[$ (specific angular momentum $\ell\in [\ell_{mso},\ell_{mbo}]$).
 \item[--]
The center of a cusped closed configuration is at  $r_{cent} \in]r_{mso},r_{mbo}^b[$, where $r_{mbo}^b=2 \left(\sqrt{5}+3\right)M\approx10.4721M$ (solution of $\ell(r_{mbo})=\ell(r_{mbo}^b)$.
\item[--]Proto-jet configuration centers are in  $]r_{mbo}^b,r_{\gamma}^{b}]$  where  $r_{\gamma}^{b}=6 \left(\sqrt{3}+2\right)M=22.3923M$.
    \item[--]Finally configurations with center in
$r >r_{\gamma}^{b}$ are quiescent and closed, with specific angular momentum $\ell_{K}>\ell_{\gamma}$.
 Radii $r_{mbo}^{b}$ and $r_{\gamma}^{b}$ are solutions of $\ell_K(r)=\ell_{mbo}$ and $\ell_K(r)=\ell_{\gamma}$ respectively--Fig\il\ref{Fig:SIGNS}.
 \end{description}
All tori with equal $\ell$ have the same center and, eventually, same location of the cusp, therefore they orbit in the same spherical shell across the radii $r_{cent}$ and limiting\footnote{We also note that there is a maximum in the distribution of angular momentum in the \textbf{RAD}  occurring at the  radius   $r_{\mathcal{M}}=2 \left(2 \sqrt{3}+3\right)M=12.9282M\in]r_{mbo}^b,r_\gamma^b[$, affecting the maximum density of tori and playing a probable role in the formation of \textbf{RAD}.
A torus centered in   $r_{\mathcal{M}}$ has constant specific angular momentum in $\ell_{\mathcal{M}}=4.25362M\in]\ell_{mbo},\ell_{\gamma}[$,  therefore this configuration corresponds to a   quiescent  torus or   proto-jet.
 The energy function distribution, $K(r)$ function,  has a saddle point in $r_\mathcal{M}^K=8.079M$ with  $K_\mathcal{M}^K=0.948996M$,  maximum point of the gradient  $\partial_r K(r)$. This  point is related to the density within each torus of  the agglomerate. {These extremes can therefore represent indications on a further constraint regarding the region of formation of the ringed disks--see discussion in \cite{dsystem}}} $r_{\times}$.
On different planes (different polar $\theta$ angles), two tori, $T_1$ and $T_2$, having  equal specific angular momentum $\ell$ but  with different  {inner and outer edges}   radii, $r=r_{in}$ and $r=r_{out}$, are in  equal center  $r_{cent}$ spheres where
$r_{in}^1<r_{in}^2 <r_{cent}<r_{out}^2<r_{out}^1$, that is, they are concentric. Location of  $(r_{in},r_{out})$ is determined by the $K$ function.
For quiescent (not cusped)  $T_i<T_o$ (i.e. $T_i$ is the torus closest to the central \textbf{BH})
   there is  $\ell_i< \ell_o$ and $r_{out}^i<r_{in}^o$.
Considering then that cusped  configurations  are   fixed  by   the   $\ell$ parameter only, where
 for  $\ell_i < \ell_o$,  there is
$r_{\times}^o < r_{\times}^i<r_{cent}^i<
 r_{cent}^o$  and  it is  $r_{out}^i<r_{out}^o$. Therefore
the outer  cusped  torus "incorporates"  the inner cusped torus, similarly  to the \textbf{eRAD} case. Nevertheless, the  tori inclination can reduce, for hight relative inclination angles, up to the limiting  orthogonal case, the  flow of impacting accreting material
from an outer to the inner torus of the agglomerate. The  tori misalignment can effectively  reduce  the  collisional effect.

\medskip

\textbf{{The function $r_p(r)$ (from the  energy function $K(r)$)}}

\medskip

Similarly to the considerations leading to the  function $r_p^{\ell}$  of equation\il\ref{Eq:schaubl},   the condition $K(r)=K(r_p)$, determines a  radius  $r_p(r)$: %
\bea\label{Eq:Krrp}
&&
r_p(r)\equiv r_{mbo}\left[\frac{{x}+1}{{x}}\right],
\quad \mbox{where}\quad {x}\equiv r-r_{mbo},\quad
K(r)=K(r_p)\equiv K_p,
\eea
{Condition $K(r)=$constant, here considered as $K(r)=K(r_p)$, determines the radii $r_p\neq r$, connecting  two different tori, $T_1$ and $T_2$,  with   $K_{cent}(T_1)=K_{\times}(T_2)$ i.e. the $K$ parameter (related to  tori density) evaluated  at the center (maximum density point) of torus $T_1$ coincides with the $K$ parameter evaluated  at the cusp (minimum density point) of torus $T_2$. (For a  cusped  torus,  condition  $K(r,\ell_{cent})=K(r_p)$ identifies the  cusp $r_{\times}$ and the   torus center, and moreover also the  \emph{outer edge} $r_{out}^{\times}$ of the accreting tori when we fix $\ell$ for each solution). }

The  limit $\lim\limits_{r\rightarrow \mathcal{Q}}r_p=\mathcal{K}$ holds  connecting the couples
\bea (\mathcal{Q},\mathcal{K})=\{(r_+,r_+),(r_{\gamma},0),(0,r_{\gamma}), (r_{mbo}, +\infty),(+\infty,r_{mbo}),(r_{mso},r_{mso}), (r_{mbo}^b, r_{mbo}^K),(r_{mbo}^K, r_{mbo}^b),(r_{\gamma}^b,r_{\gamma}^K),(r_{\gamma}^K,r_{\gamma}^b)\}\eea
{ correlating  the notable points of the \textbf{RAD} constraints. It is clear that the points are symmetrically correlated,  the fixed point of the transformation represented by the function $r_p(r)$ are, as expected, the horizon $r_+$ and  the marginally stable orbit  $r_{mso}$ (limit of  torus with center, maximum density point, $r_{cent}\gtrapprox r_{mso}$ and negligible elongation and pressure being
approximated to a free dust particles string). The couples $(r_{mbo},\infty)$ is particularly notable,  considering that $r_{mbo}$  is  the marginally bounded  orbit for test particle and limit for inner edge of the torus  ($r_{in}\eqslantgtr r_{mbo}$) to the open proto-jet configuration (with no closed outer boundary)\footnote{Of course, in the presence of repulsive cosmological constant there is naturally also the outer limit  on tori extension located  at $r_{mso}^o$  called  the static radius of the spacetime \citep{1983BAICz..34..129S,[68],2000A&A...363..425S,compar} }.}
 Function $r_p$ identifies  a pair of tori  ($T_1$, $T_2$) with  $K_{cent}(T_1)=K_{crit}(T_2)=K(\hat{r})$  where $r_p=\hat{r}=$constant.  Here we intend with  $r_{crit}$  possibly $r_{\times}$ where $K\in K_{\max}\in]K_{mso},1[$, or $r_j$ where $K\geq1$,
 with  radii $r_1>r_2$ satisfying the conditions $K(r)=K(r_p)$.  There is    $r_p\in[r_{mbo},r_{mso}]$ for the  cusped tori and
$r_{mbo}^k=\left(\sqrt{5}+7\right)M/2\approx4.61803M$ such that $K(r_{mbo}^k)=K(r_{mbo}^b)$, and   $r_{\gamma}^k={6}\left(\sqrt{3}+6\right)M/{11} \approx4.21748M$ such that $K(r_{\gamma}^k)=K(r_{\gamma}^b)$.
%
%
%

At fixed $\ell$, the  torus    reaches its  maximum elongation   $\lambda_{\times}$ on the symmetry plane for  cusped surface. The outer tori  have  larger magnitude of the  specific angular momentum   leading  in general  to  a larger elongation $\lambda$. Further notes on the {morphological constraints on  \textbf{RADs} stability} can be found in Appendix\il(\ref{Sec:doc-ready}), as well as exact expression of many characteristic functions governing properties   of the tori, discussed in \cite{mis(a)}.
\section{Relating  twin peak quasi-periodic oscillations
with  RADs structure}\label{Sec:epi}
\begin{figure*}
  \includegraphics[width=4.cm]{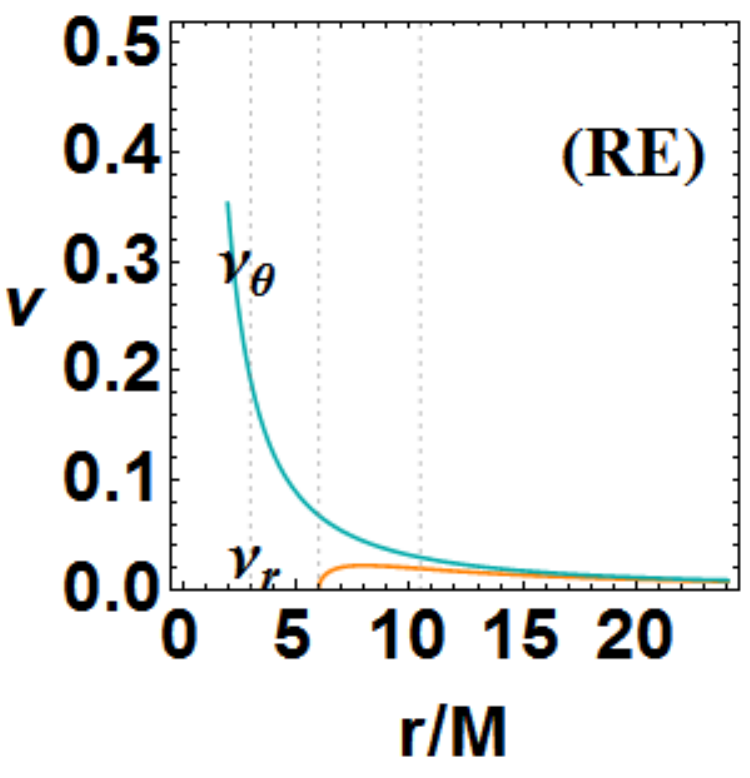}
  \includegraphics[width=4.5cm]{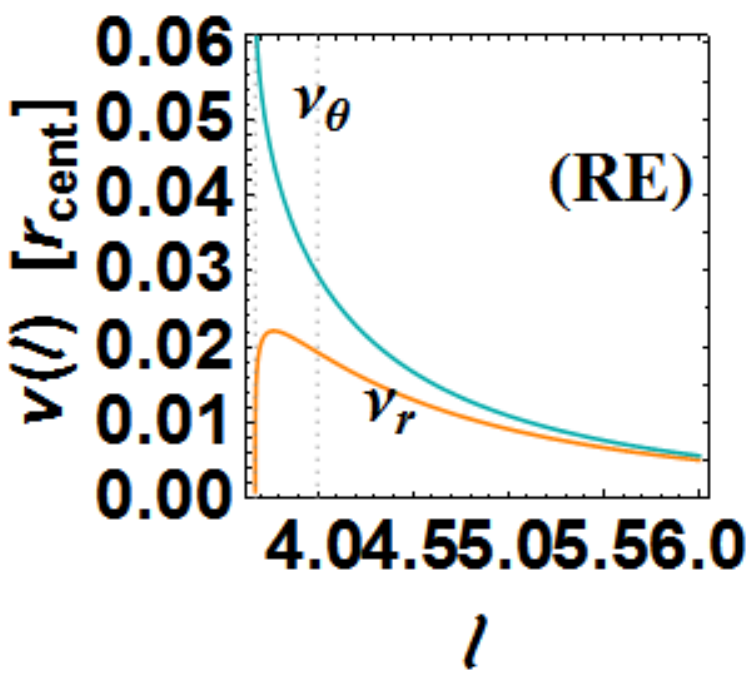}
  \includegraphics[width=4.5cm]{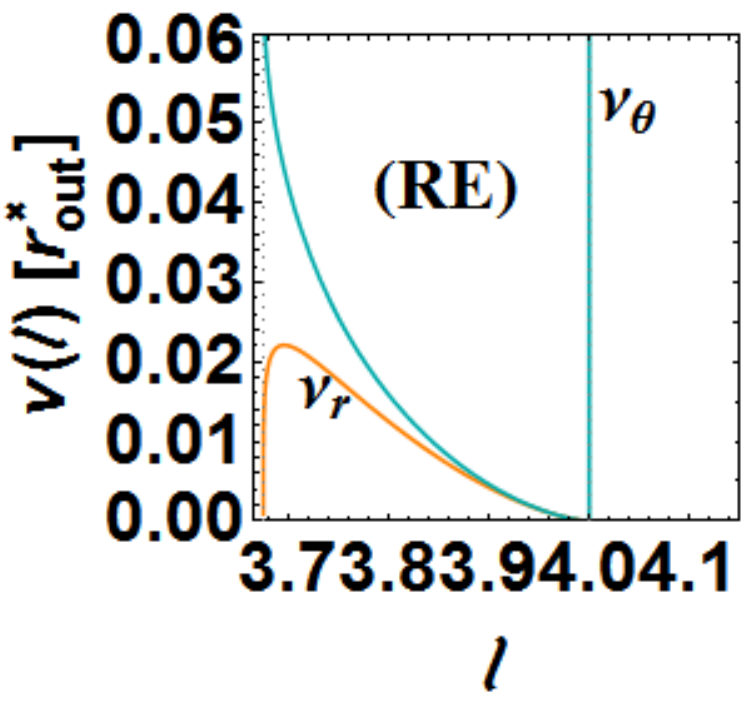}
  \includegraphics[width=4.5cm]{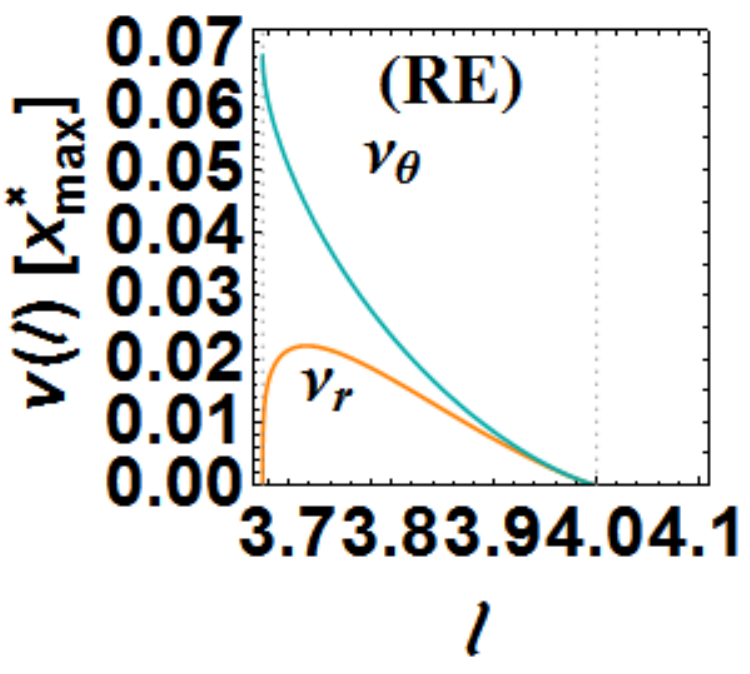}\\
  \includegraphics[width=4.5cm]{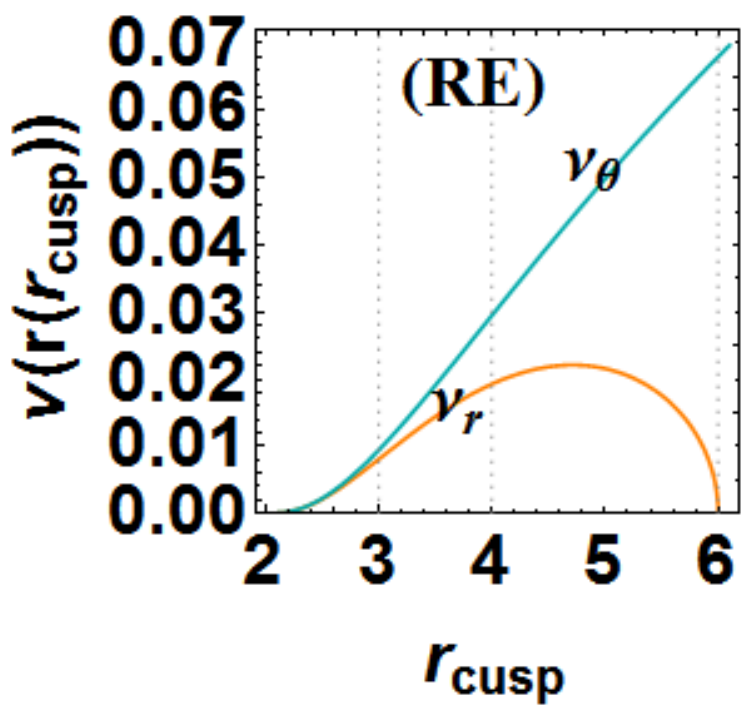}
  \includegraphics[width=4.5cm]{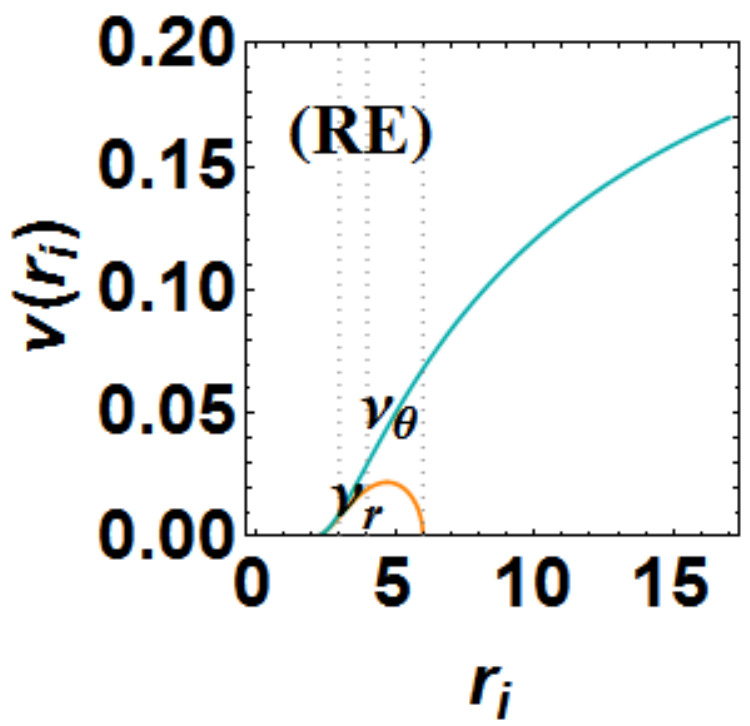}
  \includegraphics[width=4.5cm]{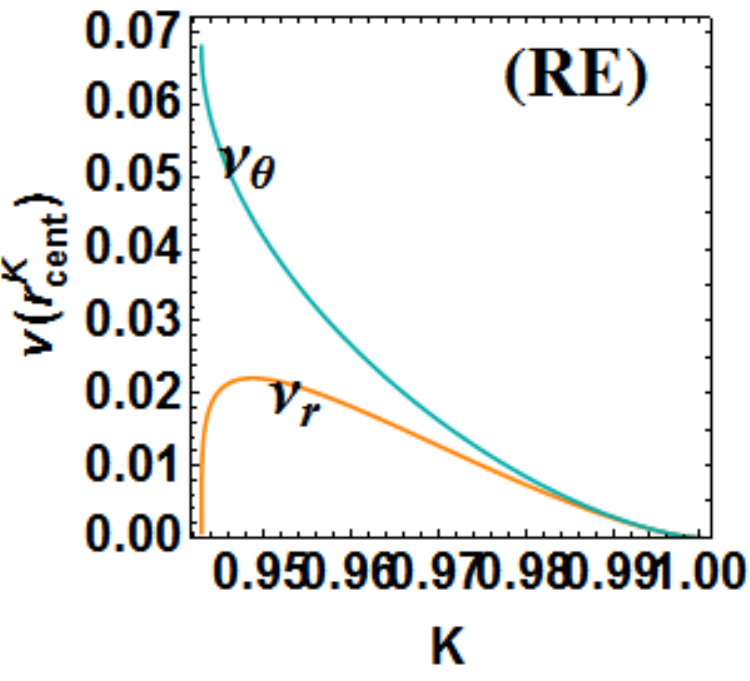}
  \caption{Radial profiles of the  radial and vertical epicyclic frequencies
$\nu_r(r)$  (orange curve) and $
\nu_\theta(r)=\nu_K(r)$(darker cyan curve)
of equations\il\ref{Eq:frenurnurthe}, and  (\textbf{RE}) frequencies for  simple resonance epicyclic  models where  $\nu_U =\nu_{\theta}$ and $\nu_L = \nu_r$, see Figures\il\ref{Fig:RE} for different tori models. Radii $r_{mso}=6M$, $r_{mbo}=4M$, $r_{\gamma}=3M$ and  $r_{mbo}^b\approx10.4721M$ as the fluid specific angular momentum $\ell_{mso}$ and $\ell_{mso}$ are shown. Upper left panel \textbf{(a)-model} $ (\nu_{\theta},\nu_{r})$ as functions of $r/M$. Second panel: \textbf{(b)-model} $ (\nu_{\theta},\nu_{r})$ for $r=r_{cent}(\ell)$ in equation\il\ref{Eq:rcentro} as function of the fluids angular momentum $\ell\in[\ell_{mso},\ell_{mbo}]$. Third panel: \textbf{(c)-model} $ (\nu_{L},\nu_{U})$ for $r=r_{out}(\ell)$ in equation\il\ref{Eq:over-top} as function of  $\ell$. Fourth panel:  \textbf{(d)-model} $ (\nu_{L},\nu_{U})$ for $x^{\times}_{\max}(\ell)$ in equation\il\ref{Eqs:rssrcitt} as function of  $\ell$. Fifth panel: \textbf{(e)-model}
$ (\nu_{L},\nu_{U})$
for $r_{\times}^{\varepsilon}$ of equation\il\ref{Eq:worrre-stowo} as function of $r/M$.  Sixth panel: \textbf{(f)-model}
 $ (\nu_{L},\nu_{U})$
for $\bar{r}(r_i)$ in equation\il\ref{Eq:ririoutcenter} as function of  $r/M$. Seventh panel: \textbf{(g)-model}
 $ (\nu_{L},\nu_{U})$
for $r=r_{cent}(K)$ in equation\il\ref{Eq:nicergerplto} as function of  $K$.}\label{Fig:vsound}
\end{figure*}

The aim of this  section is to investigate the possibility that the twin peak quasi-periodic oscillations (QPOs) could be  related to the \textbf{RAD} inner structure,  linking therefore the \textbf{RAD} discrete  geometry  to the QPOs emission\footnote{{QPOs have been also related to tilted disks as consequential to Lense--Thirring effect. In  AGN accretion for example or in the binary (X-ray) systems. The mechanism in brief is as follows:
 a  tilted accretion disks  orbiting    Kerr {\textbf{BH}s} undergoes a   break in its  central part for  Lense--Thirring effect \citep{Pict-to}. Aa s consequence of this, the  disk is   splitted into different planes, giving raise to  QPOs emission. Clearly the break would depend on the tilt angels (and the \textbf{BH} spin).}}.
The \textbf{RAD} and the  \textbf{eRAD} tori   are characterized by a special and distinctive ringed structure that, as  pointed out  in \cite{ringed,long,dsystem},  could be evidenced in  the $X$-ray emission spectrum, and as an imprint of the discrete inner \textbf{RAD}  composition  in the combined  oscillatory phenomena  associated to the  tori models.
Here we exploit this conjecture, considering  the \textbf{eRAD} and \textbf{RAD} case studying the  expected  epicyclic frequencies in the case of ringed structures.
The idea is that the discrete structure of a \textbf{RAD} could be related  to  QPOs  emission associated to the accretion torus structure, particularly with  the respect to the inner edges $r_{in}$
of (quiescent or) accreting torus\footnote{Eventually in the aggregate \textbf{RAD} model considered here we might   consider a full thick tori model for the   oscillation  analysis including dependence  on the outer edge $r_{out}$ and the maximum point $y_{\max}$ of the torus surface. On regards of the  particle model approximation of the oscillations considered  here, we note  that in this model  the inner margins of the closed surfaces are related to constat and zero pressure point.}.
As for the \textbf{eRAD} discussed in \cite{ringed,open}, the \textbf{RAD} is characterized by  relation between  $r_{in}/r_{out}$  (or elongation $\lambda=r_{out}-r_{in}$) and  height  $ y_{\max} $.  Tori spacing $\bar{\lambda}=r_{in}^o-r_{out}^i$ (for inner $T^i$ and outer $T^o$ tori) is strongly dependent on the background geometry.
  More specifically, the issues we  address in the frame of the  QPO-\textbf{RAD}  hypothesis are \textbf{(1)} the  interpretation of multiple combined signals as  expression of the ringed structure,  \textbf{(2)} the recognition of the role of each torus edges and of  the \textbf{RAD} internal structure in the emission frequencies. However,  in this analysis  we propose  a first comparative investigation on the problem of the QPO interpretation in the \textbf{RAD}  frame, we also recognize that
 this hypothesis  should be discussed by   considering   properly disko-seismology effects for each toroid, which is in many aspects dependent on its geometrical features
\footnote{{A non trivial  topic in this discussion  concerns the actual role  of the disk morphology in the disk oscillation and emission, both in the most specific sense of the \textbf{RAD} morphology, which  we address  extensively,  and more generally in each   disk  component. This issue regards the disk  oscillations and the disk  emission spectra. Here we have dealt with this argument through the evaluation of $\beta$ geometric thickness parameter, which is a relevant aspect of the  disk perturbation theory. However, a second aspect concerns the disk as extended object. More precisely, we could say that   oscillations originate in the entire disk, but it may be  reasonable to  assume that their traces (for example as QPOs) would be evident  in the emitted radiation that is usually associated  to the disk more active part which is, in many senses, the  disk boundary (and particularly the disk inner edge).  It is clear that this specific issue turns to be  extremely relevant in the ringed composite structure of \textbf{RAD} misaligned tori (which  has lead also to the idea of luminous anuli), where  each torus boundary is significant in the model. In this respect  the  accreting flow, especially in the case of  cusp (Roche lobe) overflow, can have a great relevance.}}.
 Therefore, a part of this section is also dedicated to an evaluation of the impact of the disk geometry (specifically its thickness) in this analysis.
Here, we mainly focus on the so called geodesics  oscillation models \citep{2013A&A...552A..10S} analyzing the  radial profiles and  assuming  specific oscillation models of the \textbf{RADs} constituents.
%
%
%
In the
test particle limit, the
frequencies of the epicyclic oscillations  in the Schwarzschild spacetime  are \cite{2013A&A...552A..10S}
\bea\label{Eq:frenurnurthe}
\nu_r(r)=\nu_K(r)\sqrt{1-\frac{r_{mso}}{r}},\quad
\nu_\theta(r)=\nu_K(r)\equiv\frac{1}{r^{3/2}},
\eea
 shown in Figs\il(\ref{Fig:vsound}) as associated to different tori models. Furthermore, we consider also
the modifications of the geodesic oscillation models due to the tori structure{\footnote{
 Notice that the frequencies are given as dimensionless, also the radius is constructed dimensionless ($r=r/M$) being expressed in units of mass parameter $M$. In order to obtain frequency in standard units of $Hz=1/s$ we have to use the correction factor $c^3/2\pi GM$.}}.

\begin{figure*}
  \includegraphics[width=5cm]{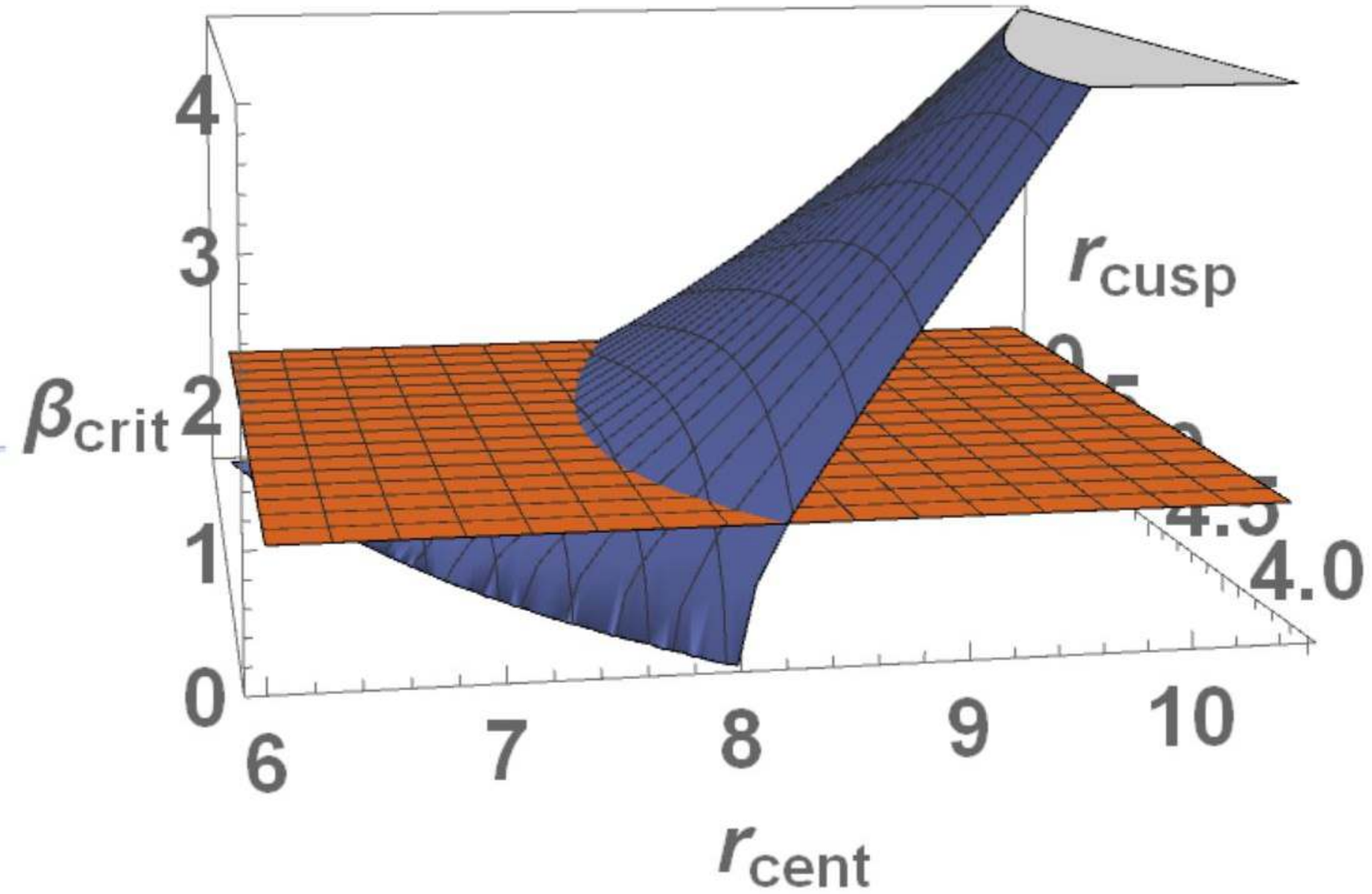}
   \includegraphics[width=6.cm]{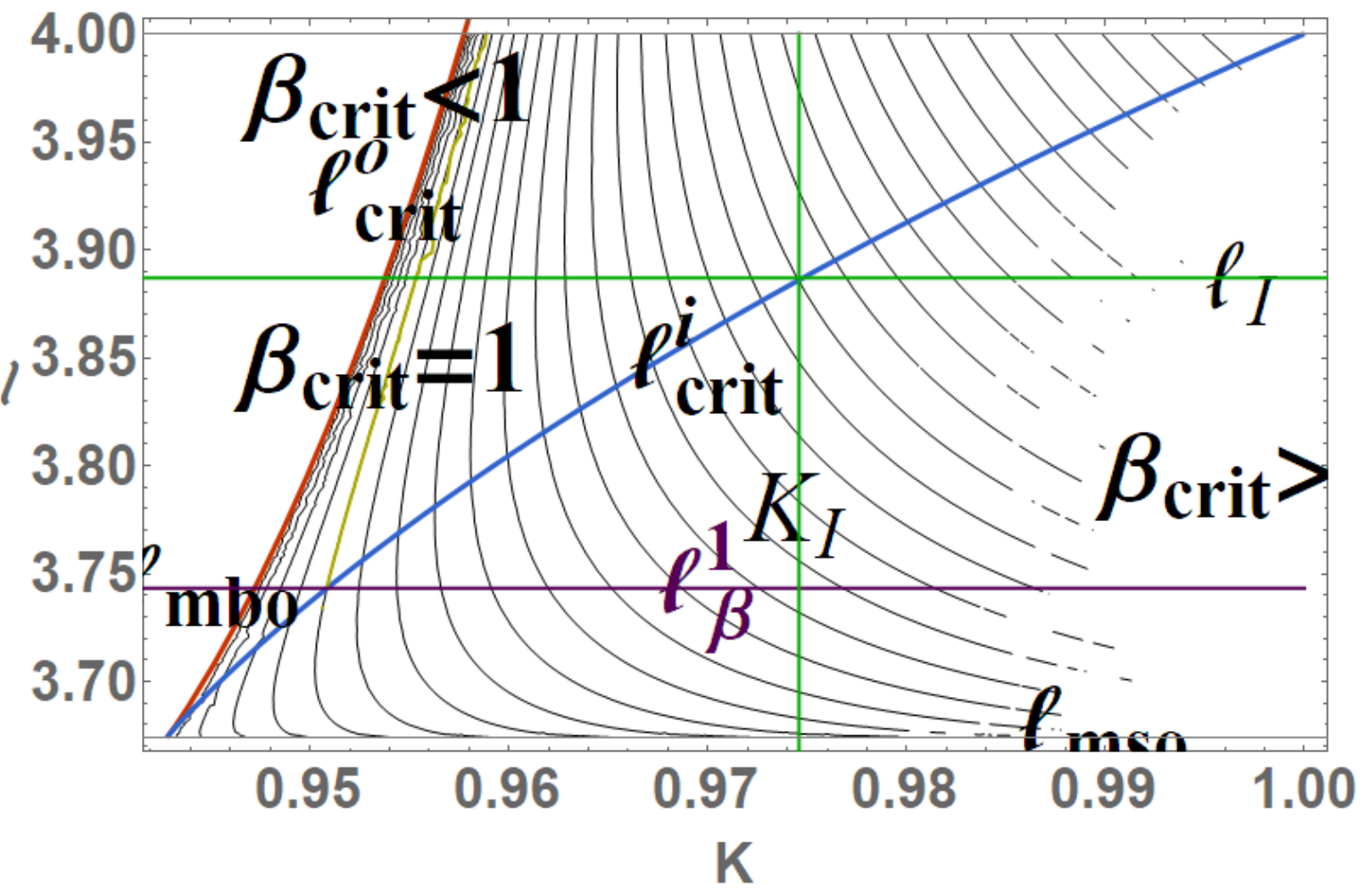}
     \includegraphics[width=5.6cm]{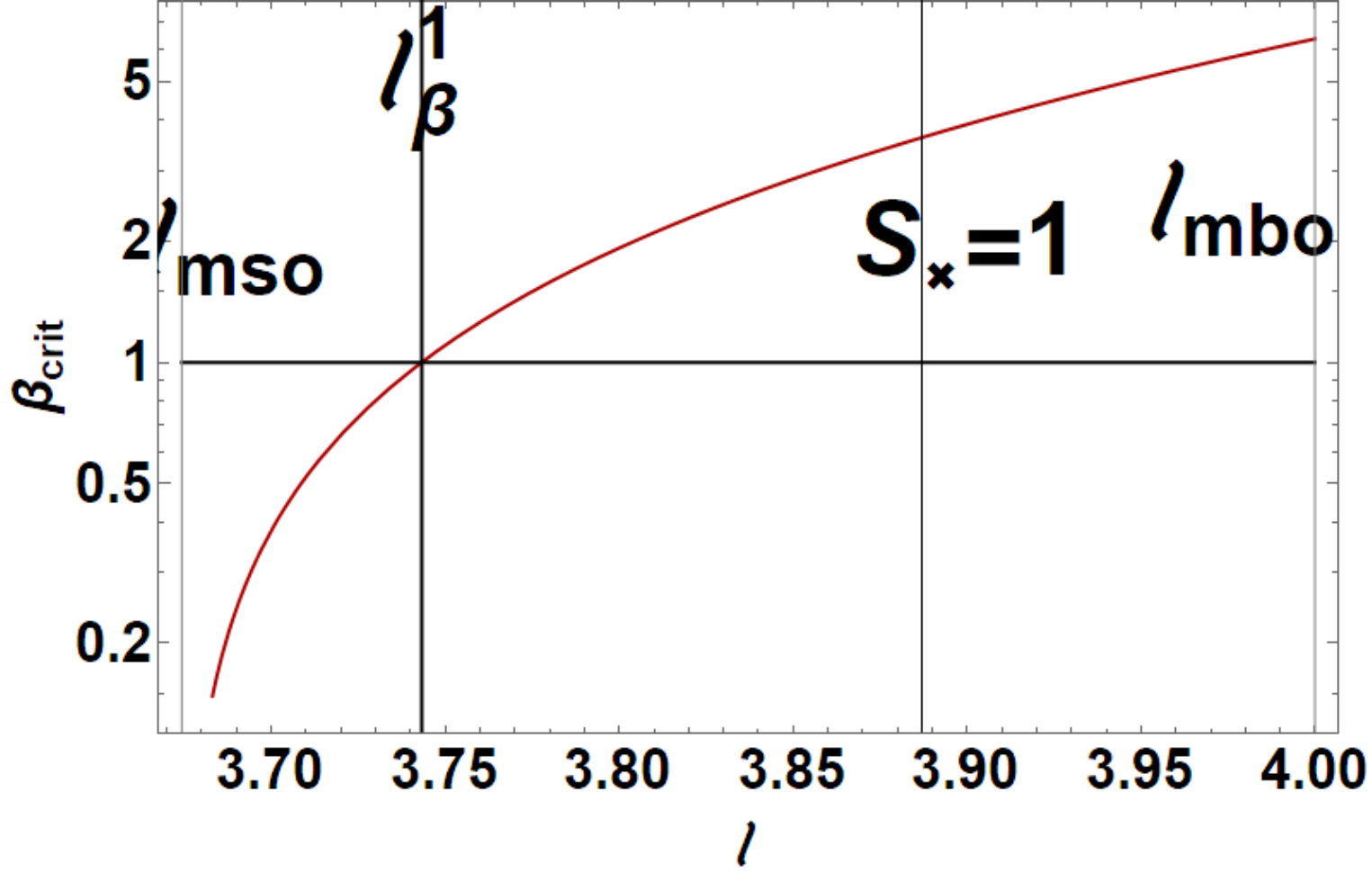}\\
   \includegraphics[width=8.cm]{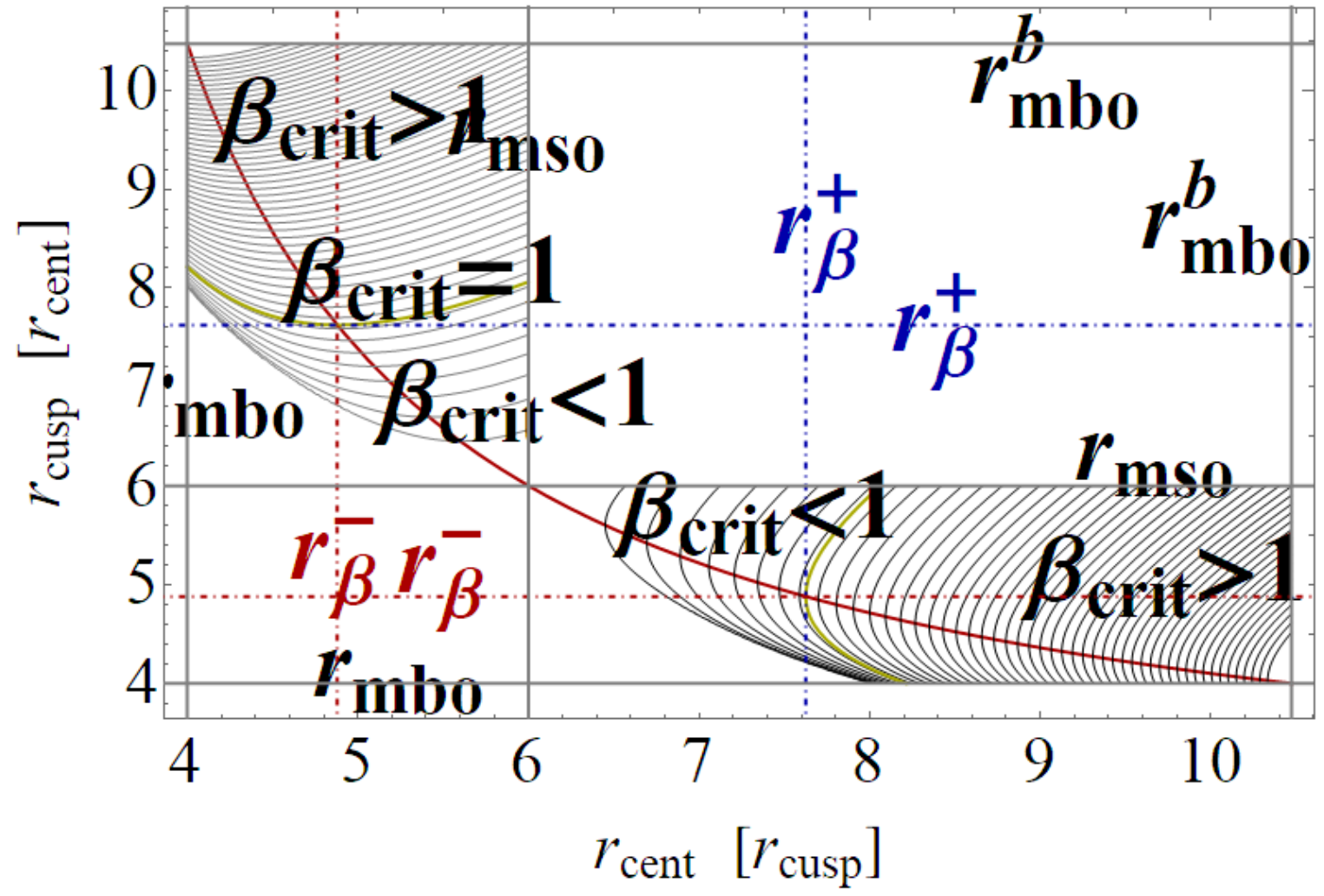}
     \includegraphics[width=8cm]{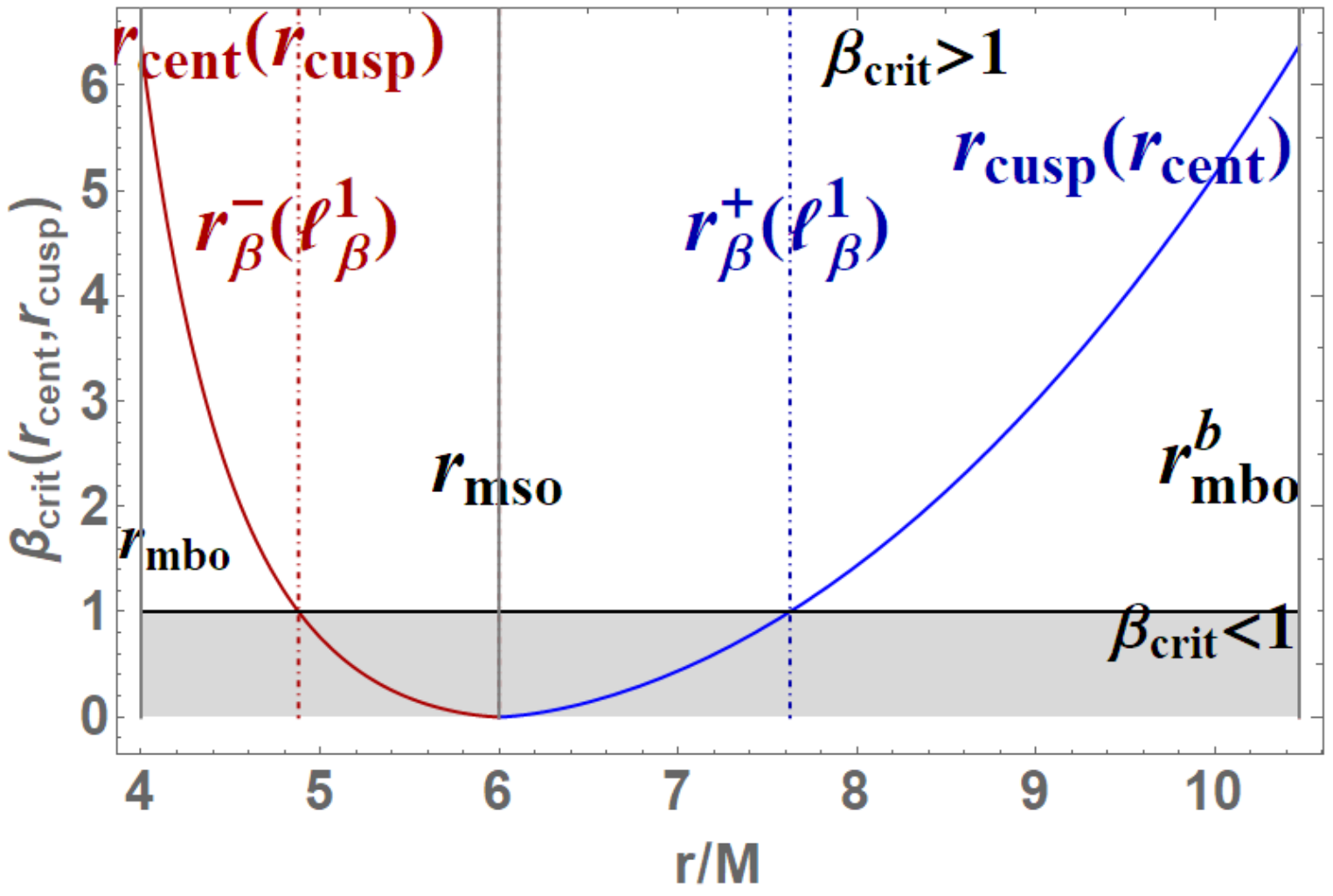}
  \caption{Study of \textbf{RAD} $\beta_{crit}$ in equation\il\ref{Eq:betacrittico}. Upper panels: \emph{left-} $\beta_{crit}$ as function of the torus center $r_{cent}$ and cusp $r_{cusp}$, the limiting value $\beta_{crit}=1$ is also shown. \emph{Upper right}: curves $\beta_{crit}=$constant in the plane $(\ell,K)$  for $r_{cent}(\ell)$ in equation\il\ref{Eq:rcentro} and $r_{cusp}=r_{in}(\ell,K)$ in equation\il\ref{Eq:outer-inner-l-A1}.  This plot shows therefore  sets of tori with equal $\beta_{crit}$.   The curve $\beta_{crit}=1$ distinguishes  tori with equal  $\beta_{crit}>1$  from  tori with  $\beta_{crit}<1$. Limiting $\ell_{crit}^o$ (red curve) and also $\ell_{crit}^i$ as functions of $K$ are defined in equations\il\ref{Eq:crititLdeval}. $\ell_I$ and $K_I$ are parameter values for torus thickness $\Sa=1$ (and $\Sa_{\times}=1$). \emph{Upper right panel}: $\beta_{crit}$ as function of the fluid specific angular momentum $\ell$, where we  made use of $r_{cent}(\ell)$ and $r_{cusp}(\ell)=r_{\times}(\ell)$  of equation\il\ref{Eq:rcentro}.  The discriminant value $\beta_{crit}=1$ is also shown this occurs for the torus with fluid specific angular momentum $\ell=\ell_{\beta}^1$ with cusp in $r_{\beta}^-$ and center in $r_{\beta}^1$. The value of the fluid specific angular momentum for the cusped torus with geometrical thickness  $\Sa_{\times}=1$. \emph{Bottom left  panel:} red curve is the radius $r_i=r_{cent}$ ($r_{cusp}$) as function of $r_{cusp}$  ($r_{cent}$) defined in equation\il\ref{Eq:ririoutcenter}. Radii $r_{mso}$, $r_{mbo}$ and $r_{mbo}^b$  are also shown. Curves are tori families in the plane $(r_{cusp},r_{cent})$ with equal $\beta_{crit}$ defined in the classes $\beta_{crit}>1$ and $\beta_{crit}<1$ with the discriminant curve $\beta_{crit}=1$ (dark-yellow) . \emph{Bottom right panel}: $\beta_{crit}$ as function of $r_{cusp}$ (red curve) and $r_{cent}$ (blue curve). In  $\beta_{crit}$ of  equation\il\ref{Eq:betacrittico} we made use of $r_i=r_{cent}$  or $r_{cusp}$ respectively defined in equation\il\ref{Eq:ririoutcenter}. The regions $\beta_{crit}>1$ and $\beta_{crit}<1$ are shown, the discriminant $\beta_{crit}=1$ is shown with radii $r_{\beta}^{\pm}$. Role of $r_{mso}$, $r_{mbo}$ and $r_{mbo}^b$ is also clear.
}\label{Fig:PlotGold}
\end{figure*}

\textbf{The torus  geometrical  thickness and the ``$\beta$-parameter'' }
In the case of a thick  disk,  the role of the toroidal geometry  and especially its  geometrical thickness is predominant  in the resolution of the oscillatory problem.
For this purpose  we evaluate  the  dimensionless
 $\beta_{crit}$ parameter for \emph{cusped tori} of our \textbf{RAD}   models
  %
\bea\label{Eq:betacrittico}
\beta_{crit}=\frac{(r_{cent}-2)^2 (r_{cent}-r_{cusp}) \sqrt{r_{cent} r_{cusp}-2 (r_{cent}+2 r_{cusp})}}{r_{cent} \sqrt{r_{cent}-3} r_{cusp} \sqrt{r_{cusp}-2}},
\eea
introduced in  \cite{2016MNRAS.457L..19T}, and derived in the more general  form
 in \cite{Straub&Sramkova(2009)} (see also \cite{Abramowiczetal.(2006)}) that    characterize the size of the torus. This quantity is roughly proportional to the flow  Mach number at the torus center and  to the ratio of the
radial (or vertical) extension of the torus to its central radius (i.e., as pointed out in \cite{2016MNRAS.457L..19T}, in this situation the sound-crossing time and the dynamical
timescale of the torus are  similar). We consider the parameter $\beta_{crit}$ within the constraints discussed in  Appendix\il(\ref{Sec:doc-ready}): then there is   $r_{cent}\in]r_{mbo},r_{mso}[$ and $r_{cusp}\in]r_{mso},r_{mbo}^b[$ (note that $\beta_{crit}$,  specified in  \cite{2016MNRAS.457L..19T},  should be considered for  $r_{cent}\lesssim10.47M\approx r_{mbo}^b$ having $\beta = \beta_{\infty}$.)
The frequencies have been therefore related to  the  epicyclic oscillations  with the
radial and vertical epicyclic modes describing eventually a global
motion of the torus--\cite{Abramowiczetal.(2006)}.
In fact it  has been shown then that for
 slender tori having   $\beta \approx 0$, frequencies of these tori modes,
$\nu_r$ and $\nu_\theta$, as measured in the fluid reference frame are equal to
 the  test particles geodesic,  epicyclic frequencies.
In a  more realistic case, where there is $\beta>0$, the relevant  pressure gradients in the disk force balance are expected to  induce a frequencies shift (i.e. for  non-slender tori the  epicyclic modes frequencies are modified by pressure).  As perturbations are generally of the order of  $\beta^2$, we will evaluate this parameter in the various tori models.
Thus, according to Eq.\il(\ref{Eq:betacrittico}) the QPO frequencies will  depend  on the location of  center and cusp of the torus  and its thickness (here the $\beta_{crit}$ parameter). We also compare in  Fig.\il\ref{Fig:PlotGold}  the  $\beta_{crit}$ parameter for toroids considered here with thickness  $\Sa_{\times}=2h_{\times}/\lambda$ as defined in Appendix\il(\ref{Sec:doc-ready}). To start with  we recall that the   disk elongation and height are maximum for accreting configurations (cusped toroids) and in general these  quantities grow  with $\ell$ and $K$. The thickness $\Sa_{\times}$ is higher than 1 only for in particular ranges of the model parameters highlighted in Appendix\il(\ref{Sec:doc-ready}). Moreover, the location of the disk center and cusp depend on the  model parameters  $\ell$ and $K$ and, as clear from the analysis in Appedix\il(\ref{Sec:doc-ready}) and Sec.\il(\ref{Sec:Misal}), there can not be two toroids with the same cusp, or  with the same center, this fact  constitutes a way to  strongly   relate  QPOs and  \textbf{RAD} structure.
 There can be  however more toroids  with equal geometrical thickness $\Sa_{\times}$ (and eventually $\beta_{crit}$), thus  relating  not uniquely  but to more tori  models the   $\beta_{crit}$ regulating the oscillation modes. (We should  then also consider that the  \textbf{eRAD} is a geometrically thin  disk even  with   geometrically thick components.)
In \cite{mis(a)} we have studied  this possibility extensively, and in  particular considering the limiting  case of the curve in the plane $(\ell,K)$ for toroids with  thickness $\Sa_{\times} = 1$, discriminating  geometrically  thick disks from geometrically thin disks.
Specifically, we have represented the situation for the cusped tori (thickness of the disk in accretion).
The $\beta_{crit}$ parameter  in Eq.\il(\ref{Eq:betacrittico}) has been studied  in Fig.\il\ref{Fig:PlotGold} considering  $(r_{cent}, r_{cusp})$ in equations\il\ref{Eq:rcentro},\ref{Eq:outer-inner-l-A1} and
 equations\il\ref{Eq:crititLdeval},\ref{Eq:ririoutcenter}.
Finally, we note that we take full advantage of \textbf{RAD} symmetry in the static case and consider
toroids oscillation on each symmetry plane. We are actually considering the problem for misaligned \textbf{RAD} tori in static spacetime as \textbf{eRAD}, nevertheless we expect that the  combination  of the oscillator models for toroids  could  depend on the  tori inclination angles  $\theta_{ij}$.
For a discussion on QPO in  titled accretion disks we mention \cite{QPOTILTER1,Banerjee:2018jlz}.

Then, there is $\beta_{crit}(\ell)=1$ for  $\ell=\ell_{\beta}^1\equiv3.7432<\ell_{I}$, where   $\ell_I=
3.887\in]\ell_{mso},\ell_{mbo}[$ and $K_I=
0.975$  define the $\Sa=1$ case. Note that  curve
$\ell(r)=\ell_{\beta}^1$ provides also the two radii $r_{\beta}^-=4.87956M$ (a cusp) and $r_{\beta}^+=7.6259M<r_{mbo}^b$  (a center) for a cusped torus, corresponding  to $\ell=\ell_{\beta}^1$  such that $\beta_{crit}(r_{\beta}^-)=\beta_{crit}(r_{\beta}^+)=1$ having expressed  $r_{cent}$ as function of $r_{cusp}$ and, viceversa, $r_{cusp}$ as function of $r_{cent}$ in Eq.\il\ref{Eq:betacrittico}. This relation between the torus  center and the cusp can be written in  a compact form as
\bea\nonumber
&&
\bar{r}(r_i)=\frac{2 \left(\sqrt{r_i^2 (2 r_i-3)}+(r_i-1) r_i\right)}{(r_i-2)^2},
 \quad\mbox{for}\quad r_{i}\in[r_{mbo},r_{mbo}^b]\quad\mbox{and}\quad \bar{r}(r_i)\in[r_{mbo},r_{mbo}^b],
\mbox{where}\quad r_{i}=r_{cusp}\in[r_{mbo},r_{mso}]\\\label{Eq:ririoutcenter}
&&  \mbox{for}
\quad \bar{r}(r_i)=r_{cent}\in[r_{mso},r_{mbo}^b],\quad \mbox{and}
\quad
r_{i}=r_{cent}\in[r_{mso},r_{mbo}^b] \quad \mbox{for}\quad  \bar{r}(r_i)=r_{cusp}\in[r_{mbo},r_{mso}]
\eea
as shown in  Figs\il\ref{Fig:PlotGold}. For convenience we also report to the following relation
\bea\label{Eq:worrre-stowo}
r_{\times}^{\varepsilon}\equiv\frac{r_{cent} \left(\sqrt{(2r_{cent}-3)}+1\right)^2}{(r_{cent}-2)^2}
\eea
which is actually a specialization of $r_p^{\ell}$ of Eq.\il\ref{Eq:schaubl} and
${\bar{r}(r_i)}$, Eq.(\ref{Eq:ririoutcenter}).
It is also clear from Figure\il\ref{Fig:PlotGold} that  \textbf{RADs} toroidal components have a prevalent $\beta_{crit}>1$, and essentially it is  $\beta_{crit}\leq1$ for $r\in[r_{\beta}^-,r_{\beta}^+]$ where $r_{\beta}^\pm$ correspond  to $\ell=\ell_{\beta}^1$. (Note that for a cusped torus, where $(r_{cent},r_{cusp})$  are related by equation\il\ref{Eq:ririoutcenter}, it is  then  $\beta_{crit}=0$ only in the limiting case of $r_{cent}=r_{cusp}=r_{mso}$.)

\medskip

\begin{figure*}
  \includegraphics[width=4.0051cm]{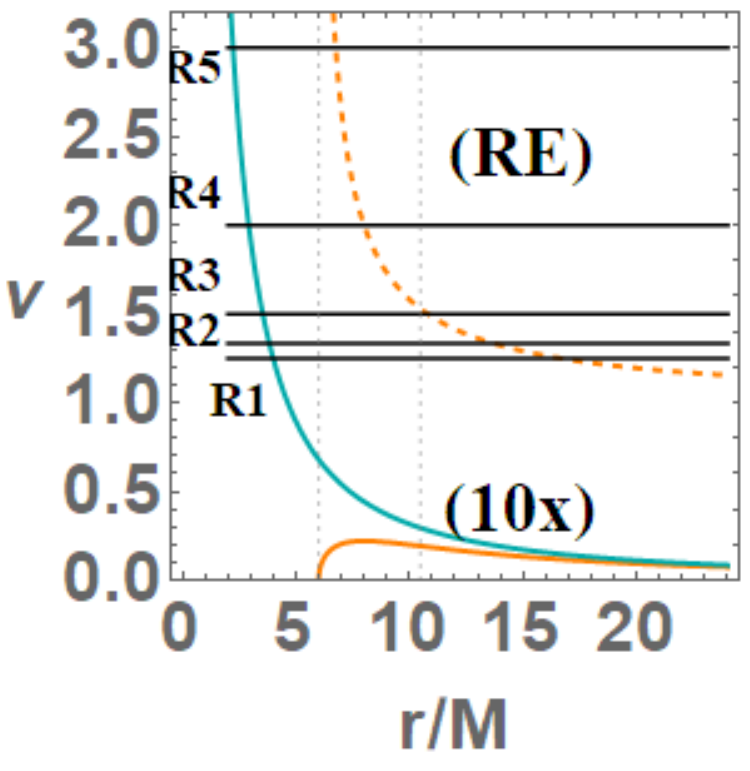}
  \includegraphics[width=4.5cm]{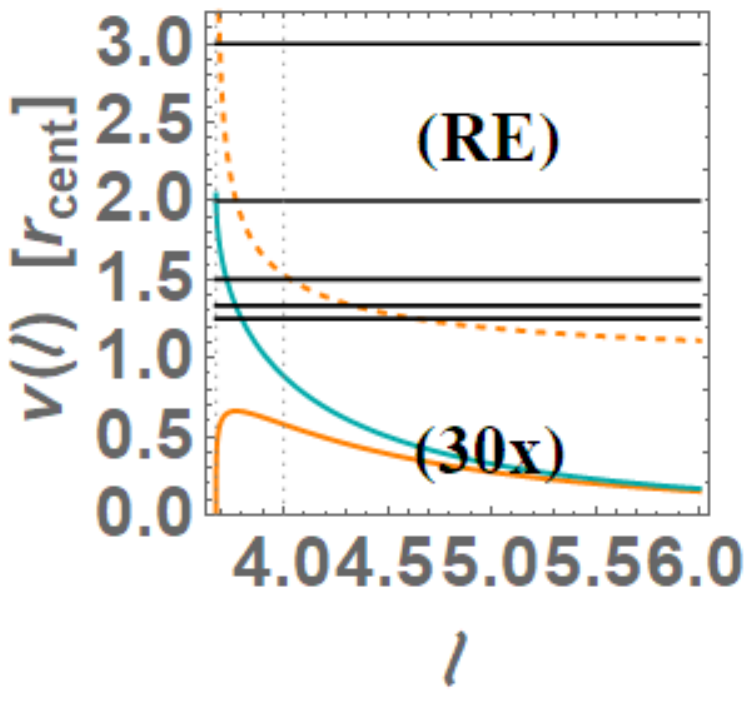}
  \includegraphics[width=4.5cm]{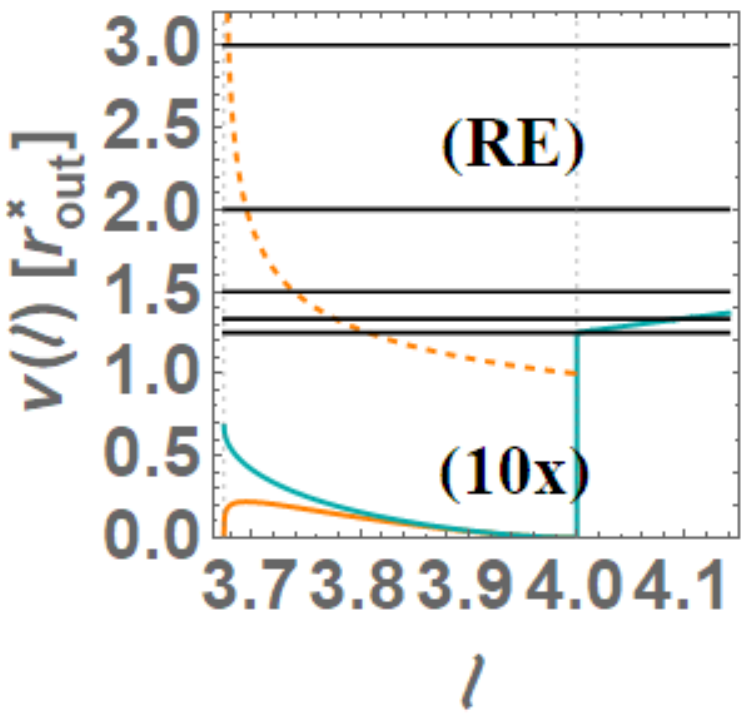}
  \includegraphics[width=4.5cm]{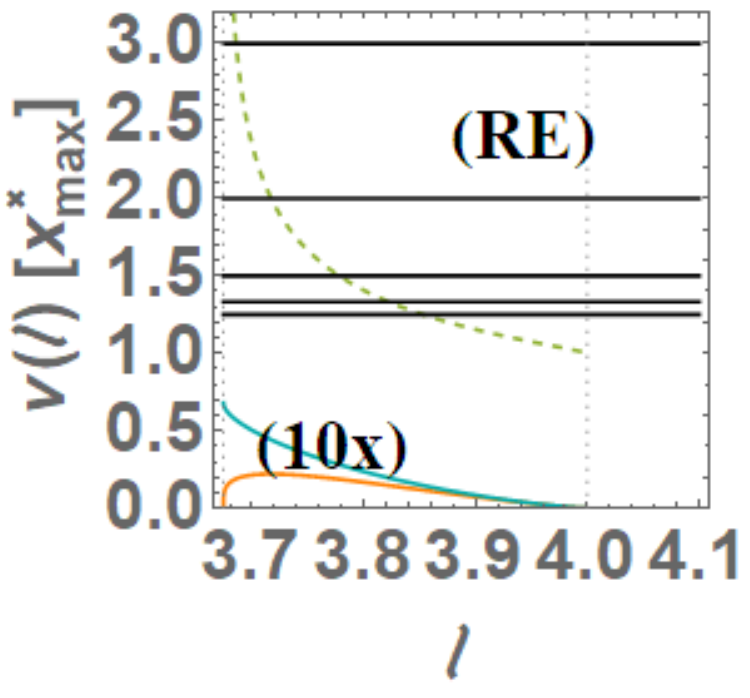}   \\
  \includegraphics[width=4.5cm]{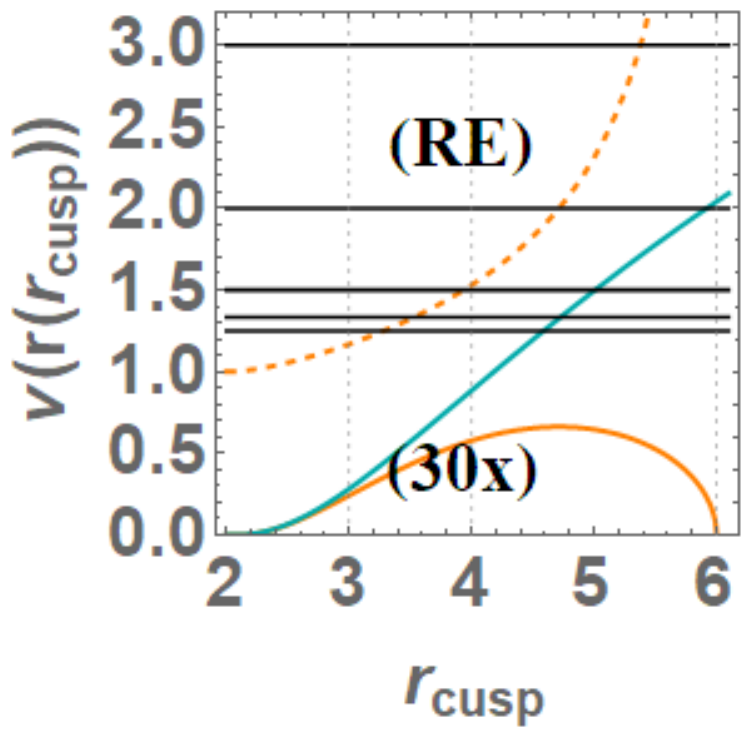}
  \includegraphics[width=4.5cm]{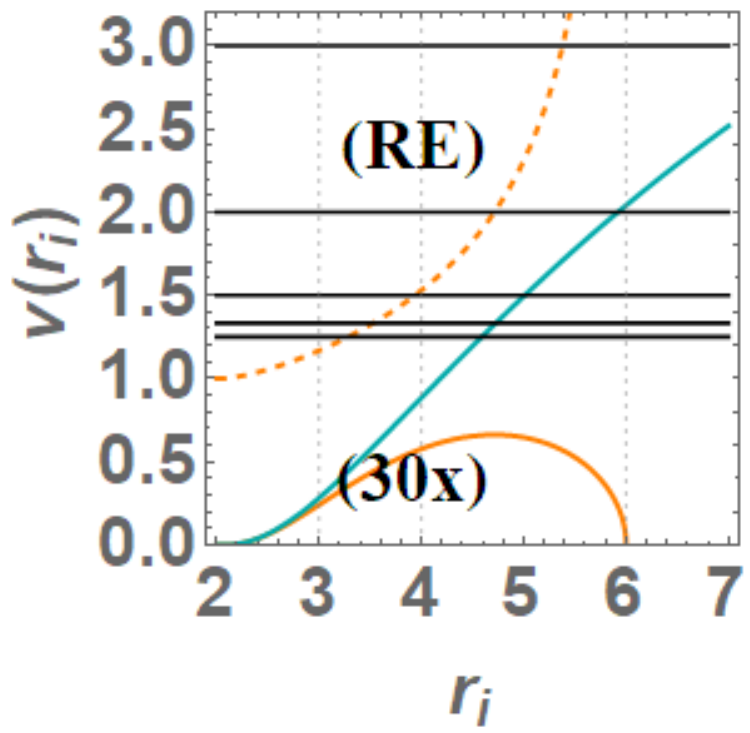}
  \includegraphics[width=4.5cm]{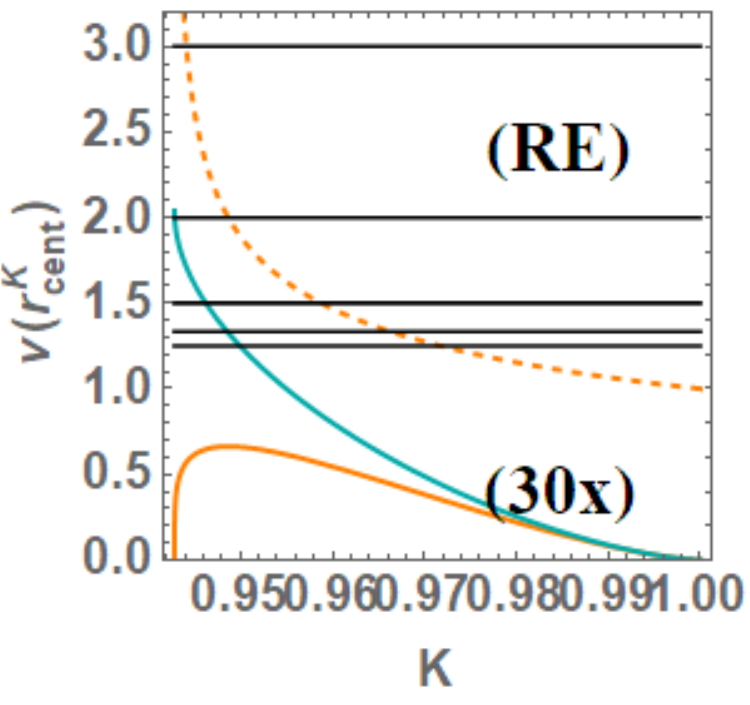}
  \caption{(\textbf{RE})-models (simple resonance epicyclic  models). Plot of $\nu_U/\nu_L$ (dashed line) in the  \textbf{(a,b,c,d,e,f,g)} models of Figures\il\ref{Fig:vsound}. There is $\nu_U =\nu_{\theta}$ and $\nu_L = \nu_r$ .  The two frequencies are also represented increased by a factor  $\mathbf{(n x)}$ present in parentheses (where $n$ is generally $\{5,10,30\}$), in dashed (orange and cyan) curves. Resonant frequency ratios $\mathbf{R1}=2:1$, $\mathbf{R2}=3:1$, $\mathbf{R3}={3}:{2}$, $\mathbf{R4}={4}:{3}$, $\mathbf{R5}={5}:{4}$ (black lines) are also shown.   See Figures\il\ref{Fig:vsound} for the radial  frequencies profiles. }\label{Fig:RE}
\end{figure*}

\textbf{Frequency models-relations of the HF QPO models}
We now focus on models that consider main epicyclical  frequencies, especially we refer to the analysis of  \cite{2017AcA....67..181S,2016A&A...586A.130S,2007A&A...463..807S,
2005A&A...437..775T,
2017A&A...607A..69K,
2016ApJ...833..273T,
2015A&A...578A..90S,
2013A&A...552A..10S,
2011A&A...531A..59T,2011A&A...525A..82S,2008CQGra..25v5016K,2007A&A...470..401S}.
Here we want to test the \textbf{RAD} as a frame for  QPO models  assuming  the geodesic frequencies governed by the background geometry  but determined by the constraints imposed on the \textbf{RAD}. Each  frequency model (\textbf{TD},\textbf{RP},\textbf{RE},\textbf{WD}) we consider  is borrowed  from a specific context from which they are derived including  slender tori and  hot spot models (assuming radiating spots in thin accretion disks)--we refer to the mentioned literature for  further details on these models relevant for both accreting systems orbiting a \textbf{BH} or a neutron stat.
Here we provide, for each of the considered (geodesic) oscillation models, only the frequency relations corresponding to the twin high-frequency oscillations, giving the upper  $\nu_U$ and lower $\nu_L$ frequency of the pair in terms of the radial and vertical oscillations and their combinations with the azimuthal  frequency-for more details see \cite{2013A&A...552A..10S,2016A&A...586A.130S}.
Results of our analysis are shown in
 Figures\il\ref{Fig:TD},\ref{Fig:RE},\ref{Fig:RP},\ref{Fig:WD}, under the assumption $\beta\approx0$ (slender tori) while we leave the case of  thick torus  with $\beta_{crit}\neq0$ to future analysis\footnote{Note that this analysis is in fact independent by the fluid equation of state, nevertheless as noted in \cite{2016MNRAS.457L..19T}, when   $\beta> 0$ (non zero  thickness), there is the possibility that, in some cases, they also depend on  the polytropic index.}.
Particular attention is given to recognize the emergence of the twin HF QPOs with resonant frequency ratios  $\nu_U/\nu_L={{3:2}}$,
${{4:3}}$ and  ${{5:4}}$ or ${{2:1}}$ and ${{3:1}}$ assumed in \textbf{BH} systems.
For the evaluation of the underlying circular orbit,  we use  in Figures\il\ref{Fig:TD},\ref{Fig:RE},\ref{Fig:RP},\ref{Fig:WD} the analysis of the  inner edge and centers of   the toroid in Appendix\il(\ref{Sec:doc-ready}).
 Figure\il\ref{Fig:vsound} shows the radial profile of the  radial and vertical epicyclic frequencies
$\nu_r(r)$ and $
\nu_\theta(r)=\nu_K(r)$
of Eqs\il\ref{Eq:frenurnurthe} in different tori models.
\begin{figure*}
  \includegraphics[width=4cm]{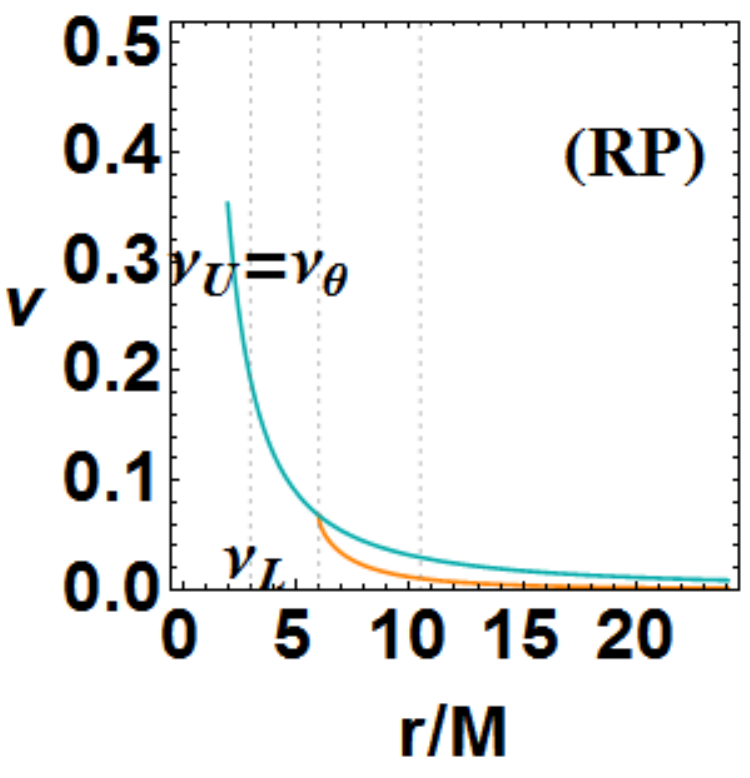}
  \includegraphics[width=4.5cm]{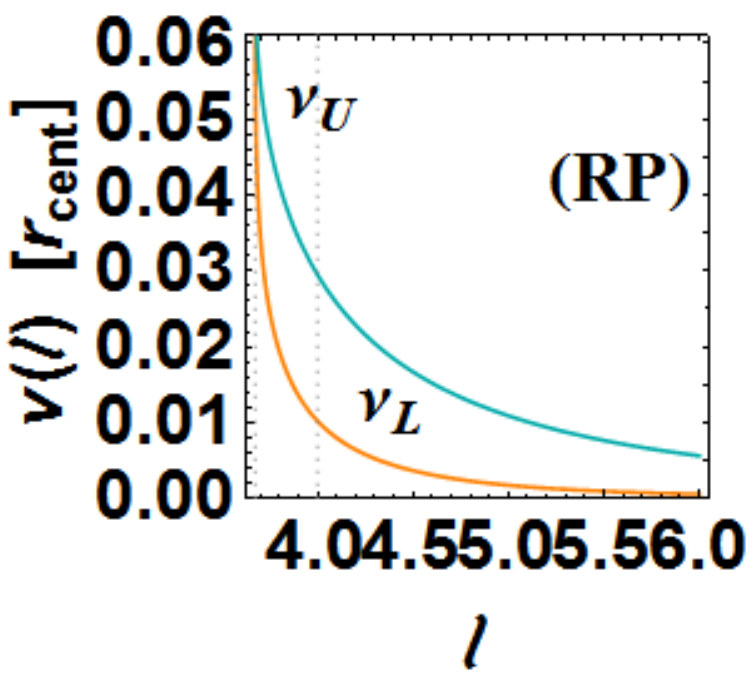}
  \includegraphics[width=4.5cm]{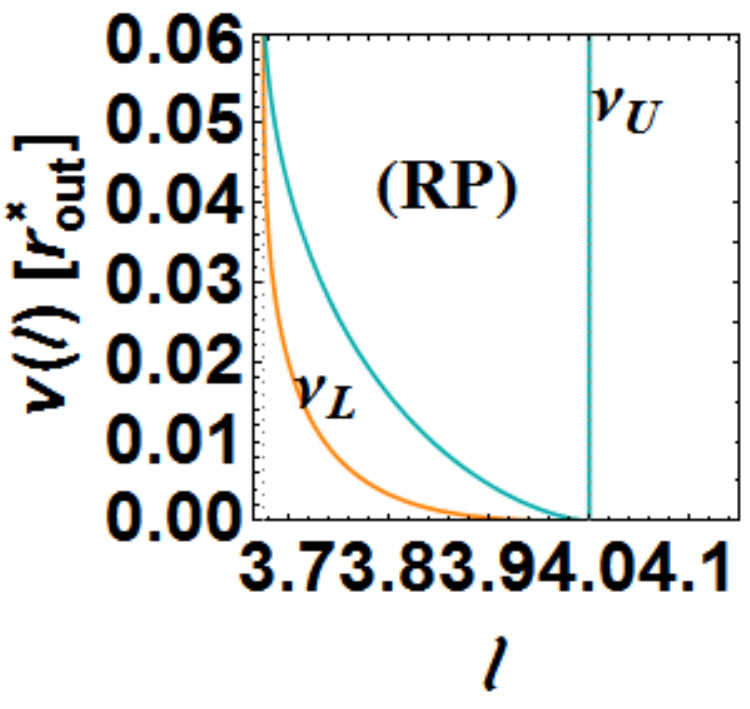}
  \includegraphics[width=4.5cm]{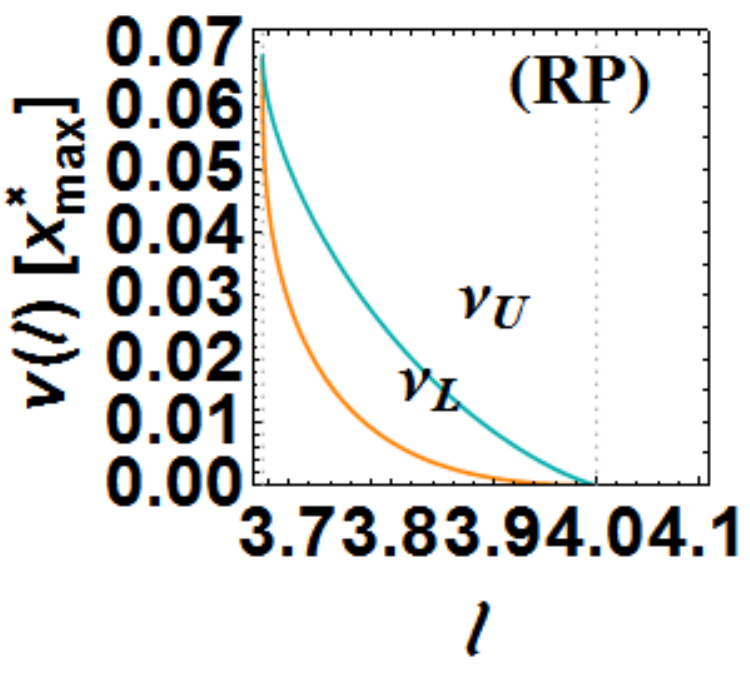} \\
  \includegraphics[width=4.5cm]{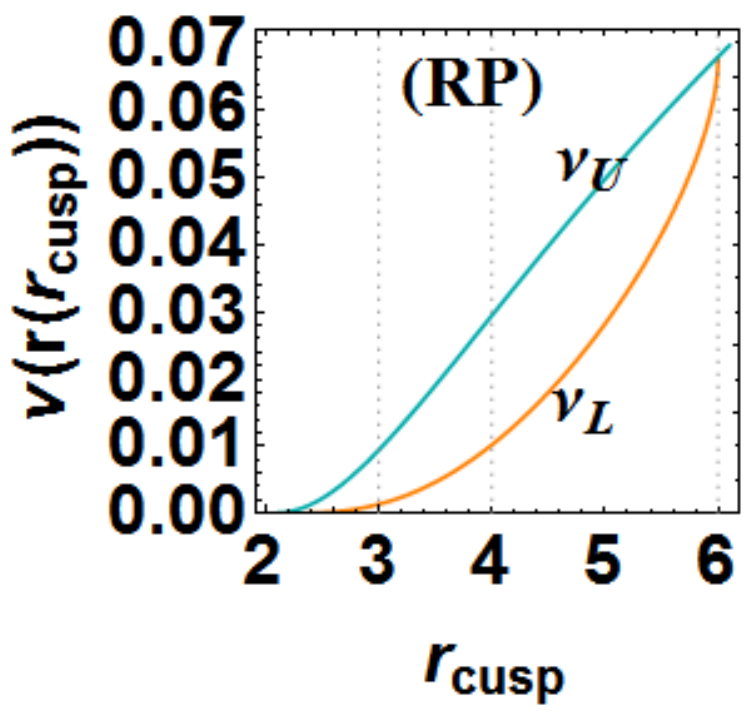}
  \includegraphics[width=4.5cm]{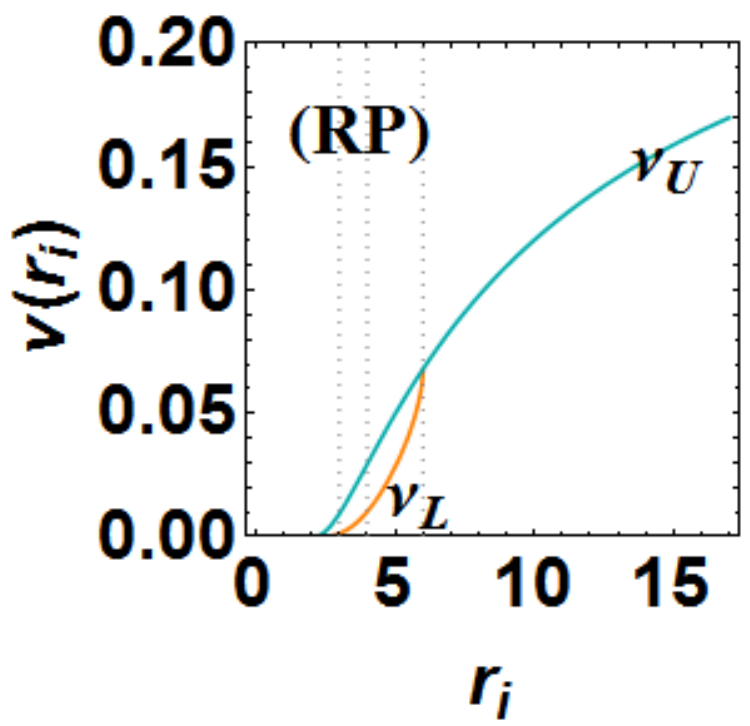}
  \includegraphics[width=4.5cm]{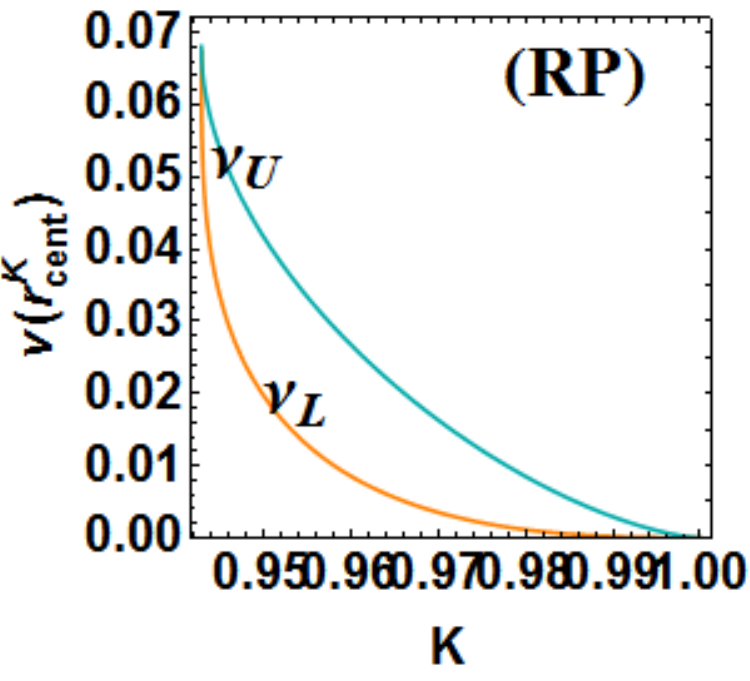}
  \caption{Frequencies $\nu_U= \nu_K$ (\emph{upper}) and $\nu_L= \nu_{per} \equiv \nu_K - \nu_r$ (\emph{lower})   for the \textbf{RP}, relativistic-precession, model, in the  \textbf{(a,b,c,d,e,f,g)} tori  models of Figures\il\ref{Fig:vsound}. The two frequencies in dashed (orange and cyan) curves are   increased by a factor present in parentheses. }\label{Fig:RPvsound}
\end{figure*}
\begin{figure*}
  \includegraphics[width=4cm]{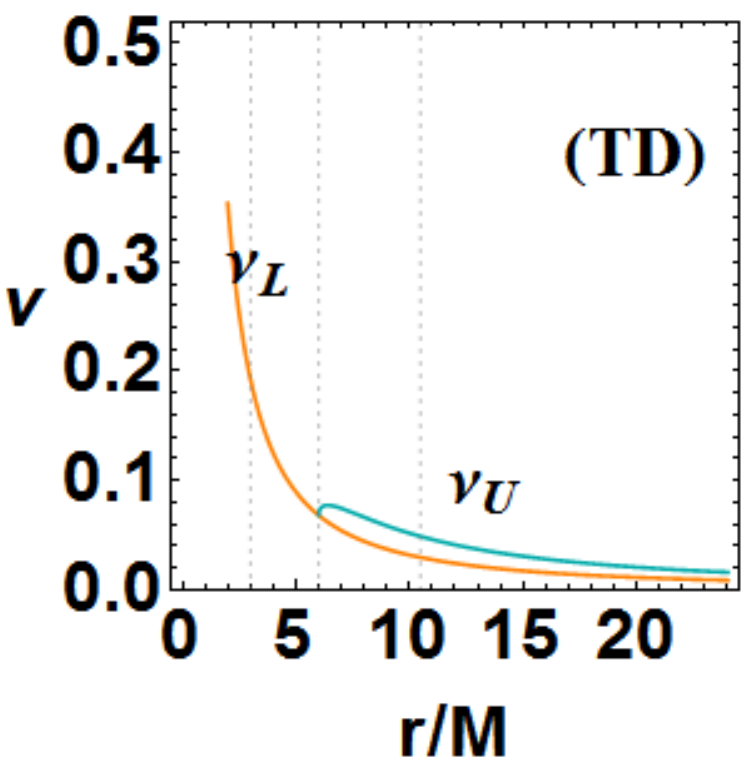}
  \includegraphics[width=4.5cm]{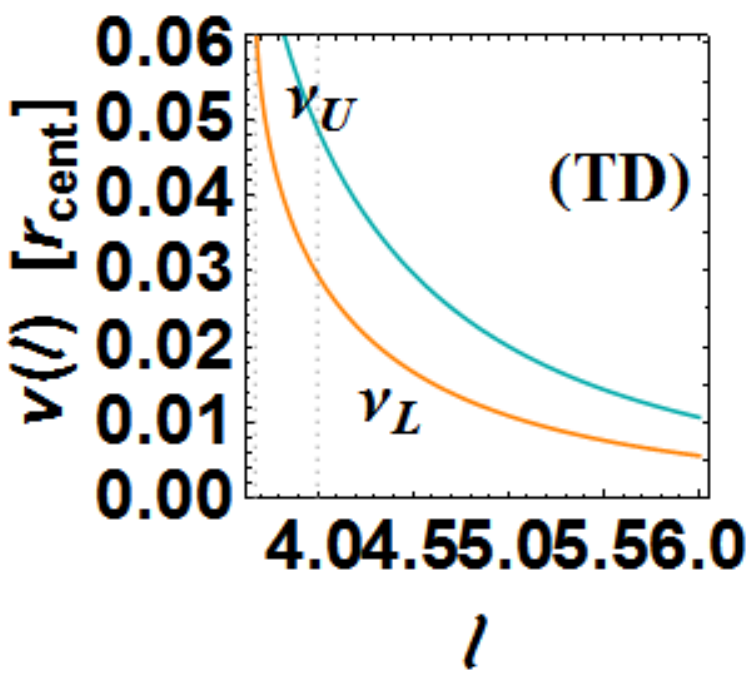}
  \includegraphics[width=4.5cm]{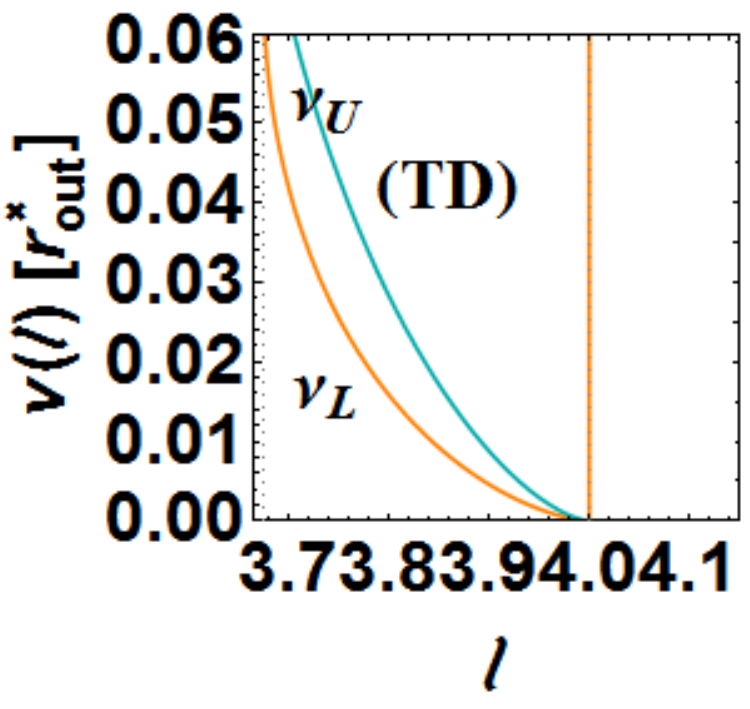}
  \includegraphics[width=4.5cm]{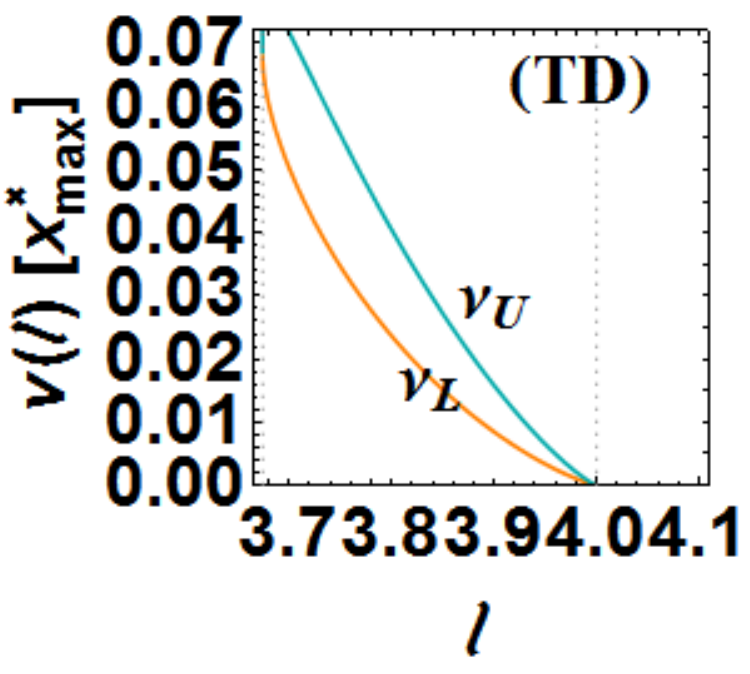}\\
  \includegraphics[width=4.5cm]{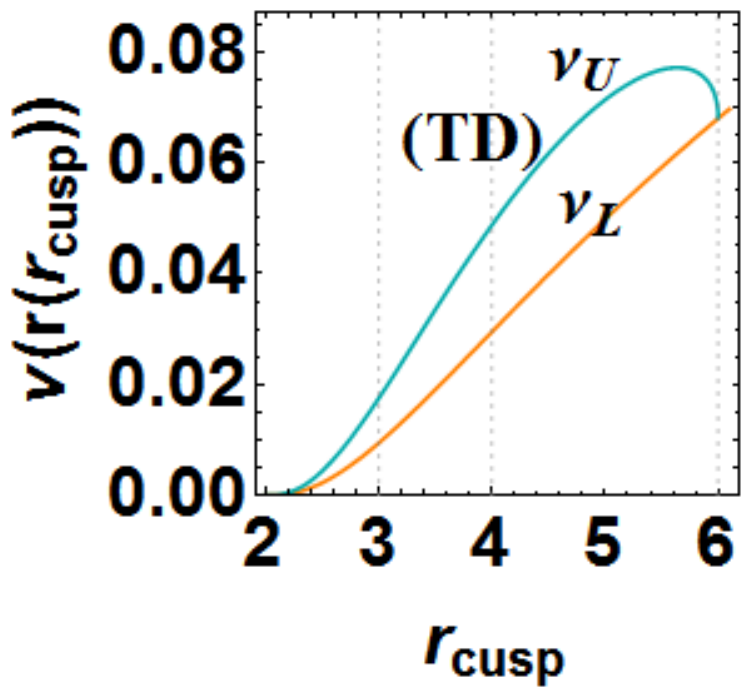}
  \includegraphics[width=4.5cm]{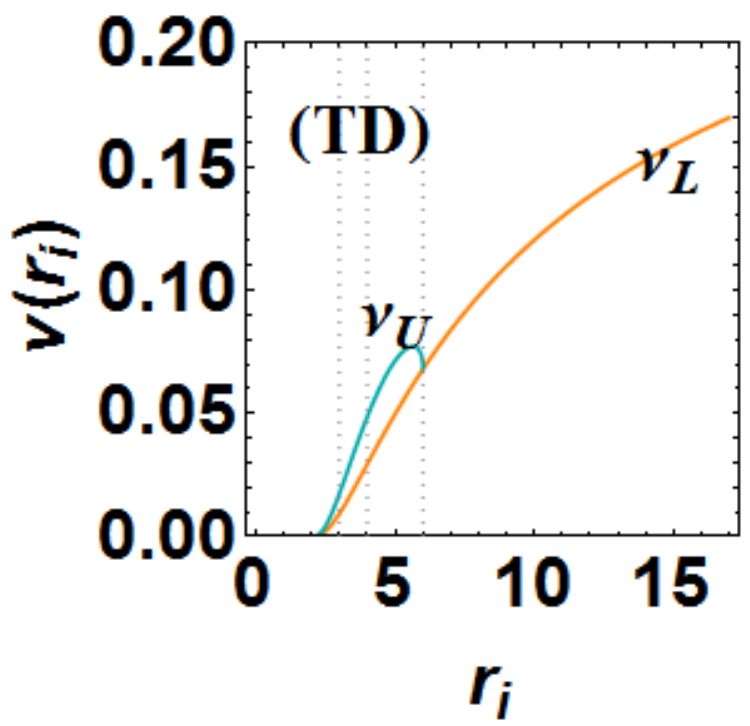}
  \includegraphics[width=4.5cm]{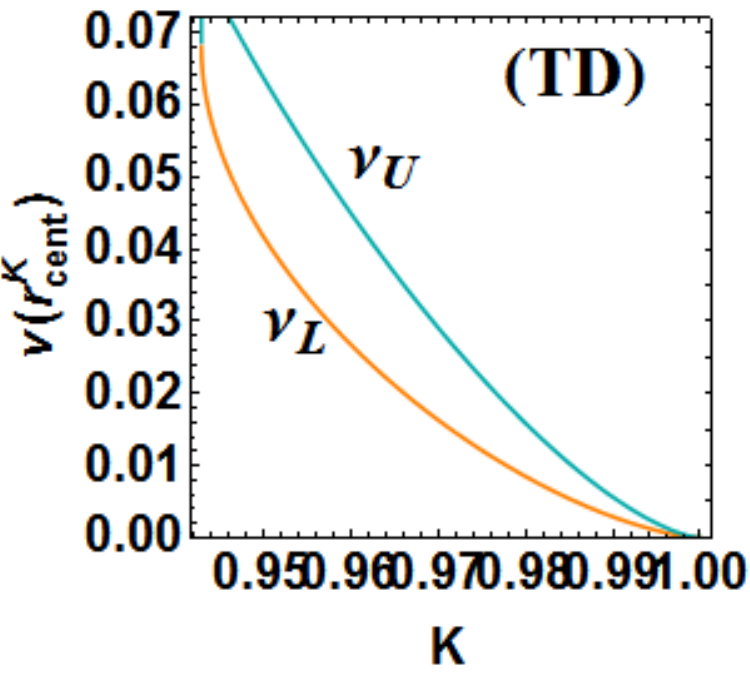}
  \caption{Frequencies $\nu_U=\nu_K$ (\emph{upper}) and $\nu_L= (\nu_K + \nu_r)$ (\emph{lower}) in the (\textbf{TD}) models for  the  \textbf{(a,b,c,d,e,f,g)} tori models of Figures\il\ref{Fig:vsound}.   Frequencies are increased by a factor present in parentheses, in dashed (orange and cyan) curves.}\label{Fig:TDvsound}
\end{figure*}
\begin{figure*}
  \includegraphics[width=4.0051cm]{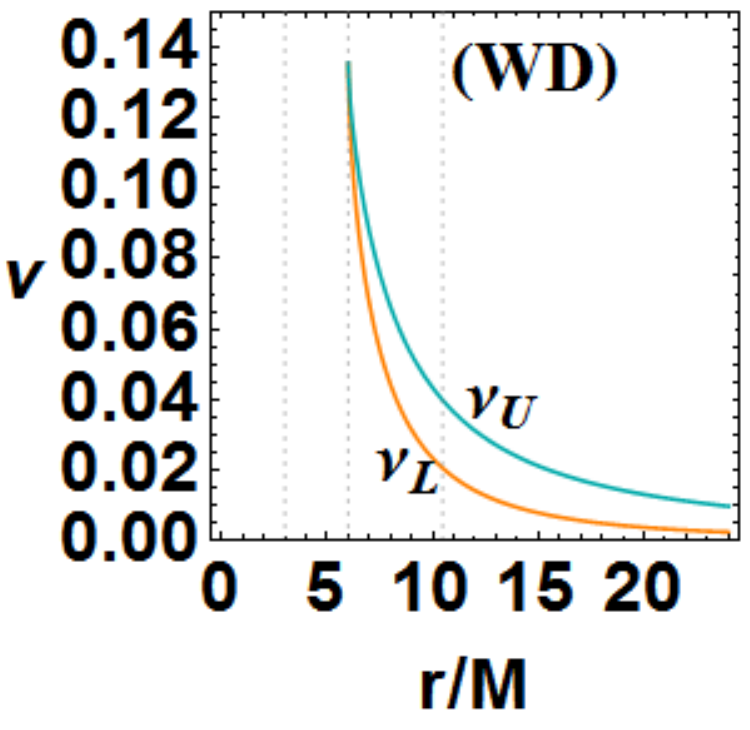}
  \includegraphics[width=4.5cm]{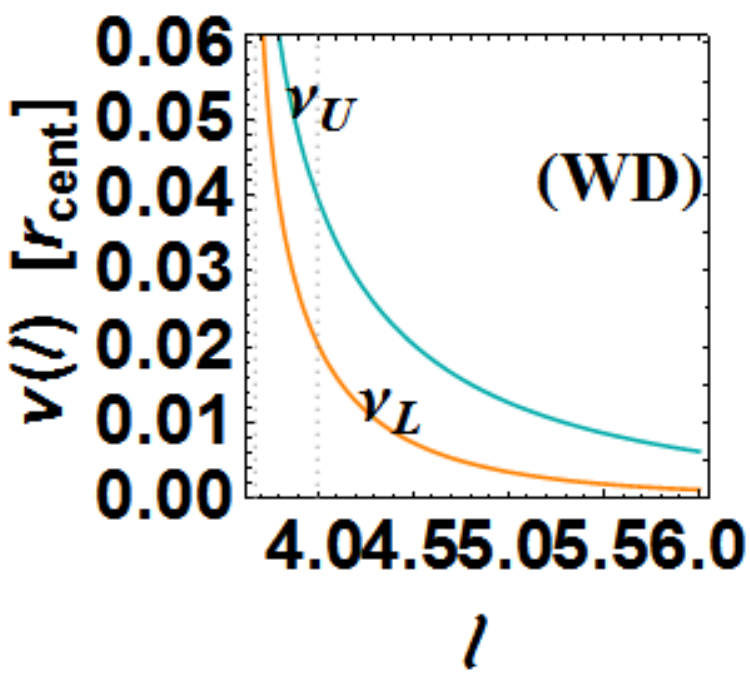}
  \includegraphics[width=4.5cm]{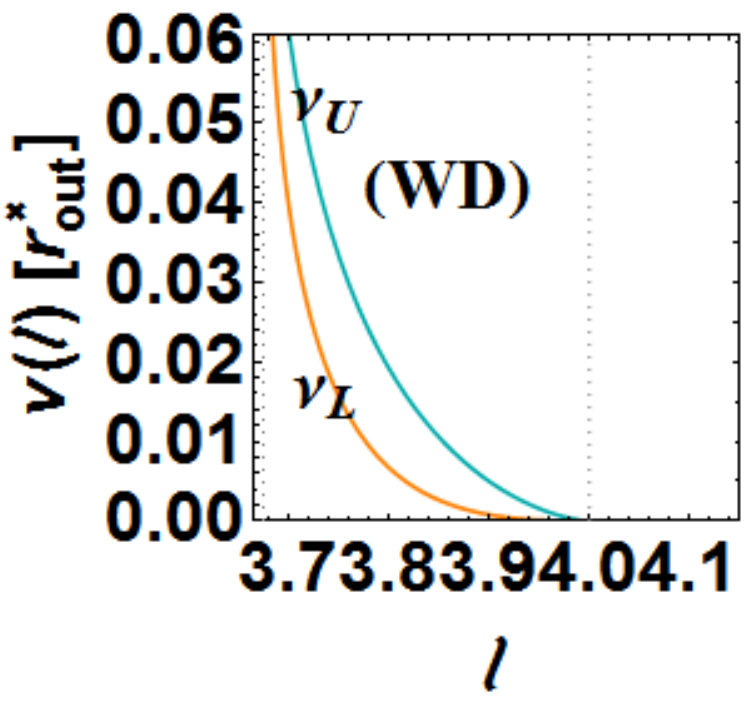}
  \includegraphics[width=4.5cm]{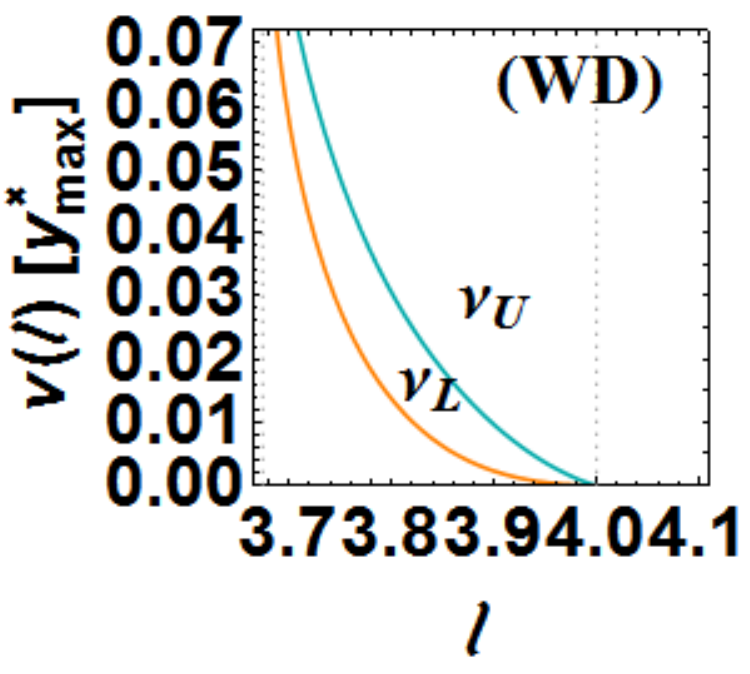} \\
  \includegraphics[width=4.5cm]{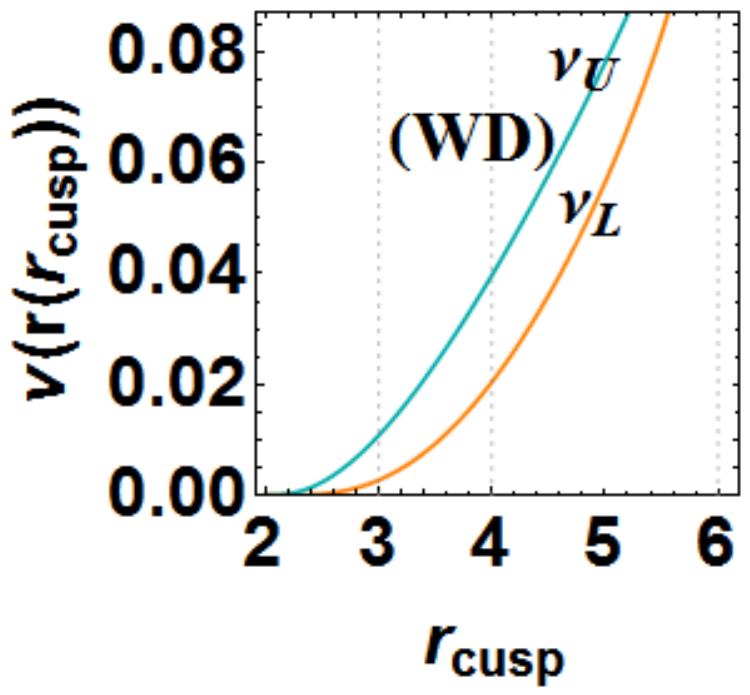}
  \includegraphics[width=4.5cm]{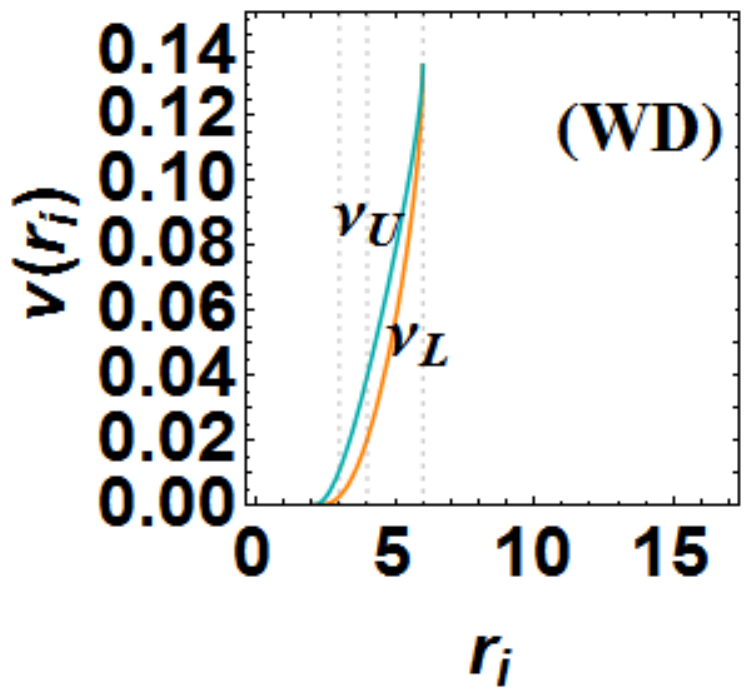}
  \includegraphics[width=4.5cm]{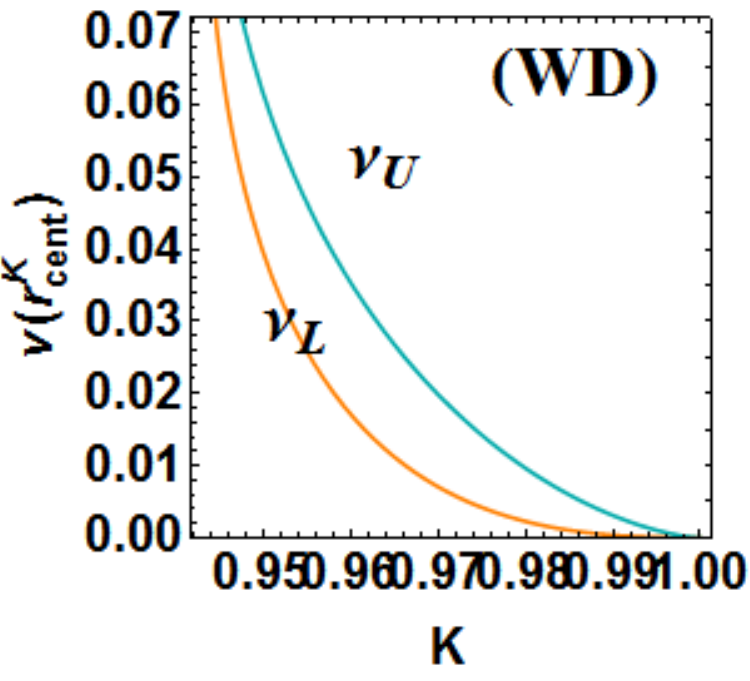}
  \caption{Frequencies in the  (\textbf{WD}) (warped disk) models  where  $\nu_L= 2 (\nu_K-\nu_r)$, and $\nu_U = (2 \nu_K- \nu_r)$ in the  \textbf{(a,b,c,d,e,f,g)} tori models of Figures\il\ref{Fig:vsound}. Frequencies are  represented increased by a factor present in parentheses, in dashed (orange and cyan) curves.}\label{Fig:WDvsound}
\end{figure*}

($\bullet$)One of the first QPO models is  the so called ``relativistic-precession model'' (\textbf{RP} model or standard relativistic
precession (\textbf{RP}) model (in our application it is  coincident also with total precession models \textbf{TP}). This  identifies the twin-peak kHz
QPO frequencies $\nu_U$ (\emph{upper}) and $\nu_L$ (\emph{lower}) with the two fundamental frequencies of a nearly circular geodesic motion: the Keplerian orbital frequency and the periastron-precession frequency, or
$\nu_U = \nu_K$ and  $\nu_L = \nu_{per} \equiv \nu_K - \nu_r$. This model is investigated in Figure\il\ref{Fig:RP}, the radial profiles of the  frequencies are shown in Figs\il\ref{Fig:RPvsound}.
\begin{figure*}
  \includegraphics[width=4.0051cm]{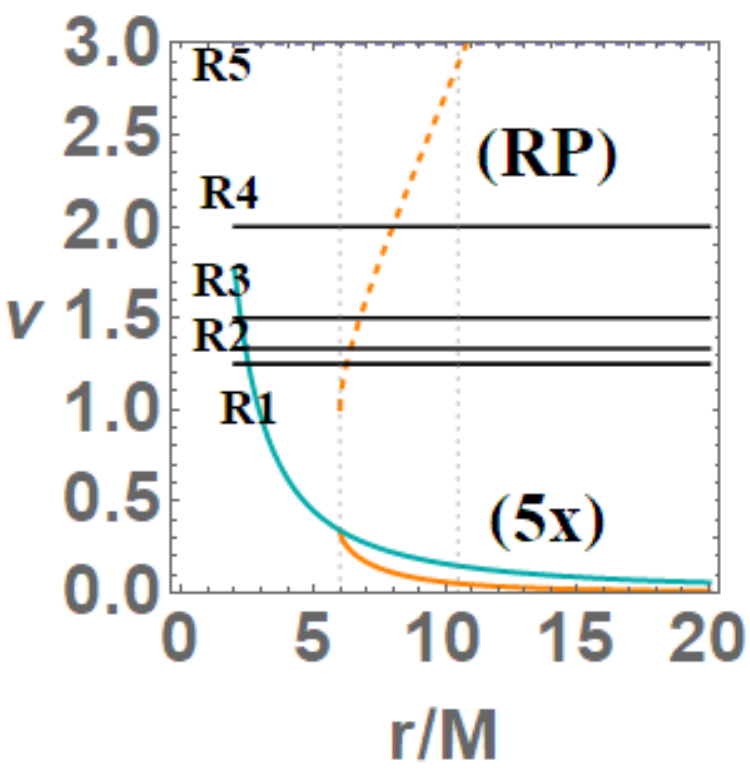}
  \includegraphics[width=4.5cm]{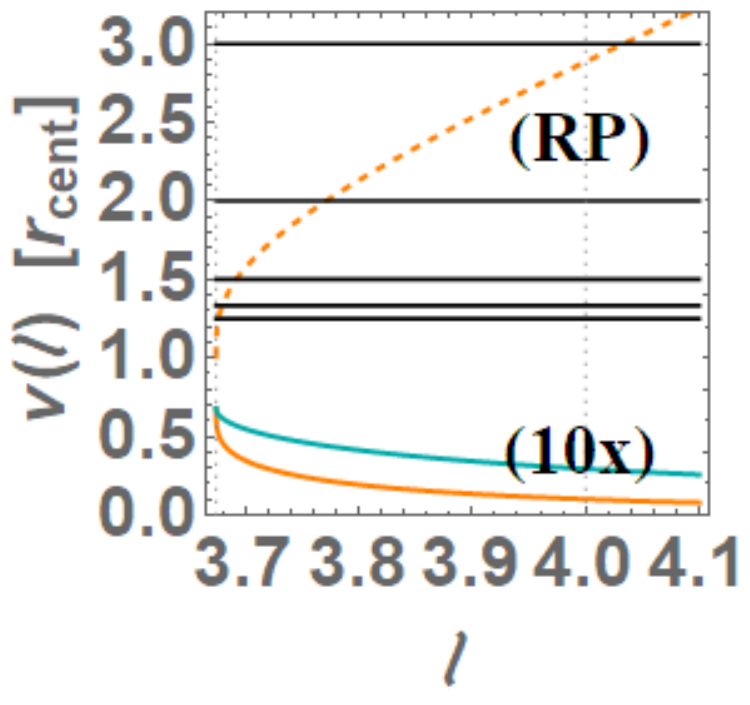}
  \includegraphics[width=4.5cm]{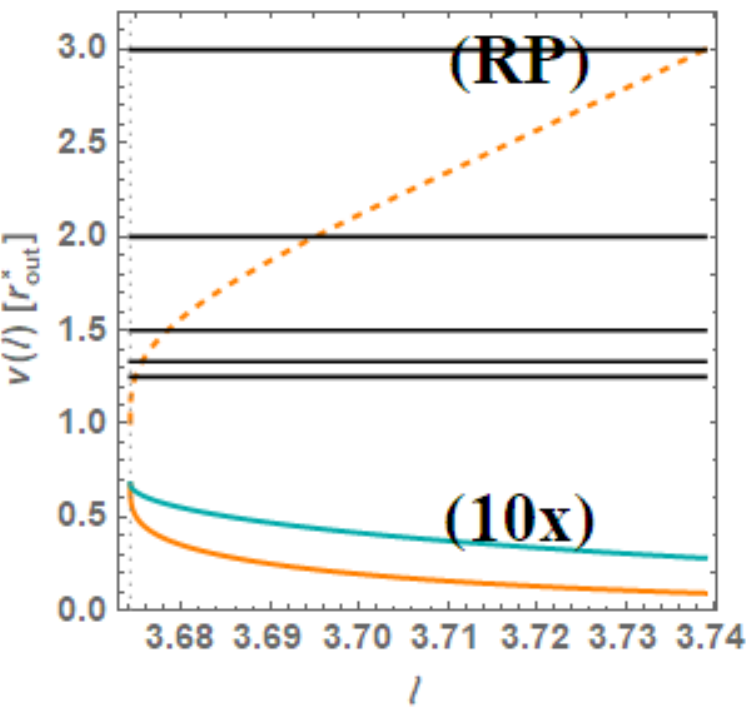}
  \includegraphics[width=4.5cm]{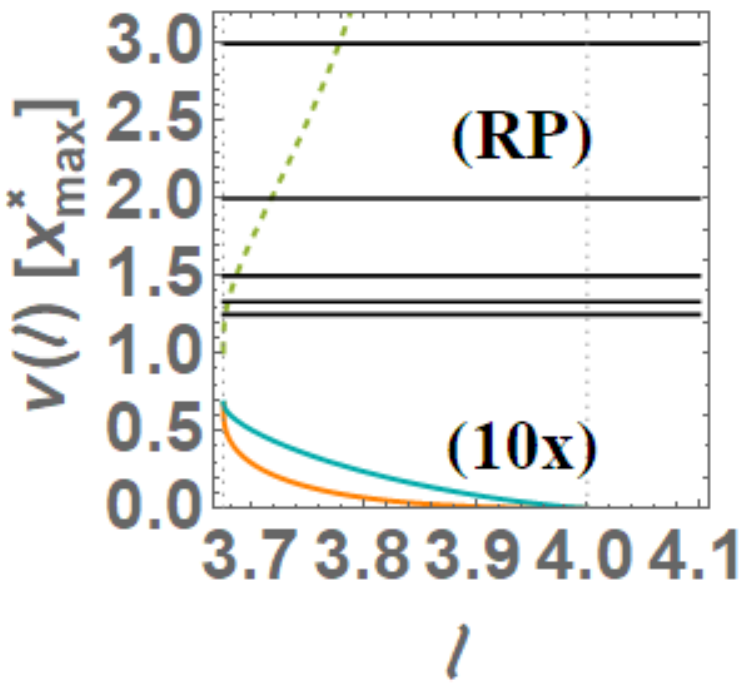}\\
  \includegraphics[width=4.5cm]{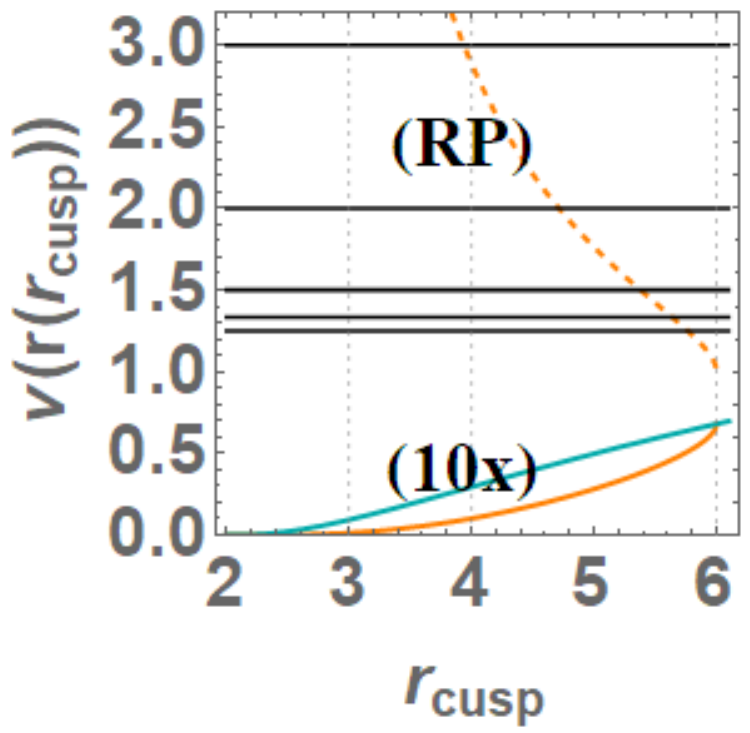}
  \includegraphics[width=4.5cm]{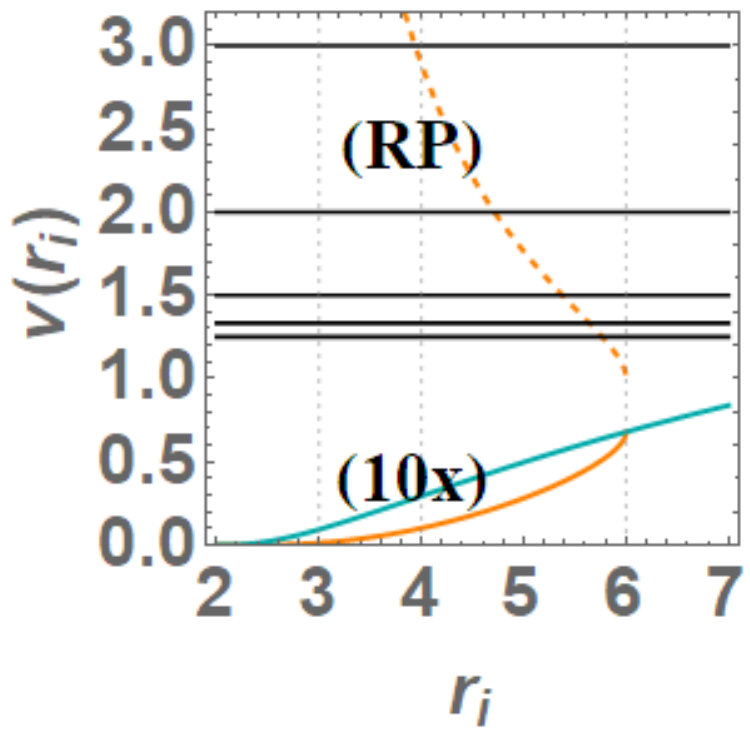}
  \includegraphics[width=4.5cm]{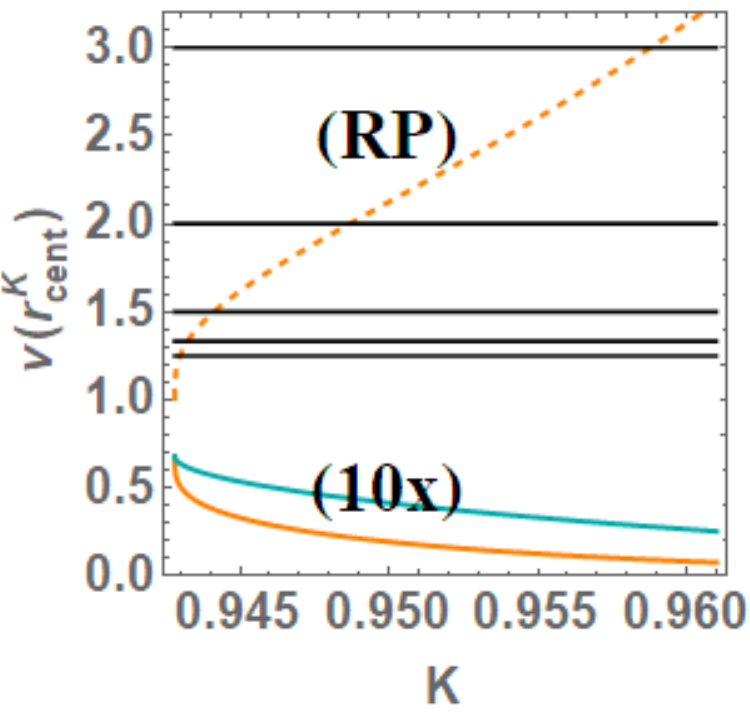}
  \caption{(\textbf{RP}) ({relativistic}-precession model) models, plot of $\nu_U/\nu_L$ (dashed line)  where $\nu_U = \nu_K$ and  $\nu_L = \nu_{per} \equiv \nu_K - \nu_r$, in the  \textbf{(a,b,c,d,e,f,g)} tori models of Figures\il\ref{Fig:vsound}. Frequencies are shown in Figures\il\ref{Fig:RPvsound}. Resonant frequency ratios $\mathbf{R1}=2:1$, $\mathbf{R2}=3:1$, $\mathbf{R3}={3}:{2}$, $\mathbf{R4}={4}:{3}$, $\mathbf{R5}={5}:{4}$ (black lines) are also shown. The two frequencies are also represented increased by a factor $\mathbf{(n x)}$ present in parentheses (where $n$ is generally $\{5,10,30\}$), in dashed (orange and cyan) curves. }\label{Fig:RP}
\end{figure*}
($\bullet$) Further  HF QPOs model,  \textbf{RE} model i.e.  simple resonance epicyclic  model, features
  resonances between epicyclic oscillation modes of the orbiting  fluids. Among the most favored  is the so-called ``3:2 epicyclic resonance model'', identifying  the resonant eigen-frequencies with
frequencies $(\nu_{\theta},\nu_r)$ of radial and vertical epicyclic axisymmetric disk  modes,  with
$\nu_U =\nu_{\theta}$ and $\nu_L = \nu_r$, particularly of the ratio  $\nu_U/\nu_L = 3/2$.  This model is considered in Fig\il\ref{Fig:RE}  while  Fig\il\ref{Fig:vsound} shows  the radial profiles of the   $\nu_L$ and $\nu_U$ frequencies.  
We also consider the following alternatives tidal distortion  (\textbf{TD}) model  where $\nu_L=\nu_K$  and  $\nu_U = (\nu_K + \nu_r)$, considered in Fig\il\ref{Fig:TD}, with radial profile of the frequencies given in Fig\il\ref{Fig:TDvsound},
 \begin{figure*}
  \includegraphics[width=4.0051cm]{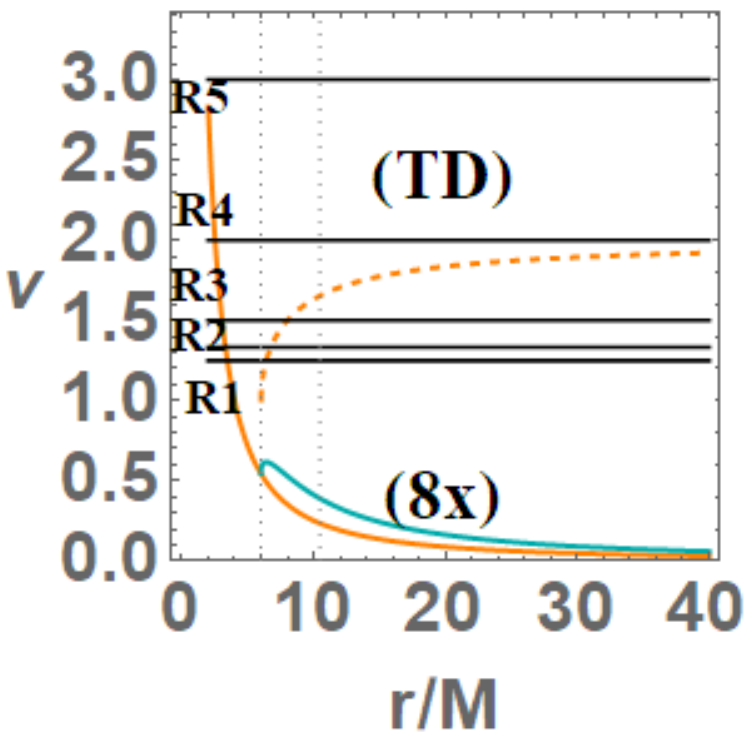}
  \includegraphics[width=4.5cm]{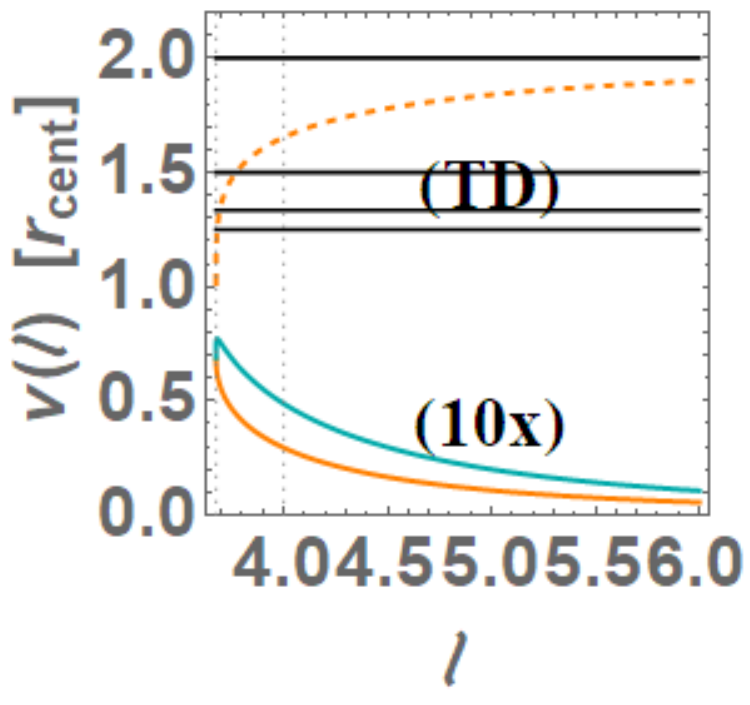}
  \includegraphics[width=4.5cm]{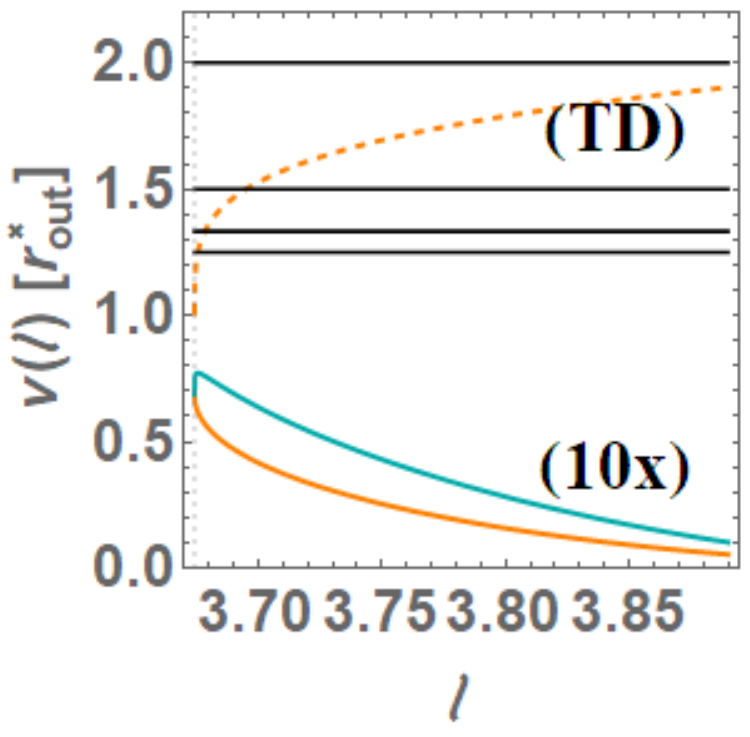}
  \includegraphics[width=4.5cm]{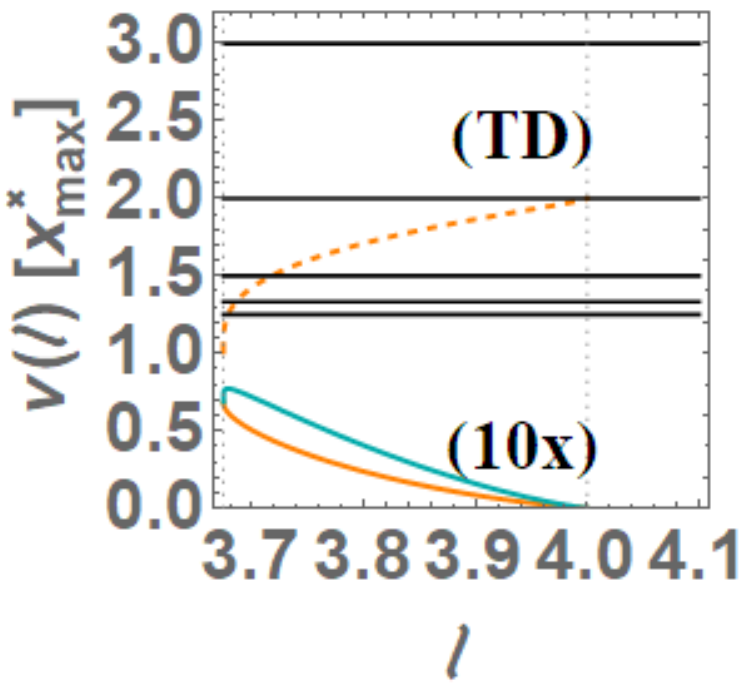}\\
  \includegraphics[width=4.5cm]{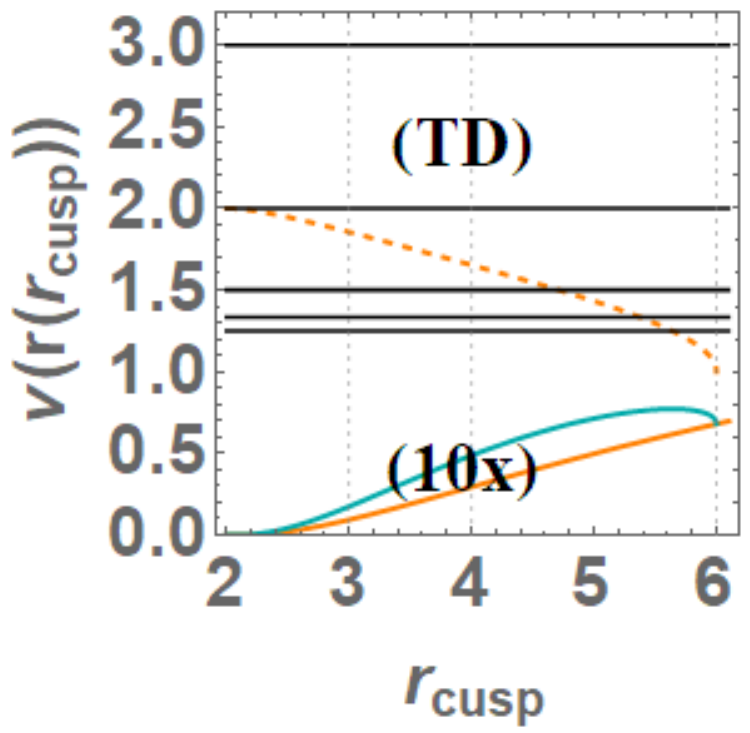}
  \includegraphics[width=4.5cm]{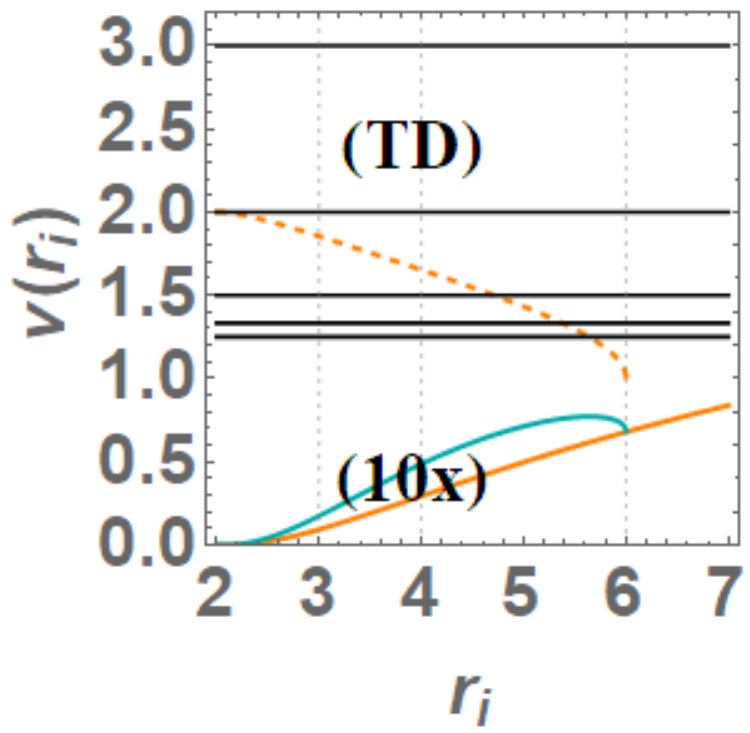}
  \includegraphics[width=4.5cm]{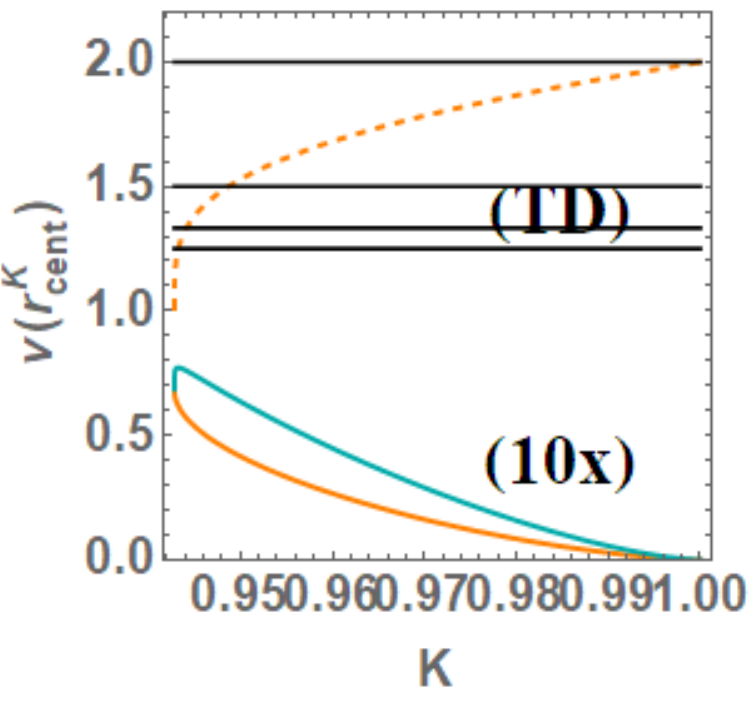}
  \caption{(\textbf{TD}) models, plot of $\nu_U/\nu_L$ (dashed line)   where $\nu_L=\nu_K$  and  $\nu_U = (\nu_K + \nu_r)$.  Models  \textbf{(a,b,c,d,e,f,g)}  of Figures\il\ref{Fig:vsound} have been considered.   Frequencies are in Figures\il\ref{Fig:TDvsound}. The two frequencies are also represented increased by a factor present in parentheses $\mathbf{(n x)}$ present in parentheses (where $n$ is generally $\{5,10,30\}$), in dashed (orange and cyan) curves. Resonant frequency ratios $\mathbf{R1}=2:1$, $\mathbf{R2}=3:1$, $\mathbf{R3}={3}:{2}$, $\mathbf{R4}={4}:{3}$, $\mathbf{R5}={5}:{4}$ (black lines) are also shown. }\label{Fig:TD}
\end{figure*}
and  warped disk
 (\textbf{WD})
model featuring a combination of the orbital and epicyclic frequencies where  $\nu_L= 2 (\nu_K-\nu_r)$, and $\nu_U = (2 \nu_K- \nu_r)$. This model is studied in Figs\il\ref{Fig:WD} and Figs\il\ref{Fig:WDvsound}.
\begin{figure*}
  \includegraphics[width=4.0051cm]{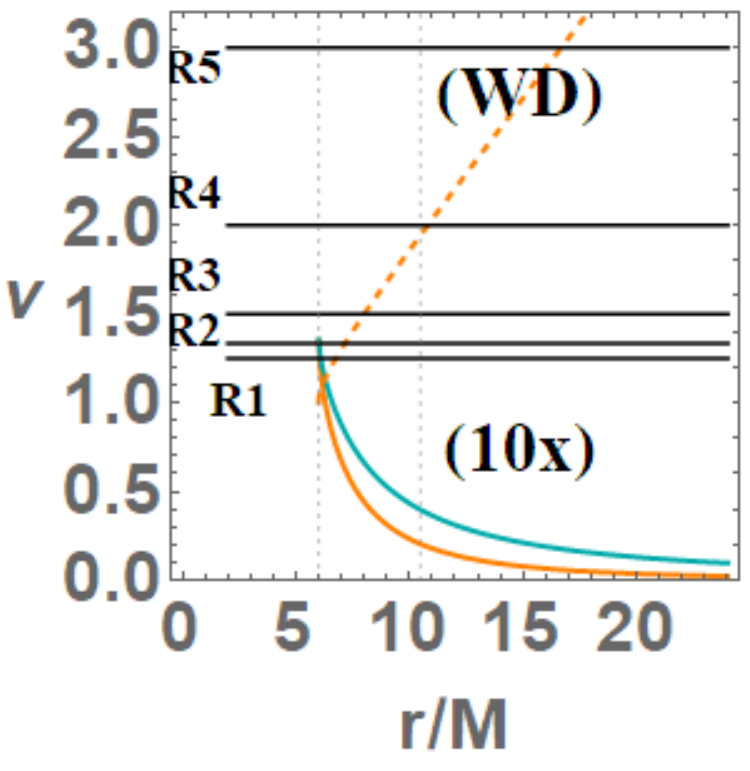}
  \includegraphics[width=4.5cm]{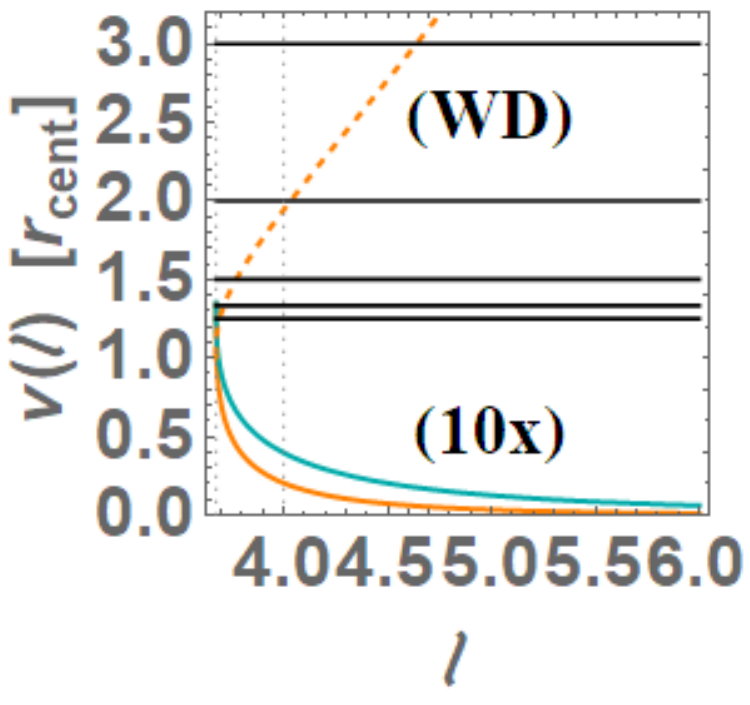}
  \includegraphics[width=4.5cm]{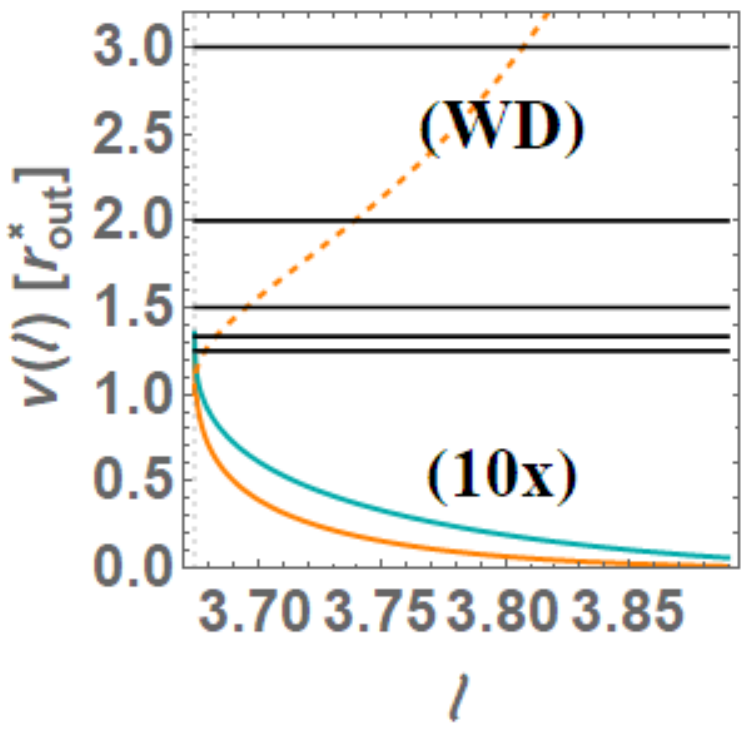}
  \includegraphics[width=4.5cm]{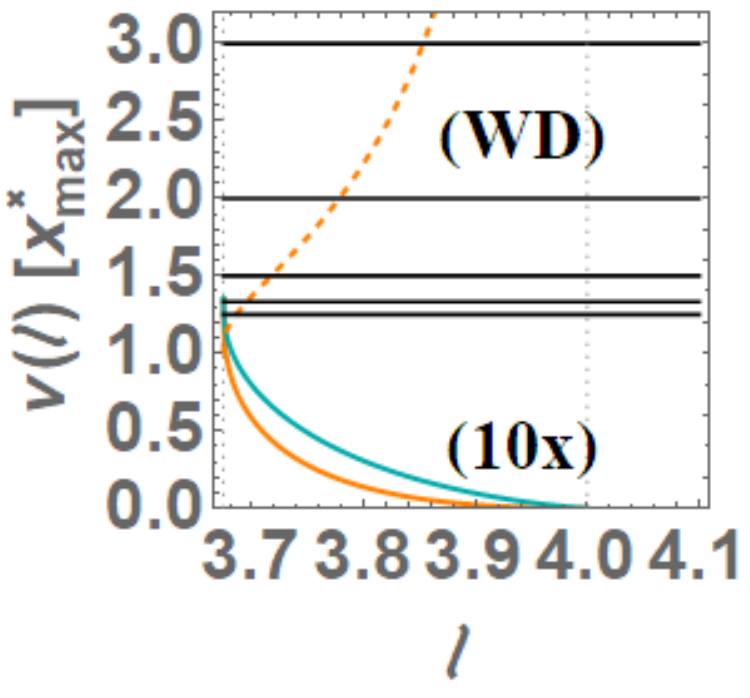}     \\
  \includegraphics[width=4.5cm]{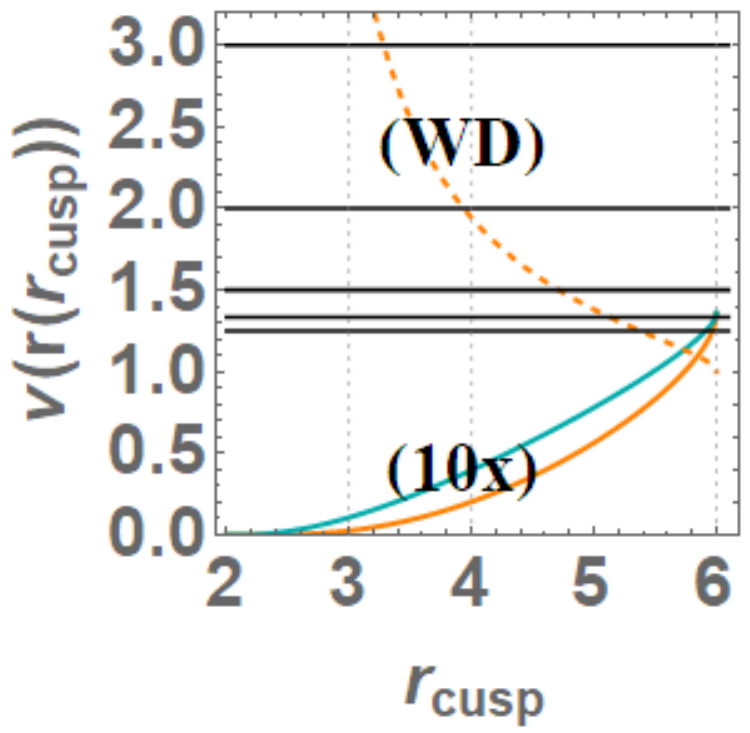}
  \includegraphics[width=4.5cm]{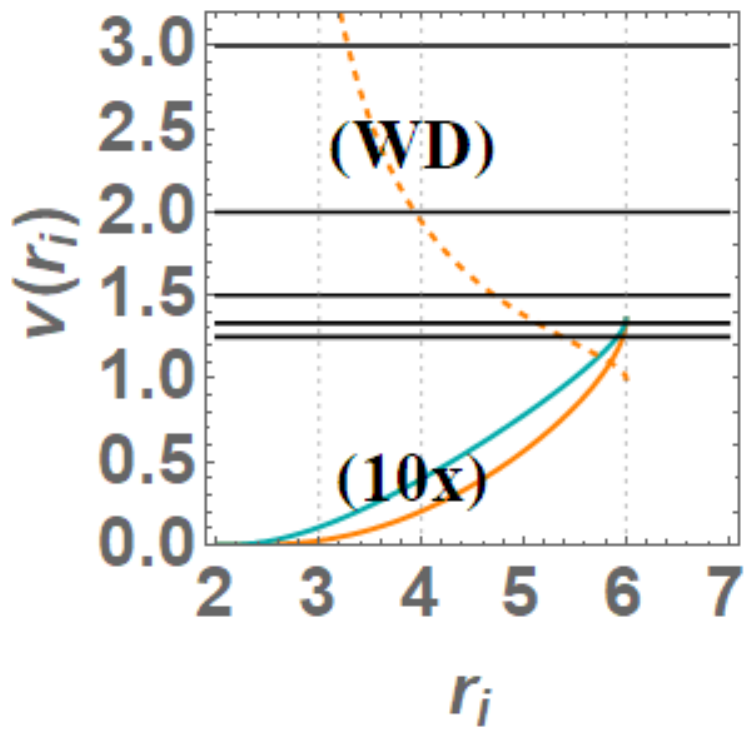}
  \includegraphics[width=4.5cm]{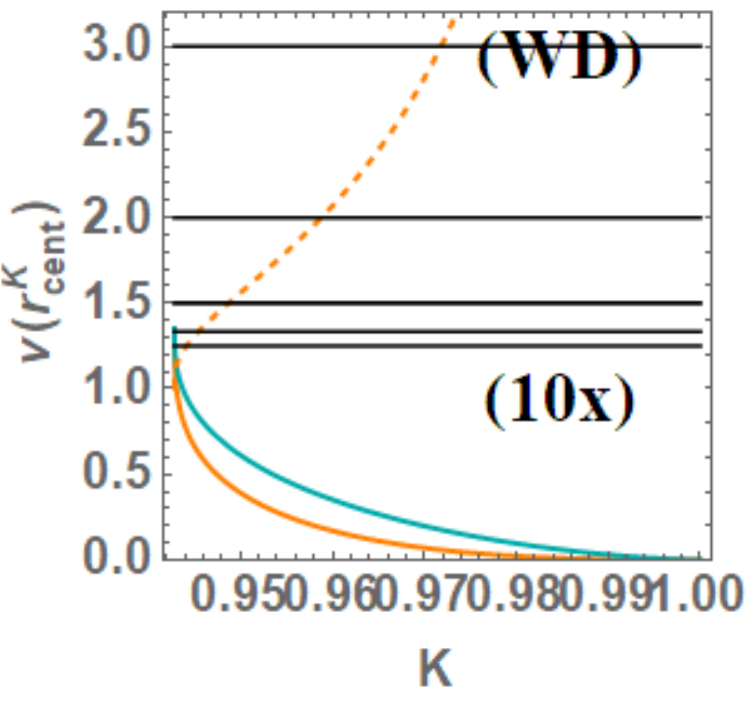}
  \caption{(\textbf{WD}) (warped disk) models, plot of $\nu_U/\nu_L$ (dashed line)  where  $\nu_L= 2 (\nu_K-\nu_r)$, and $\nu_U = (2 \nu_K- \nu_r)$.  The two frequencies are also represented increased by a factor present in parentheses  $\mathbf{(n x)}$  (where $n$ is generally $\{5,10,30\}$), , in dashed (orange and cyan) curves. Resonant frequency ratios $\mathbf{R1}=2:1$, $\mathbf{R2}=3:1$, $\mathbf{R3}={3}:{2}$, $\mathbf{R4}={4}:{3}$, $\mathbf{R5}={5}:{4}$ (black lines) are also shown. Frequencies are in Figures\il\ref{Fig:WDvsound}.  Models  \textbf{(a,b,c,d,e,f,g)}  of Figures\il\ref{Fig:vsound} have been considered.}\label{Fig:WD}
\end{figure*}
We consider the frequencies in different models  introduced  in  Figs\il\ref{Fig:vsound}:  in the \textbf{(a)-model} as functions of $r/M$. In the   \textbf{(b)-model} frequencies are evaluated in  $r=r_{cent}(\ell)$ in equation\il\ref{Eq:rcentro} as function of the fluids angular momentum $\ell\in[\ell_{mso},\ell_{mbo}]$. In  this way the frequencies will be  parameterized by the specific moment of the fluid setting the orbits. In the   \textbf{(c)-model} $ (\nu_{L},\nu_{U})$ are evaluated for $r=r_{out}(\ell)$ in equation\il\ref{Eq:over-top} as functions of  $\ell$. We use the outer edge of the torus  as a test according to the analysis  on the marginally collisional sphere considered in \cite{mis(a)}. In the   \textbf{(d)-model} $ (\nu_{L},\nu_{U})$  are evaluated in {$r=r^{\times}_{\max}(\ell)$ (the geometric maximum of the torus) of  Eq.\il(\ref{Eqs:rssrcitt}) as function of  $\ell$.} The  \textbf{(e)-model} considerers
$ (\nu_{L},\nu_{U})$
for $r=r_{\times}^{\varepsilon}$ of  Eq.\il\ref{Eq:worrre-stowo} as function of $r/M$ torus cusp or center.  The  \textbf{(f)-model} features
 $ (\nu_{L},\nu_{U})$ as function of
 $\bar{r}(r_i)$ in Eq.\il\ref{Eq:ririoutcenter}, therefore depending on the cusp or center  if considered in the center or cusp respectively. In  the  \textbf{(g)-model},
 $ (\nu_{L},\nu_{U})$ are considered
for $r=r_{cent}(K)$ in  Eq.\il\ref{Eq:nicergerplto} as function of  $K$.
{The profiles of the  two-peaks frequencies $(\nu_L,\nu_U)$ and their ratios in the oscillation models (\textbf{TD},\textbf{RP},\textbf{RE},\textbf{WD})   of  Figures\il\ref{Fig:TD},\ref{Fig:RE},\ref{Fig:RP},\ref{Fig:WD}, can be interpreted,  according to  toroidal models  $(\mathbf{a,b,c,d,e,f,g})$  of Figures\il\ref{Fig:vsound},  taking into account the radial dependence of the epicyclic geodesic  frequencies  $(\nu_r,\nu_{\theta})$  of  Eq.\il\ref{Eq:frenurnurthe},   and the  radial profile of leading functions  $\ell_K(r)$ in Eq.\il\ref{Eq:lqkp} and energy function $K(r)$ in equation\il\ref{Eq:Kdir}. The radial dependence of $\ell_K(r)$ and $K(r)$  are  clearly reflected in the frequency analysis. In fact,  the increase  of $\ell_K(r)$ corresponds in general to an increase of the radius  $r>r_{mso}$  and parameter $K=K(r)$, corresponding  also to  an increase of the torus  outer edge  $r_{out}$ and the torus center  $r_{cent}$ and in general the torus height  $x_{\max}$. Viceversa,  the increase  of  the specific  angular momentum $\ell_K(r)$ corresponds to a decrease of radius     $r<r_{mso}$,   therefore it corresponds  to a decrease of the cusp $r_{cusp}$  location,  which corresponds particularly to the   toroid cusp $r_{cusp}=r_{\times}\in ]r_{mbo},r_{mso}[$ in accreting models, present also in $r_i$ definition   of equations\il\ref{Eq:ririoutcenter},\ref{Eq:worrre-stowo}--see also Figures\il\ref{Fig:PlotGold}. In   Figures\il\ref{Fig:vsound},\ref{Fig:TD},\ref{Fig:RE},\ref{Fig:RP},\ref{Fig:WD} we pointed out   the topological status of each \textbf{RAD} component,  describing  their   stability, as   associated with a given frequency radial profile, by showing the  limiting  values $\Qa_{\bullet}$, for any quantity $\Qa_{\bullet}=\{r,K,\ell\}$, evaluated in   $r_{\bullet}\in\{r_{mso},r_{mbo},r_{\gamma}\}$. For the frequency models considered here, $\Qa_{\bullet}$ corresponds generally to the   limiting frequency values.}
\subsection{{Comments on the \textbf{RAD} structures and  the    outcomes  of  the analysis}}\label{Sec:pen-comme}
 In our investigation  we  focus on a clustered set of misaligned tori  framed in the   \textbf{RAD} context firstly  developed for the case of equatorial tori in Kerr spacetime (\textbf{eRAD})  in \cite{ringed}. By considering  the tori agglomerate  as one orbiting object  we  use  a global approach to the characterization of the structure  singling out a  leading function for the distribution  of tori in the \textbf{RAD} rather than focusing on the details of the physics of each specific toroid components hence enhancing a macro-structure approach. {We are thus mapping the Schwarzschild geometry in relation to complex structures of arbitrarily inclined perfect fluid tori. The mapping is reflected by a leading function connected to the distribution of the specific angular momentum of the orbiting fluid; its generalization to the perfect fluid combined with internal toroidal magnetic field is also introduced. The mapping can be useful in establishing the starting point in studies of a variety of dynamical situations, including collisional effects, jets, etc. Here we have considered the case of small oscillations of slender toroidal configurations that can be related to HF QPOs}. As a consequence of this analysis  we provide a set of initial data for dynamical situation,  and a procedure to select the initial configuration for dynamical simulations adapted also to more complex situations where toroidal components  of the \textbf{RAD} have diversified nature. We provide constrains on morphology and stability on the \textbf{RAD} and  first evaluation of multiple tori associate  QPOs emission.
We also discussed the   possible  correlated observational properties.
We propose the \textbf{RAD} of clustered set of inclined tori as base for a different series of phenomena while we indicate possible observational outcomes related to different phases of activity of these structures.
In fact, the \textbf{RADs} are therefore characterized by a typical  ringed structure that could be evidenced  primely for example in  the X-ray emission spectrum and as an imprint of the discrete inner \textbf{RAD}  composition, or in the combined  oscillatory phenomena  associated to the  tori model observable for example by the  X-ray observatory ATHENA\footnote{http://the-athena-x-ray-observatory.eu/}.

In details  we  summarize below    some  key points of this analysis and the main  results.

-\textbf{The clusters: an overview} In this article we provide  the main characteristics of a set of accreting misaligned tori  around  the  central \textbf{BH}. Within this analysis therefore
we  provided limitations on \textbf{RAD} existence and stability  and an overview of the possible emissions   associated with  these structures as  the QPOs.
From methodological point of view  we believe relevant  the
approach shift   which characterizes the \textbf{RAD} frame, where the cluster is studied as one gravitating composite (macro-)structure orbiting around one central Schwarzschild attractor, hence the identification of a leading function for the tori distribution. This analysis  also considers  the misaligned tori collision emergence.
Constrains on existence of such configurations are discussed in  Sec.\il(\ref{Sec:Misal}),  particularly against  tori collision   and Paczynski instabilities (instability associated to installing of  accretion phases, the cusped tori, or open cusped configurations, the so-called proto-jet ) and developed more extensively in \cite{mis(a)}. Many morphological properties of the torus  are discussed in relation to the stability problem for the \textbf{RAD} and each of its gravitating components. We  discuss as  possible QPOs emerging  from the  \textbf{RAD} ringed  structure in  Sec.\il(\ref{Sec:epi}), proposing the \textbf{RAD} to be considering for entangled emission from each of its misaligned component.
  Part of our analysis was   dedicated to the evaluation of the torus considered for  its thickness which is  crucially  significant in many aspects of the accretion disk physics and phenomenology. In particular
we  focus on a more specific analysis of the role of geometrical thickness in relation to the  disco-seismology effects for each toroid. In Sec.\il(\ref{Sec:epi}) we  provided  the conditions for which these can be considered geometrically thick---- Figs\il(\ref{Fig:PlotGold}).
We have therefore characterized  the model   on the basis of the  parameters determining  the particular configurations according to their stability as related to the toroids morphology. The evaluation of the toroids geometrical thickness  is  indication  of the predicted  \textbf{BH} accretion rates correlation\cite{mis(a)}.
The analysis identifies  the    sets of  \textbf{RAD} inclined  toroids having  equal characteristics as the torus thickness. Therefore we provided the  tori distribution in the  \textbf{RAD}  in Figs\il(\ref{Fig:PlotGold}) by considering classes depending on the  geometric thickness $\Sa$ and an evaluation of the geometric thickness of the disks considered  in the \textbf{RAD} frame establishing conditions under  which disks are geometrically thick according to the model parameters.

\textbf{The  phenomenological outcomes:}
The \textbf{RAD} in any stage of its life, we believe can offer an interesting set of complex observational aftereffects.
    The  \textbf{RAD} frame  implies  some relevant  consequences from phenomenological view-point. We could  relate different phenomenological  aspects to the presence of a  \textbf{RAD}  structure, we briefly discuss some of these here. \emph{\textbf{(i.)}} Firstly we mention obscuration in the emission spectrum induced by an inner torus of the agglomerate, the analysis of this case  requires also the understanding  of main morphological characteristics of the  torus closest to the central  attractor.  \textbf{\emph{(ii.)}} \textbf{RAD} implies the   possibility of an globulus depending on tori number thickness and inclination angle, the \textbf{BH} horizon in this case would be "covered"  in a multipole embedding composed by different tori having different orientation rotation . \emph{\textbf{(iii.)}} The \textbf{RAD} may be considered as a model to increasing accretion  mass rates and  to explain hight masses    for the \textbf{SMBHs},   due to multiple accreting tori emerging from the typical ringed structure, which also  ensures the possibility of interrupted phases of accretion; \textbf{\emph{(iv.)}}  In Sec.\il(\ref{Sec:epi}) we investigated the possibility of  QPOs emission enabled in complicated structure of  the cluster  described  within different emission models and assumptions. \textbf{\emph{ (v.)}}The inner composite structure of \textbf{RAD} implies an inner activity which could end also  in  violent, catastrophic outburst with ejection of matter and possible destruction  of the  \textbf{RAD} and  formation of a large \textbf{SMBH}. This situation  implies  the existence of periods characterized by different   levels of activity for the \textbf{BH} and the \textbf{RAD} which would relate them ultimately also to  the  host characteristics. \textbf{\emph{(vi}).} One of the possible outcomes of these unstable phases could be  jet-emission which, in the \textbf{RAD}, would be launched from different points internal to the cluster and with different inclinations.
More in details:
a \textbf{BH} and especially a \textbf{SMBH} can be  characterized by different  periods of activity, defined by  enhanced accretion rates or  interrupted accretion periods and jet emissions. These phases ultimately could be  detected  in the alteration of the mass accretion rates, or  recognizable as mechanism at base for  high masses considered in the \textbf{SMBHs}.
The main interesting aspect of these  clusters    relies in their   internal activity,  particularly the internal exchanges of energy and matter between the tori as well as the  tori and central \textbf{BH}.
Tori accretion, in the case of a globular model,  could ends also into a    relatively fast  collapse  of the entire  structure into the central \textbf{BH}  contributing therefore with a huge  mass and  spin and  a great release of energy  and matter.
Instability in one point of the structure could initiate a sequence  of   associated, complicated \textbf{RAD} phenomenology.
 \textbf{RAD}  would be recognizable by an  articulated internal life triggered by each torus dynamics and empowered by its    inner structure. Some of these phenomena are for example the  tori collisions,  internal jet emission and  accretions,  tori oscillation modes, eventually related to QPOs observed in non-thermal X-ray emission from compact objects.  Concerning the possible correlation with QPOs emission, which is considered Sec.\il(\ref{Sec:epi}) the oscillations  of each component are added to others and are pulsations of the \textbf{RAD}, and  possibly the  globule creating   eventually a rather distinct detectable emission spectra.

 \textbf{The model setup:}
 In this work we propose also an adapted model setup.
  The novelty  of our approach, resides  on the
   {methodological} view point, proposing a  conceptual global setup that constitutes the \textbf{RAD} frame pursuing the existence of a \emph{leading function} to represent and constraint the tori distribution around the central attractor.
 We then identified also an energy function $K(r)$  regulating  the \textbf{RAD} stability (cusp emergence) and     defining relevant quantities as the  mass accretion rate and cusp luminosity.  Our analysis places constraints on the existence and stability of misaligned tori which can be used in  dynamical (time-dependent, evolutive)  analysis of a similar system with these initial configurations.
 This analysis can be compared with similar studies in  \cite{Martin:2014wja,2012ApJ...757L..24N,2012MNRAS.422.2547N,2015MNRAS.448.1526N,2006MNRAS.368.1196L,Feiler}.
The current literature considers  similar  objects   within a numerical approach, fixing very specific initial data and physical setup for each torus.
       Our approach , with respect to other studies has the advantage  to be an  exact  analysis of different morphological characteristics and emergence of tori collision conditions, in a non dynamical frame. The key  element consists in the fact that we privileged a global approach focusing on the tori distribution around the \textbf{RAD} (where the general relativists effects are relevant) rather than on the analysis  the details of physics and evolutions of  each specific toroidal components, which would narrow the analysis to the selected system.
      Results therefore can be  interpreted as initial configurations for  dynamic simulations,
focusing on the global issue the \textbf{RAD} structure,  as the tori location in the cluster, the location of the  maximum and minimum pressure points and other morphological characteristics,  hence
the choice $ \ell =$constant for  each torus. Therefore the torus  parametrization with the $\ell$ value  is a very convenient choice for  the  HD \textbf{RAD} macrostructure  scenario.  In fact we   provided the classes  toroidal components and therefore \textbf{RAD}  which can be used to the  match with the  subsequent phases of development of the  fluid dynamics\footnote{
 On the other hand, although we focus on a non-dynamical structure, we note that we could  follow the evolution of a torus considering as a sequences of tori at differences stages  having  different  values of the defining  model parameters for example
choosing    the specific  angular momentum as evolutive parameter--\cite{pugtot}.}.
Our results completely constraint the possible initial configurations with multiple tori considering both the possibility of tori collision and accretion emergence, or  their morphological characteristics.

\textbf{Globuli as  limiting cases and embedded BHs:}
Conceiving the \textbf{RAD} as a  whole macro-structure leads also  to focus on interesting limit situations, both limiting  from the point of view of the conditions on the configurations, for example the  thickness, and regarding conditions imposed on the activity that can be inferred from the very imposed  constraints. In this sense
 a particular interesting limiting case of the \textbf{RAD} proposed in this work consists in   the possibility of an embedded  \textbf{BH}  or a globulus. The formation of these   objects  would refer to periods of low activity,  (cold-globuli), giving rise, when activated, to  catastrophic  outbursts    distinguished by a huge  release of energy and matter. The end of this process would  ultimately go  into a   different \textbf{SMBHs} or with the  formation of   a different \textbf{RAD} configuration.
 This is a limiting case where the \textbf{BH}  horizon is expected to be    ``covered''  to an observer at   infinity--Figs \il(\ref{Fig:solidy}).  Figure \il(\ref{Fig:SIGNS})--\emph{right}  depicts this situation the emergence of luminous anuli and the complexity of the \textbf{RAD} case.
 However, a key aspect to focus on the observation of this  situation remains the  stability of the   set of accreting \textbf{RAD} tori especially in case of attractor with spin. Primarily in this case it is necessary to  evaluate  the  maximum torus distance from the central attractor considering the dependence from the model parameter of the outer edge of the torus, the geometrical thickness  considered in Sec.\il(\ref{Sec:epi}) and the  density of tori in the \textbf{RAD} here discussed in Sec.\il(\ref{Sec:Misal}).

\textbf{Morphology:}
A large part of our analysis concerns the  investigation of the geometry of \textbf{RAD} of orbiting misaligned tori, which  is in fact related to the emergence of instable phases of \textbf{RAD}, such as  tori collision and accretion or open cusped configurations  which are variously related to proto-jet emission.
Even in the simplest case of static background, the \textbf{RAD}  toroidal components are characterized by  boundary conditions dependent  on  the distance, in the clusters, from the central attractor.
 These systems, have therefore  several  restrictions on the possibility  of formation,  their evolutions and  related  observational characteristics.
 The occurrence of  accretion and collision  are here regulated by the model parameters which in turn determines the disk morphology. We  provide the conditions determining these cases.
 We mention the tori distance from the central attractor, here considered  in Appendix\il(\ref{Sec:doc-ready}), especially the  characteristics of  the outer  and the inner  toroids.
  The analysis of outer edge of the outer  tori of the clusters, which is an aspect deepened in \cite{mis(a)}, results in  constrains on the inner structure of the \textbf{RAD} and therefore the tori collision but also the  radius of the \textbf{RAD}, therefore setting the radial dimension of the globulus, which is relevant also for the  formation, evolution  and stability of the \textbf{RAD}.
A further  significant quantity for the \textbf{RAD} systems affecting   both stability, including  the  \textbf{BH} tori energetics as the accretion rates or the cusp luminosity, observation and oscillation, for example in the evaluation of the effects of disc-seismology,  is  the torus and the \textbf{RAD}  geometrical thickness.
 Two  tasks of the \textbf{RAD} investigation  was  therefore to establish  conditions  of geometrically thickness by considering the $(\Sa,\beta)$ parameters of the model and the limiting  value  $\Sa=1$ and to characterize the   tori distribution in the  \textbf{RAD} considering the characteristic of the geometric thickness-- Figs\il(\ref{Fig:PlotGold}).
  This analysis is shown in Figs.\il(\ref{Fig:PlotGold}) where we  analyze also the geometrical thickness parameter $\Sa$ of the  tori especially in the range of parameter values  adapted to  the onset of accretion phases (from cusped toroidal configurations) as well as the
$\beta_{crit}$ thickness parameters of Eq.\il(\ref{Eq:betacrittico})  which is used to establish the approach for the oscillation analysis. A comparing between the classes of tori and \textbf{RAD} with equal  $\Sa$  or $\beta_{crit}$ is therefore shown in Figs.\il(\ref{Fig:PlotGold}).
 We can see that the farthest the torus  of the cluster is from the central \textbf{BH} attractor, in a limiting spherical region  of  Figures\il(\ref{Fig:SIGNS})--\emph{left} and the largest can be  its thicknesses. However  $\beta_{crit}$- parameter becomes relevant  for smaller values of $\ell$ and $K$  then the values of $(K,\ell)$ for geometrically thick tori according to $\Sa$, therefore for thinner  tori  (according to $\Sa$ or $\Sa_\times$ specifically for the cusped configurations) and for tori close to the \textbf{BH}, in other words for tori whose parameter of geometrical thickness would be significantly smaller than the reference values  for thick tori. Figs\il(\ref{Fig:PlotGold})--\emph{Upper right}  and \emph{Bottom-right} ultimately show the region where the  QPOs models featuring   an adapted particle oscillation model can be applied distinguishing also the main parameters $(\ell, K)$ and the relative location of the critical pressure points in the tori and their geometrical thickness ($\Sa$).
Figs.\il(\ref{Fig:PlotGold})--\emph{Bottom-left} and \emph{upper --center}  show  the classes of tori having similar conditions, in terms of $\beta_{crit}$-parameter,  for the QPOs  emergence  analysis,  considering the torus cusp $r_{\times}$, torus center (maximum density point) and the parameters  $\ell$ and  $K$,  essential  for the identification of the toroidal components. A comparative analysis of these features would start from the assessment of one or more of  the following elements: the $\beta_{crit}$ (through QPOs detection) or $K$ and $\ell$ by means of  one morphological characteristic  reported in Sec.\il(\ref{Sec:doc-ready}) and from the  torus state, accreting or quiescent (closed, not cusped tori ). A more comprehensive view of $\beta_{crit}$ in terms of cusp and center location  is given by  the three dimensional plot of the Figs.\il(\ref{Fig:PlotGold})--\emph{upper-center}.
Figures\il (\ref{Fig:vsound},\ref{Fig:RPvsound},\ref{Fig:TDvsound},\ref{Fig:WDvsound},\ref{Fig:RP},\ref{Fig:RE},\ref{Fig:TD},\ref{Fig:WD}) show the frequencies, relevant in our approximation, for a set of the QPOs emission models, to be composed  for each torus of the agglomerated, confronted with the observational data and compared with the data  of  different morphological properties of the torus, as the  inner edge or center of maximum pressure, which can be  translated into the model parameters according to the analysis of Sec.\il(\ref{Sec:Misal}) and Appendix(\ref{Sec:doc-ready}).
Here we are interested particularly in the identification  of the trend  of the frequencies with the location of cusps orbits,  clearly defining, with  location of the  torus inner edge  (or the center),   the torus and  consequently the \textbf{RAD} inner structure and recognizing   the possible status of the torus by considering  for example Eq.\il(\ref{Eq:ririoutcenter}) and Eq.\il(\ref{Eq:worrre-stowo})
or, more explicitly, from the quantities listed  in Sec.\il(\ref{Sec:doc-ready}).
Our results  ultimately would  serve as a guideline for  possible observational identification of a \textbf{RAD}.
  This morphological aspect is in fact a a key  element  for  the observation and recognition   for   the  \textbf{RAD} and the establishments of its  stability conditions.
  We investigate the  \textbf{RAD} geometry and morphology considering the parameters $(\ell,K)$  ($2n$ parameters for $n$ \textbf{RAD} tori) introduced in Sec.\il(\ref{Sec:Misal}),   points on the curves $\ell(r)$ and $K(r)$.
One or more of these characteristics   may correspond to an entire  class of objects,  determining a  class of components rather than one specific \textbf{RAD}.
Therefore we considered a number of morphological characteristics, here listed in Appendix\il(\ref{Sec:doc-ready}) and thoroughly considered in \cite{mis(a)}, correlating  in a comparative analysis the   different \textbf{RAD} aspects,
 narrowing these classes through the combinations of  the  analysis on the  other quantities.

\textbf{Stability:} A further outcome of this analysis   is the  characterization  of the stability of these structures which however is firstly crucially related   to the tori distance of  the central \textbf{BH} and  the eventual spin of central attractor \footnote{In this case obviously  the system instability   would be  dependent on the   torus   inclination  angle and fluid rotation as well as the eventual presence of magnetic field. In this scenario the   Lense--Thirring  effect from the central spinning  attractor,  could induce even a torus break,  with  the consequent formation of two equatorial disks, and the emerging, especially  for viscous disks,  of the so called Bardeen\&Petterson effect--\cite{[BP75],2015MNRAS.448.1526N}.}. However, in  the stationary (steady) frame developed in this work where    the spacetime is spherical symmetric, \textbf{RAD}  analysis of the instability emergence corresponds  in the  establishment of the occurrence of tori collisions, and
conditions for tori accretion into \textbf{BH}, or a combinations of these,  constraining   the torus dimensions and the specific fluid  angular momentum\footnote{There is a further  mechanism of instability for the \textbf{RAD}  which  originates from a  typical instability  of geometrically thick tori  orbiting  a central \textbf{SMBH}.  The \emph{Runaway--runaway} instability, introduced first in \cite{dsystem}   consists in the combination of runaway instability, involving the inner edge of the inner accreting torus of the \textbf{RAD} with the consequent destabilization of the aggregate. The accretion   induces  a change of the inner torus morphology,   and   the change of background geometry which has repercussions in the all  \textbf{RAD} structure establishing a sequence of events having different possible outcomes--\cite{Font02}.}--Sec.\il(\ref{Sec:Misal})--Appendix\il(\ref{Sec:doc-ready})--\cite{mis(a)}.

\section{Concluding Remarks}\label{Sec:conclu}
{The investigation of stationary axisymmetric  toroids orbiting  a central \textbf{SMBH} is a  timely issue especially   for  the physics of super-Eddington accretion onto very compact objects, being related to several high energy phenomenological environments as  Active Galactic Nuclei, X-ray sources and  Gamma-Ray Bursts.
In general, accretion tori are  related to a huge variety of   physical phenomena, and   the analysis  of these objects is   important also  for the identification of   the central attractor features, being  the disks   directly involved,  especially in the case of misaligned or  tilted disks,   in   \textbf{SMBHs} dynamics (as for example the spin/mass evolution).
In this view, however, several  notable  aspects of  the disks   structure and morphology are still unclear, for example
the location of the innermost boundary of the disk (inner edge of accreting disk), or the disks  dynamics as   the accretion process and  the QPOs  associated to these structures.
The existence and picture  of a  complete  theoretical interpretation of  the  associated phenomenology in an  unique  satisfactory framework  (for example covering both jets emission and accretion)   remains still to be proved.}
In this work we considered  a model of aggregated misaligned (inclined) tori orbiting one central Schwarzschild attractor, using the approach  developed  in   \cite{ringed,open,dsystem}  for the   \textbf{eRAD}.
Such orbiting  aggregates were first considered in \cite{mis(a)} where
 constraints on the existence and stability of misaligned tori were discussed, also as possible initial data for  dynamical (time-dependent)  analysis of  related systems. Then geometry of \textbf{RAD} accreting tori, stability and collision emergence were also  constrained.
     Special sets of  \textbf{RAD} misaligned  toroids were identified having  equal values of one or more of model characteristics $\mathbf{P}\equiv\left(r_{out},r_{in},r_{cent},\lambda,\Sa\right)$ , where $(r_{out},r_{in})$ are the outer and inner edges of the orbiting configurations, $r_{cent}$ is the center of the toroid (point of maximum density and HD pressure), $\lambda$ is the torus elongations of its symmetric plane, $\Sa$ is the torus geometrical thickness. Configurations of these classes  might   correspond to similar  observational effects depending   on the  \textbf{P}-characteristics. We  used these results in the present  analysis.   Particularly the  evaluation of the toroids geometrical thickness  $\Sa$ has an essential role  in establishing the effects of disk-seismology  discussed in  Sec.\il(\ref{Sec:epi}).
 We analyzed    the geometric thickness $\Sa$  discussing  conditions under  which these disks can be  considered  geometrically thick according to the model parameters, using  the limiting value  $\Sa=1$.
From the observational viewpoint, the  special and distinctive ringed discrete  structure of these aggregates   could be evidenced in  the X-ray emission spectrum and as a track of the  inner \textbf{RAD}  composition. In Sec.\il(\ref{Sec:epi}) we  explored   this possibility  studying the  expected  epicyclic frequencies in the case of ringed structures with misaligned tori, for a first approximate description of  the twin peak quasi-periodic oscillations (QPOs) in the context of the   \textbf{RAD} tori structure.  Particularly we showed the  possibility that the twin peak quasi-periodic oscillations  could reflect  \textbf{RAD} inner structure, particularly with  respect to the inner edges $r_{in}$
of the cusped tori.
 However,   this analysis  is   a first comparative investigation on the problem of the QPOs interpretation in the \textbf{RAD}  frame,  which focuses  on geodesics  oscillation models with the analysis of the  radial profiles and  assuming  specific oscillation models of the tori.  The geodesic frequencies are governed by the background geometry  and additionally determined by the constraints imposed on the \textbf{RAD}.
Nevertheless, the   disko-seismology effect, for each toroid is in many aspects dependent on its geometrical features and particularly its  geometrical thickness. Therefore,  part of our analysis was also  dedicated to an evaluation of the impact of the disk geometry (specifically its thickness) in this investigation. In future we plan to investigate the influence of the tori thickness on the modifications of the oscillatory frequencies  related to HFQPOs in the RAD framework. Another possibility is the study of the role of  the toroidal magnetic fields and  the relation  to oscillating string  loops \citep{2012JCAP...10..008S}.

As a sideline of this study,
we explored  in  Appendix \il(\ref{Sec:magne}) the case of \textbf{RADs} where some of their components are  magnetized tori with toroidal magnetic field  \citep{Komissarov:2006nz,epl,mnras}, in the approach considered in \cite{epl,mnras}.
In this Appendix we also discuss  the case when
 {leading \textbf{RAD} function}, defining the distribution of tori in the \textbf{RAD} with  misaligned disks   has changed to an alternative definition.

 In \cite{mis(a)}  we also pointed out the possibility  that such \textbf{RAD} systems could form {accreting globuli} of matter  ``embedding'' the central
 static \textbf{BH}    in a set of orbiting \textbf{RAD} tori where  the \textbf{BH}  horizon would be partially or totally  ``covered'' to an observer at   infinity\footnote{{Note that in \cite{23,263,22,2013ApJS..209...15C,2014PhRvD..90d4029K,24,25} were considered  electrically , off-equatorial, charged configurations, i.e. levitating  tori or clouds,  existing in addition to  extended toroidal structures  crossing the equatorial plane of the central massive attractor. In some circumstances these may  be possibly  interpreted as a  sort of  \textbf{BH} "horizon covering"  (with respect to a very wide observational angle).  In this sense similarly to  the case of globules considered here, where the  black hole could be "embedded", for  a period of  its life, in such  orbiting  configuration having several   maximum (and eventually minimum) points of  pressure (and density). Such obscuring effects were also predicted  by the case of inscribed Keplerian disk orbiting with different inclination angles \citep{ido2019}}}. The presence of combined oscillatory modes  from different orbiting tori could possibly be understood as the "pulsation" of such globules and therefore give track for their existence. Luminous anuli were traces of the presence of a complex structure for inner edges of orbiting  accretion disks that could be traced (for example as intertwined luminous profiles) from the observations of Event Horizon Telescope-- \cite{Akiyama:2019cqa,Akiyama:2019brx,Akiyama:2019sww,Akiyama:2019bqs,Akiyama:2019fyp,Akiyama:2019eap}.
 Further notes  on these issues are also discussed in Appendix\il(\ref{Sec:doc-ready}). In  Fig.\il\ref{Fig:solidy} we have  shown different significant  view angles for orbiting \textbf{RAD}    found from  the integration of the  Euler equations for the misaligned tori. These orbiting configurations  should be recognizable by their peculiar ringed  structure, therefore in  Fig.\il\ref{Fig:fondam} we report also  schemes for different   \textbf{RAD} view angles.  Eventually they should be also recognizable  from absorbtion due to the  presence of an  inert or active  (covered)   inner orbiting torus.  Further optical effects associated to geometrically thick tori  that can be considered as the components of these orbiting structures are  discussed for example  in \cite{KS10,S11etal,Schee:2008fc,Schee:2013bya}).

 For the accreting globuli model, the issue of  the stability of these  static \textbf{BHs} which would be immersed in the  set of misaligned thick tori becomes particularly significant. In this context  the  influence of the tori self-gravity should be also discussed\footnote{{For example in \cite{Pugliese:2014yla}
  the stability of Einstein-Maxwell perfect fluid configurations
with a privileged radial direction (a self-gravitating spherically symmetric ideal plasma ball) as been constrained showing an important dependence on the sound velocity.}}.
 In future works, we expect to use the constraints given in \cite{mis(a)}, and the analysis considered here, as initial data for more complex dynamical situations.
 The analysis presented here will be extended  to the case  of Kerr attractors, where a co-evolution of central \textbf{BH} with the disks  can occur and influence of magnetic fields could be more dominant in several aspects of tori and \textbf{BH} energetics.

\medskip

D. P.  and Z. S. acknowledge the financial and institutional support of Silesian University.
Z. S. acknowledges the support of the Czech Science Foundation grant No. 19-03950S.

\appendix
\section{Review of misaligned (accreting) tori morphology}\label{Sec:doc-ready}
In this Appendix  we    present explicitly  some morphological  characteristics of the misaligned (accreting) tori,  used in Sec.\il(\ref{Sec:epi}),  especially with regard to the geometric thickness and the toroids unstable phases (relative to the cusped closed configurations). The evaluation of the geometrical thickness for instance  has an essential role in the evaluation of the effects of disk-seismology discussed  in  Sec.\il(\ref{Sec:epi}).
Details on these quantities and their  derivation are presented  in  \cite{mis(a)}.
The  \textbf{RAD} geometry and morphology can be parameterized in terms of  $(\ell,K)$, and only one  of these parameters in the case of cusped tori.
We consider especially   the  torus  elongation  $\lambda(\ell,K)$, the location of inner  and outer edge ($r_{in}$, $r_{out}$), the
location of the torus  center  $r_{cent}$ (point of maximum density and hydrostatic pressure in HD model), the location of the geometric  maximum $r_{\max}\equiv x_{\max}$  of the  \textbf{RAD} tori, and  the  torus thickness $S\equiv2h_{\max}/\lambda$, where $h_{\max}\equiv y_{\max}$ is the torus height. In the particular case of cusped disk where  these quantities depend on one parameter ($\ell_{crit}$ or $K_{crit}$ or $r_{crit}$) only. (In the following with the  notation $crit$ we indicate  quantities calculated at the critical points of the torus potential, i.e., either the  torus centers or the cusps.)

%
%
%

\medskip

\textbf{List of main toroids features}

\medskip

\begin{figure*}
\begin{center}
  \includegraphics[width=9cm]{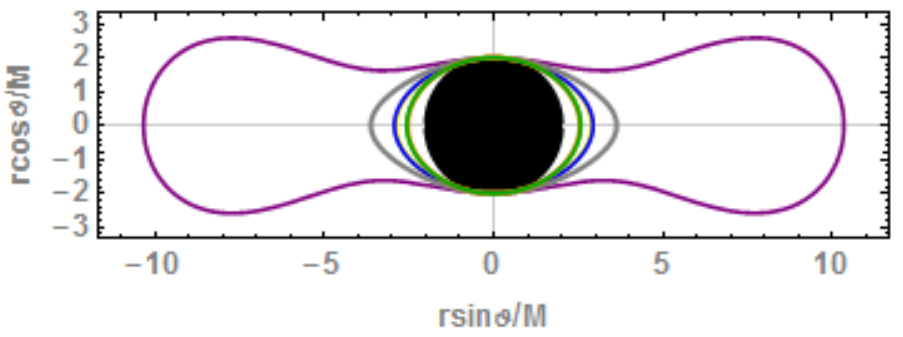}
  \includegraphics[width=4cm,angle=90]{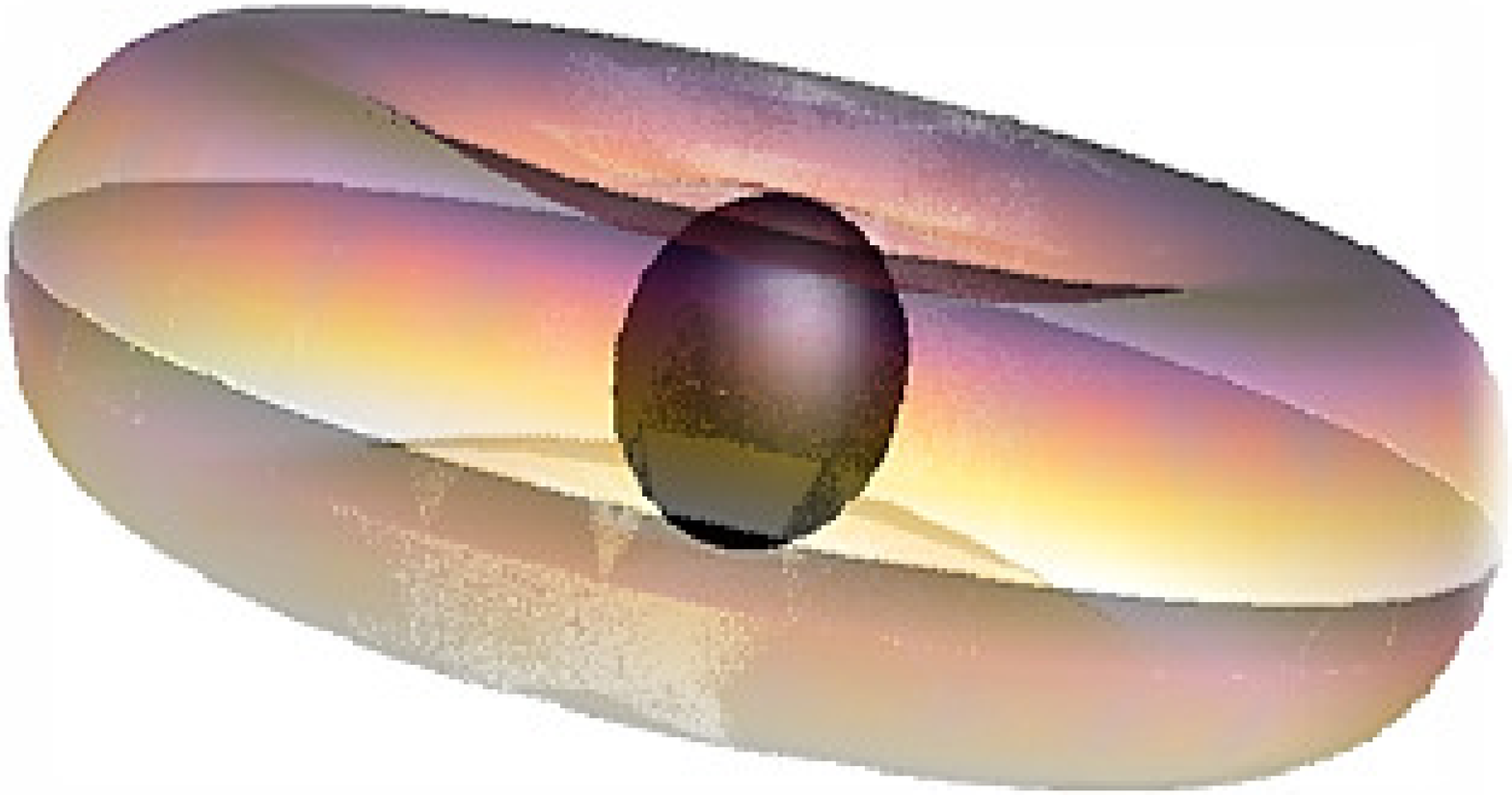}
  \caption{{\label{Fig:ecompoli}Closed, fat-disk surfaces, solutions of equation\il\ref{Eq:Eulerif0} embracing the central \textbf{BH} with inner edge $r_{in}^{BH}(\ell,K)$, correspond to different values of $\ell$ and $K$ parameters--see Appendix\il(\ref{Sec:doc-ready})  and  \protect\cite{pugtot}. Left panel shows 2D integrations, right  panel is a 3D integration, black central region is the central Schwarzschild \textbf{BH}.}}
  \end{center}
\end{figure*}

\begin{figure*}
  \includegraphics[width=9cm,angle=90]{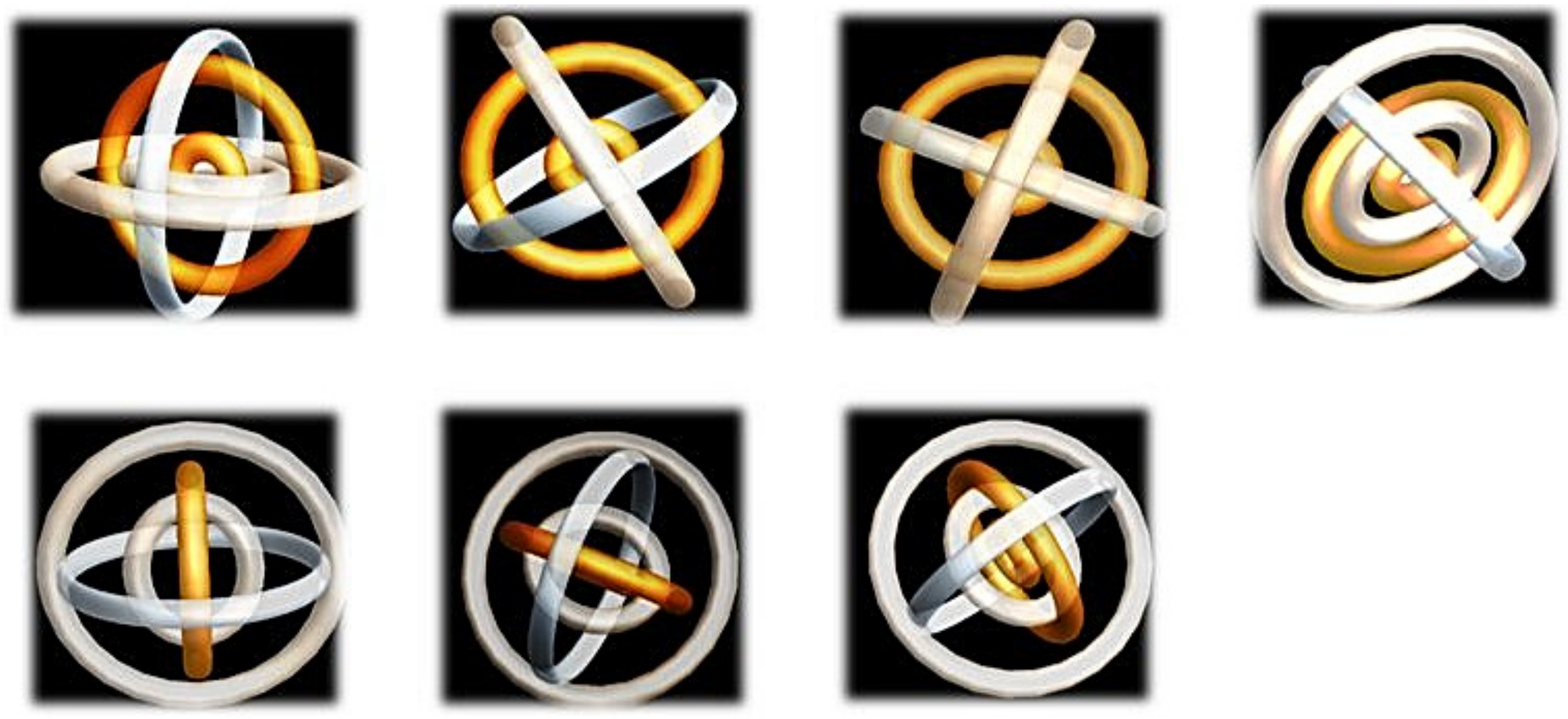}
  \caption{Schematic representation of \textbf{RAD} of order (number of orbiting disk components)  $n=5$ all orthogonal, quiescent and non-interacting  orbiting around a central attractor. The scheme represents different viewing angles where the central black hole (outer horizon) is embedded into the orbiting globule. The \textbf{eRADs} (on the order $n=2$) embedded in the \textbf{RAD} are  differently colored, the coplanar ringed  structure is evident and should also be recognized also  in the observations--see Figures\il\ref{Fig:solidy}.}\label{Fig:fondam}
\end{figure*}

\textbf{Edges and elongations} 
%
\bea\label{Eq:outer-inner-l-A1}
&&
\mbox{\textbf{Outer torus edge:}}\quad
r_{out}\equiv \frac{2 \left[1+\mathbb{K} \tau  \cos \left(\frac{1}{3} \cos ^{-1}(\alpha )\right)\right]}{3 \mathbb{K}};\quad  \mbox{\textbf{Inner torus edge:}}\quad r_{in}\equiv\frac{2 \left[1-\mathbb{K}\tau  \sin \left(\frac{1}{3} \sin ^{-1}(\alpha )\right)\right]}{3\mathbb{K}}.
 \\\label{Eq:elong-l-A1}
 &&\mbox{\textbf{Tori elongation:}}\quad
\lambda\equiv \frac{2 \tau  \cos \left(\frac{1}{6} \left[2 \cos ^{-1}(\alpha )+\pi \right]\right)}{\sqrt{3}},
\\&&\nonumber
\mbox{where}
\quad
\tau\equiv\sqrt{3} \sqrt{-\frac{\Qa}{\mathbb{K}}+\Qa+\frac{4}{3 \mathbb{K}^2}},\quad \mathbb{K}:\; K\equiv\sqrt{1-\mathbb{K}},
\quad
 \alpha\equiv\left[\frac{8-9  \Qa (\mathbb{K}-1) \mathbb{K} (3 \mathbb{K}-1)}{\mathbb{K}^3 \tau ^3}\right],
\quad\Qa\equiv \ell^2
\eea
quantity $\lambda(\ell,K)$  is the elongation on   each symmetry plane     of a cusped or quiescent torus
\footnote{It is possible to find the location of the outer radius of the innermost surface embracing the central \textbf{BH}, a closed toroidal solution  of equation\il\ref{Eq:Eulerif0} (with appropriate boundary conditions) typical of the geometrically thick disks considered here. These surfaces are for example represented in Fig\il\ref{Fig:fondam}  in the case of cusped tori--$(\mathbf{A}^0,\mathbf{A}^1)$-models. We have removed these structures  in the case of quiescent tori of  Figures\il\ref{Fig:fondam}--$(\mathbf{B},\mathbf{C})$-models. They generally appear in three different  classes of solutions whose  role still needs to be clarified. Doubled structures are associated with the quiescent disks with radius
 $r_{in}^{BH}(\ell,K)$:
 \[ r_{in}^{BH}\equiv\frac{2\left[\frac{1}{\mathbb{K}}-\tau  \sin \left(\frac{1}{6} \left[2 \cos ^{-1}(\alpha )+\pi \right]\right)\right]}{3},\] close to the horizon and coincident  with the inner ``Roche lobe'' of cusped torus, which means that the  distance \[\lambda_{in}^{BH}\equiv r_{in}-r_{in}^{BH}=\frac{2}{3} \tau  \left[\sin \left(\frac{1}{6} \left[2 \cos ^{-1}(\alpha )+\pi \right]\right)-\sin \left[\frac{1}{3} \sin ^{-1}(\alpha )\right]\right],
\]
is vanishing  for a cusped (accreting) torus. Doubled lobes  have clearly equal values of the parameters $(\ell,K)$, and  correspond therefore to the same point of the curves $\ell(r)$ and $K(r)$. An  interesting issue would be the observation of these doubled configurations in the first or later phases of the accretion disks evolutions. Another similar, not doubled, structure occurs for different values of  $\ell$ and $K$ parameters, as detailed in  \cite{pugtot} for the case of a Kerr \textbf{BH} with  a general spin $a\in[0,M]$--see also \cite{proto-jet}. It should be noted here that these disk-like solutions of the HD equations  still need to be settled in a proper interpretative frame--Figure\il\ref{Fig:ecompoli}. It is an aim of  future analysis  to investigate  observational evidences  through  the exploration of the  optical effects in the regions  close to the  \textbf{BH}, for example a clear and immediate observational channel for these structures   is provided by the  recent analyzes of the   Event Horizon Telescope \cite{Akiyama:2019cqa,Akiyama:2019brx,Akiyama:2019sww,Akiyama:2019bqs,Akiyama:2019fyp,Akiyama:2019eap}. }

\textbf{Critical points of cusped  tori:}
\bea&&\label{Eq:rcentro}
\mbox{\textbf{The center}  of maximum density (and hydrostatic pressure):}
\\
&&\nonumber
r_{cent}(\ell)\equiv\frac{1}{3} \left[\Qa+2 L_{ \ell} \cos \left(\frac{1}{3} \cos ^{-1}(L_{\mathcal{ll}})\right)\right],\quad
\mbox{where}\quad L_{ \ell}=\sqrt{\Qa \left(\Qa-12\right)}\quad\mbox{and}\quad L_{\mathcal{ll}}=\frac{\Qa \left(\ell^4-18 \Qa+54\right)}{L_{ \ell}^3}.
 \\\label{Eq:r-inner-A1}&&\mbox{\textbf{The inner edge:}}\quad
 r_{\times}(\ell)\equiv\frac{1}{3} \left[\Qa-2 L_{ \ell} \cos \left(\frac{1}{3} \left[\cos ^{-1}(L_{\mathcal{ll}})+\pi \right]\right)\right].
\eea
\textbf{The critical $K$-parameter:} At the  center of maximum density (and hydrostatic pressure), and at the  inner edge of accreting torus  there is
%
%
\bea&&\label{Eq:grod-lock}
K_{cent}(\ell)\equiv\sqrt{\frac{\left[\Qa+2 L_{ \ell} \cos \left\{\frac{1}{3} \cos ^{-1}(L_{\mathcal{ll}})\right\}-6\right] \left[ \Qa+2 L_{ \ell} \cos \left(\frac{1}{3} \cos ^{-1}[L_{\mathcal{ll}}]\right)\right]^2}{3 \Qa \left[3 \ell^4+2 \left(2  \Qa-15\right) L_{ \ell} \cos \left[\frac{1}{3} \cos ^{-1}(L_{\mathcal{ll}})\right]-39  \Qa+2 L_{ \ell}^2 \cos \left(\frac{2}{3} \cos ^{-1}[L_{\mathcal{ll}}]\right)+54\right]}},
 \\
 &&K_{\times}(\ell)\equiv\sqrt{\frac{\left[ \Qa-2 L_{ \ell} \sin \left(\frac{1}{3} \sin ^{-1}[L_{\mathcal{ll}}]\right)-6\right] \left( \Qa-2 L_{ \ell} \sin \left(\frac{1}{3} \sin ^{-1}[L_{\mathcal{ll}}]\right)\right)^2}{3 \Qa \left[3 \ell^4+2 \left(15-2  \Qa\right) L_{ \ell} \sin \left[\frac{1}{3} \sin ^{-1}(L_{\mathcal{ll}})\right]-39  \Qa-2 L_{ \ell}^2 \cos \left(\frac{2}{3} \sin ^{-1}[L_{\mathcal{ll}}]\right)+54\right]}}.
\eea
%
%
These expressions allow to explicit the  distribution of $K$-parameters in the \textbf{RAD} in terms of the leading function $\ell(r)$ in equation\il\ref{Eq:lqkp}, correlating  $K$ and $\ell$ in the  misaligned tori.

\medskip

\textbf{The critical $\ell(K)$ and $r_{K}$ dependence}

\medskip

Eliminating  also the radial dependence  there is
 \bea\label{Eq:crititLdeval}
 &&
 \ell_{crit}^o(K)\equiv\sqrt{-\frac{-27 K^4+K \left(9 K^2-8\right)^{3/2}+36 K^2-8}{2K^2 \left(K^2-1\right)}},
  \quad\ell_{crit}^i(K)\equiv\sqrt{\frac{27 K^4+K \left(9 K^2-8\right)^{3/2}-36 K^2+8}{2K^2 \left(K^2-1\right)}},
 \\\nonumber
 &&
\ell_{crit}^o(K)> \ell_{crit}^i(K)>\ell_{mso},\quad \ell_{crit}^i(K)\in[\ell_{mso},\ell_{\gamma}[,
 \eea
 note $K_{mso}^2=8/9$.
 The function  $\ell_{crit}^o(K)$,  for $K=K_{\max}$, provides the specific angular momentum of the torus whose instability accreting phase is associated  to the occurrence of the  value $K=K_{\max}$, while  $\ell_{crit}^i(K)$,  for $K=K_{cent}$, provides the specific angular momentum of the torus whose center of maximum density  corresponds to  $K=K_{cent}$.
   For a quiescent torus  $K\in [K_{\min},K_{\max}]$ is a free parameter, where  $K_{mso}<K_{\min}<K<\mathfrak{K}$,  and  $\mathfrak{K}= K_{\max}$ for torus in the accreting range of specific angular momentum values or
 $\mathfrak{K}=1$ otherwise. The couple $(K_{\min},K_{\max})$  is defined by  the $K(r)$ \textbf{RAD} energy function of equation\il\ref{Eq:lqkp}  at maximum  $r_{\max}$ or minimum $r_{\min}$ of the hydrostatic pressure.
Finally, considering  $\ell(r)=\ell_{crit}^o(K)$ and
 $\ell(r)=\ell_{crit}^i(K)$, we find an expression for the critical radii   $r_{crit}(K)$ of the tori as a function of $K_{crit}$ as
 \bea\label{Eq:nicergerplto}
r_{crit}^o(K)\equiv -\frac{8}{K \left(\sqrt{9 K^2-8}+3 K\right)-4},
 \quad r_{crit}^i(K)\equiv\frac{8}{K \left(\sqrt{9 K^2-8}-3 K\right)+4}.
 \eea
There is $r_{crit}^i(K_{\times})=r_{inner}^{\times}$ (inner edge for accreting torus), and   $r_{crit}^o(K_{cent})=r_{cent}^{\times}$ (center of cusped configurations). 

\medskip

\textbf{The tori geometric thickness}

\medskip

Geometric maximum  radius $r_{\max}^o(K,\ell)$ of the torus,   and the innermost surface $r_{\max}^i(K,\ell)$, and the maximum  height $h_{\max}^o(K,\ell)$ of the torus surface, given  as functions of $K$ and $\ell$, where $r_{\max}^i$ is the location of maximum point for the inner Roche lobe close to the central \textbf{BH},
$h_{\max}^o$ is the semi height of the torus:
\bea\label{Eq:r-quest-v}
&&
r_{\max}^o(K,\ell)\equiv\sqrt{\frac{K^2 \Qa}{K^2-1}+4 \sqrt{\frac{2}{3}} \psi  \cos \left[\frac{1}{3} \cos ^{-1}( \psi_\pi)\right]},
 \quad r_{\max}^i(K,\ell)\equiv\sqrt{\frac{K^2 \Qa}{K^2-1}-4 \sqrt{\frac{2}{3}} \psi  \sin \left[\frac{1}{3} \sin ^{-1}(\psi_\pi)\right]},\quad\mbox{and}
 \eea
 %
 \bea
\label{Eq:xit-graci}
&&
h_{\max}^o(K,\ell)\equiv\sqrt{-\frac{K^2 \Qa}{K^2-1}+\frac{\left[3 K^4 \Qa \sec \left(\frac{1}{3} \cos ^{-1}[ \psi_\pi]\right)+4 \sqrt{6} \left(K^2-1\right) \psi \right]^2}{24 \left(K^2-1\right)^4 \psi ^2}-4 \sqrt{\frac{2}{3}} \psi  \cos \left[\frac{1}{3} \cos ^{-1}(\psi_\pi )\right]},
\\
&&
 \psi \equiv\sqrt{-\frac{K^4 \Qa}{\left(K^2-1\right)^3}},\quad\mbox{and}
\quad\psi_\pi \equiv-\frac{3}{4}\sqrt{\frac{3}{2}} \left(K^2-1\right)^2 \psi.
\eea
%
%
 \textbf{RADs} with  cusped misaligned  configurations   constitute a particularly significant case.
Therefore we evaluated  also the   torus height $h_{\max}^o(r_{\times})$  and the locations  $r_{\max}^o(r_{crit})$ and $r_{\max}^i(r_{crit})$ as functions of cusp location $r=r_{\times}\in]r_{mbo},r_{mso}[$:
%
\bea\label{Eqs:rssrcitt}
&&
r_{\max}^o(r_{crit})=\sqrt{4 \sqrt{\frac{2}{3}} \psi_{\lambda } \cos \left[\frac{1}{3} \cos ^{-1}\left(-\frac{3}{4} \sqrt{\frac{3}{2}} \psi_{\lambda } \psi _{\sigma }^2\right)\right]+\frac{r^2}{(r-r_{\gamma}) \psi_{\sigma }}}
 \\\nonumber
 &&
 r_{\max}^i(r_{crit})=\sqrt{\frac{r^2}{(r-r_{\gamma}) \psi _{\sigma }}-4 \sqrt{\frac{2}{3}} \psi _{\lambda } \cos \left[\frac{1}{3} \left(\cos ^{-1}\left[-\frac{3}{4}\sqrt{\frac{3}{2}} \psi _{\lambda } \psi _{\sigma }^2\right]+\pi \right)\right]},
 \eea

 and the torus height
 \bea\label{Eq:max-prob-A1}
h_{\max}^o(r_{\times})=\left(-2 \sqrt{6} \sqrt{\frac{(r_{\times}-r_{\gamma}) (r_{\times}-r_+)^2 r_{\times}^4}{(r_{\times}-r_{mbo})^3}} \sec \left[\frac{1}{3} \cos ^{-1}(\psi_\rho )\right]+\right.
\left.\frac{9 (r_{\times}-r_+)^2 r_{\times}^2 \sec ^2\left[\frac{1}{3} \cos ^{-1}(\psi_\rho )\right]}{8 (r_{\times}-r_{mbo}) (r_{\times}-r_{\gamma})}+\frac{(r_{\times}-r_+) (5 r_{\times}-18) r_{\times}^2}{(r_{\times}-r_{mbo})^2}\right)^{1/2},
\eea
{where}
\bea
&&
 \psi _{\sigma }\equiv\frac{r_{mbo}-r}{(r-r_{\gamma}) r},\quad
 \psi_{\lambda }\equiv\sqrt{-\frac{(r-r_+)^2 r}{(r-r_{\gamma})^2 \psi _{\sigma }^3}},
 \quad \psi_\rho\equiv-\frac{3 \sqrt{\frac{3}{2}} (r_{\times}-r_{mbo})^2 \sqrt{\frac{(r_{\times}-r_{\gamma}) (r_{\times}-r_+)^2 r_{\times}^4}{(r_{\times}-r_{mbo})^3}}}{4 (r_{\times}-r_{\gamma})^2 r_{\times}^2},
\eea
\medskip

 The outer and inner edges of an accreting torus as function of $r_{\times}$ are
%
\bea&&\label{Eq:over-top}
r_{out}^{\times}(r_{\times})=\frac{2}{3} \left[\sqrt{\frac{(r_{\times}-r_{mso})^2 r_{\times}^2}{(r_{\times}-r_{mbo})^2}}\right.
 \left.\cos \left[\frac{1}{3} \cos ^{-1}\left(-\frac{(r_{\times}-r_{mso}) r}{(r_{\times}-r_{mbo}) \sqrt{\frac{(r_{\times}-r_{mso})^2 r_{\times}^2}{(r_{\times}-r_{mbo})^2}}}\right)\right]+\frac{r_{\times}}{r_{\times}-r_{mbo}}+r\right]
\eea

\bea&&\label{Eq:over-top1}
r_{inner}^{\times}(r_{\times})=\frac{1}{3} \left[\frac{r_{\times}^3}{(r_{\times}-r_+)^2}-2 \sqrt{\frac{r_{\times}^3 \left[\frac{r_{\times}^3}{(r_{\times}-r_+)^2}-12\right]}{(r_{\times}-r_+)^2}}
\right.
\left.\cos \left[\frac{1}{3} \left(\cos ^{-1}\left[\frac{r_{\times}^3 \left(\frac{r_{\times}^6}{(r_{\times}-r_+)^4}-\frac{18 r_{\times}^3}{(r_{\times}-r_+)^2}+54\right)}{(r_{\times}-r_+)^2 \left(\frac{r_{\times}^3 \left(\frac{r_{\times}^3}{(r_{\times}-r_+)^2}-12\right)}{(r_{\times}-r_+)^2}\right)^{3/2}}\right]+\pi \right)\right]\right],
\eea
from which we can obtain the critical elongation $\lambda_{\times}$, and thickness $\Sa_{\times}=2 h_{\times}/(\lambda_{\times})$ of the cusped tori where
$(r_{inner}^{\times}(r_{\times}),r_{out}^{\times}(r_{\times}))$  combine  together solutions
$r=r_{\times}$ and $r=r_{out}$.
Disk geometric thickness underlies different aspects of the physics of the accretion disks, including  the disks oscillation  modes and  accretion rates. These conditions set also the  asymptotic limits, and conditions on the quasi-sphericity of the torus discussed in \cite{mis(a)}.
 The point of maximum density and (hydrostatic) pressure in the torus is not coincident with  the geometric maximum of the torus surface, i.e., in general there is  $r_{\max}^o\neq r_{cent}$,
 Nevertheless,  there is a special class of toroids where  $r_{cent}=r_{\max}^o$.

The torus thickness is
$\Sa=1$ for  $K_I=
0.975$ and $\ell_I=
3.887\in]\ell_{mso},\ell_{mbo}[$. Schemes of different angle views of  an orbiting \textbf{RAD} are in Figures\il\ref{Fig:fondam}.
 It is clear that for  accreting configurations the  farther from the central attractor is the accreting torus    the larger it is.
 For low magnitude of the specific angular momentum the thickness is essentially determined by the  $K$ parameter (i.e. fluid density), particularly for the accreting tori, viceversa, the main governing parameter for quiescent tori at $\ell>\ell_{mbo}$ is the specific angular momentum.
\section{An alternative RAD leading function}\label{Sec:magne}
\begin{figure*}
  \includegraphics[width=5.7cm]{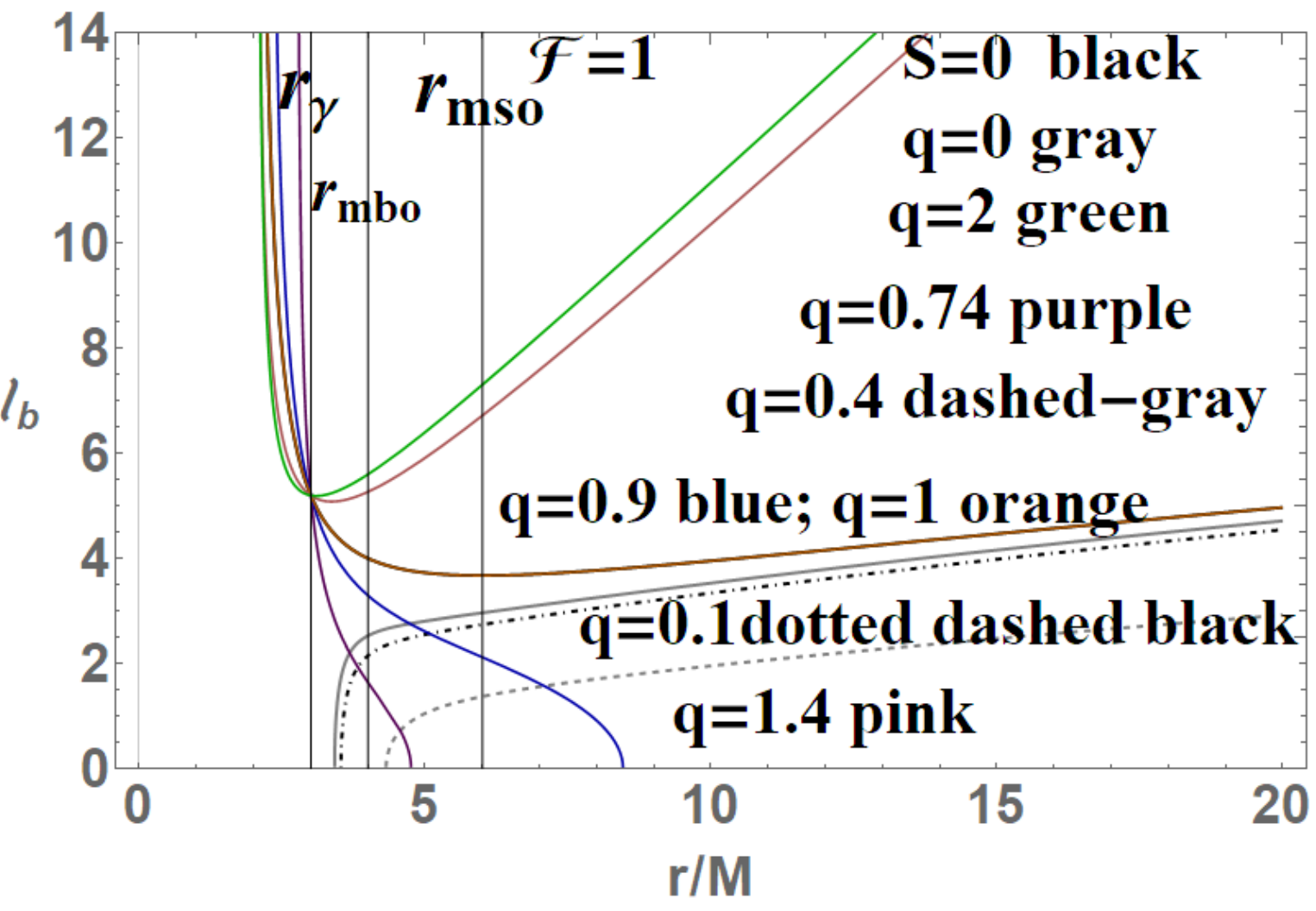}
   \includegraphics[width=5.7cm]{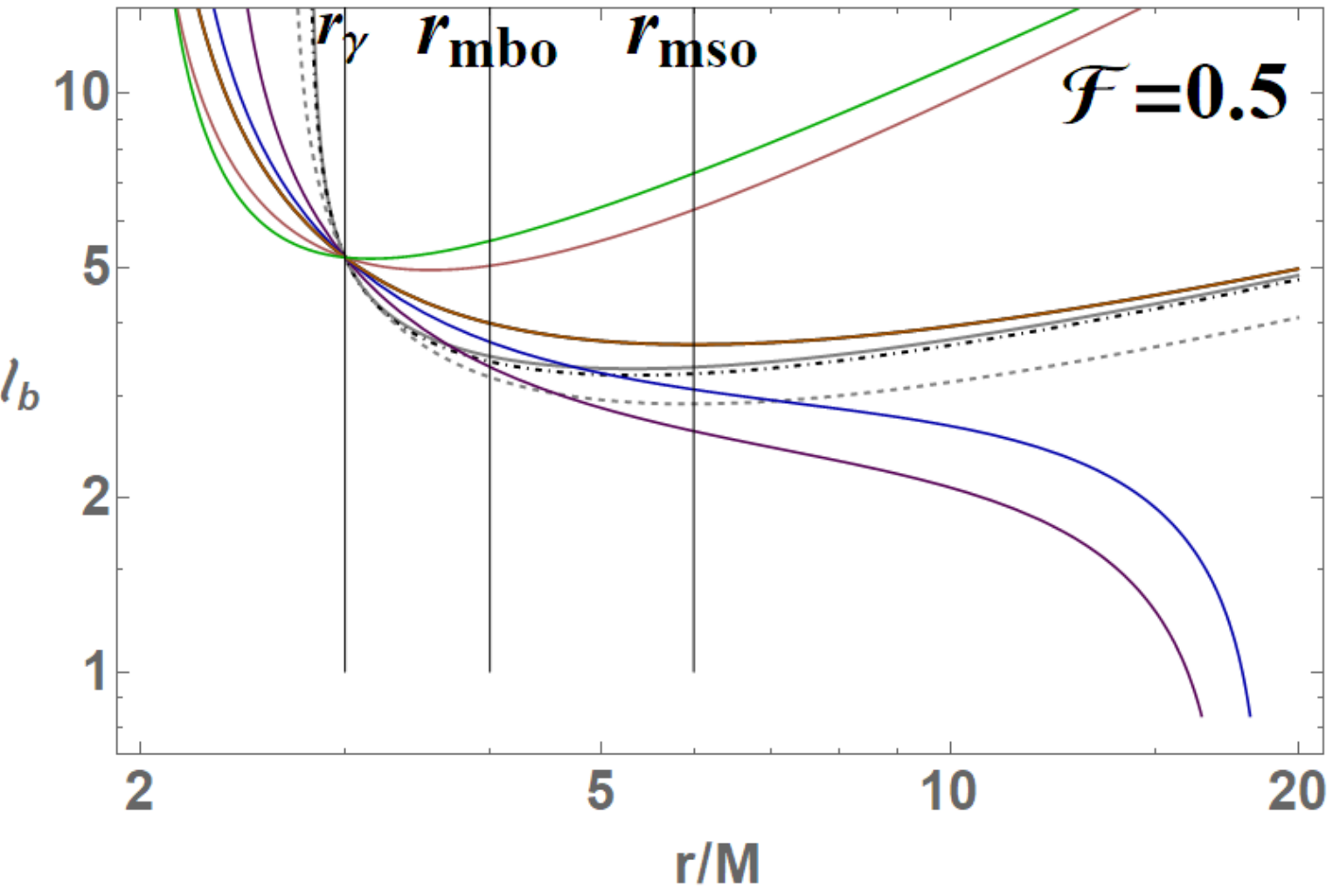}
    \includegraphics[width=5.7cm]{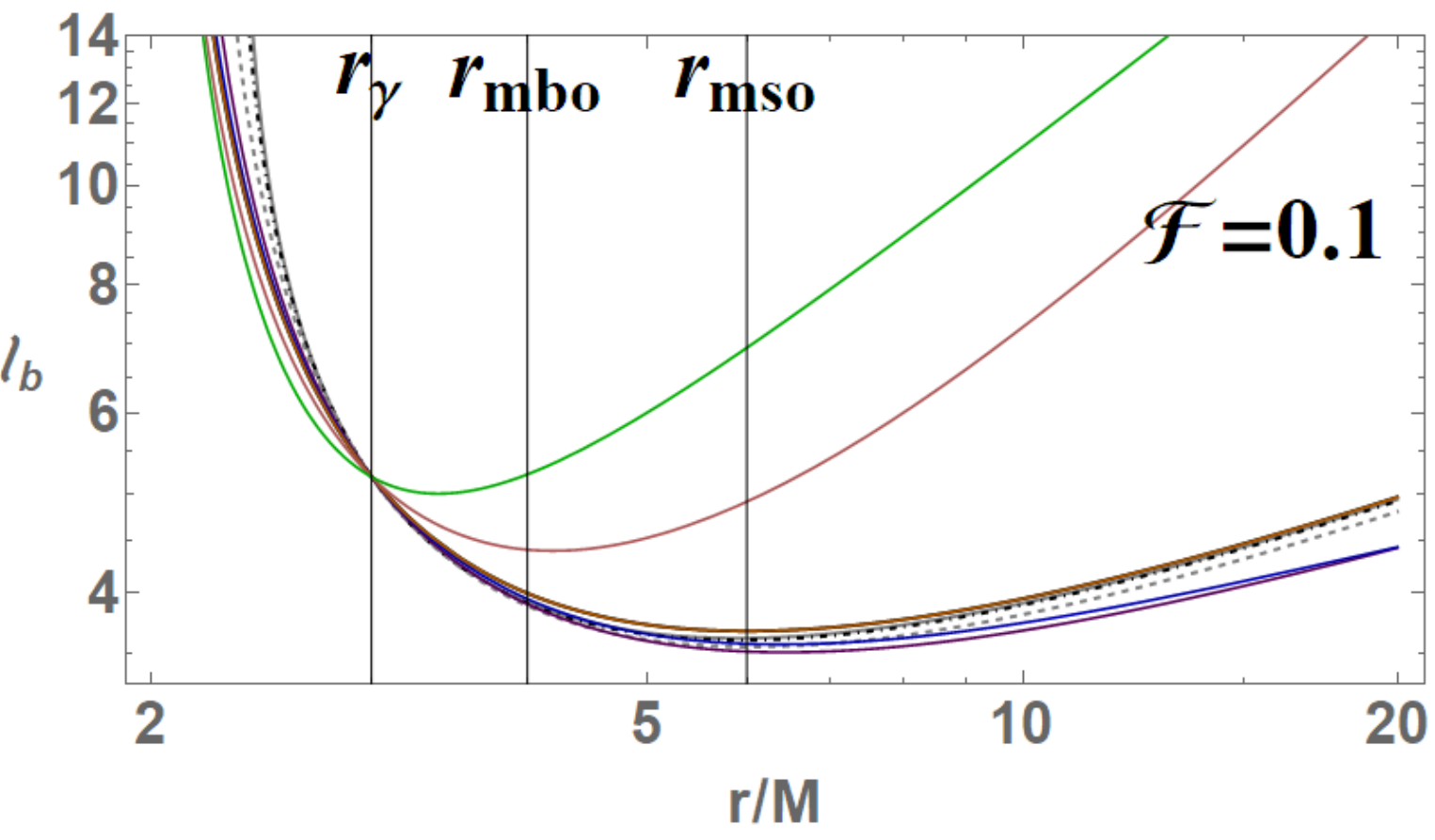}
  \caption{Fluid specific angular momentum distribution $\ell_b$ in equation\il\ref{Eq:ellb} for magnetized tori in the \textbf{RAD} with a toroidal magnetic field. Left panel: $\ell_b$ as function of $r/M$ for different values of $q$ and $\Fa$, center and right panel  $\ell_b$  for two values of $\Fa$. }\label{Fig:coxxlinas}
\end{figure*}

\begin{figure*}
   \includegraphics[width=8cm]{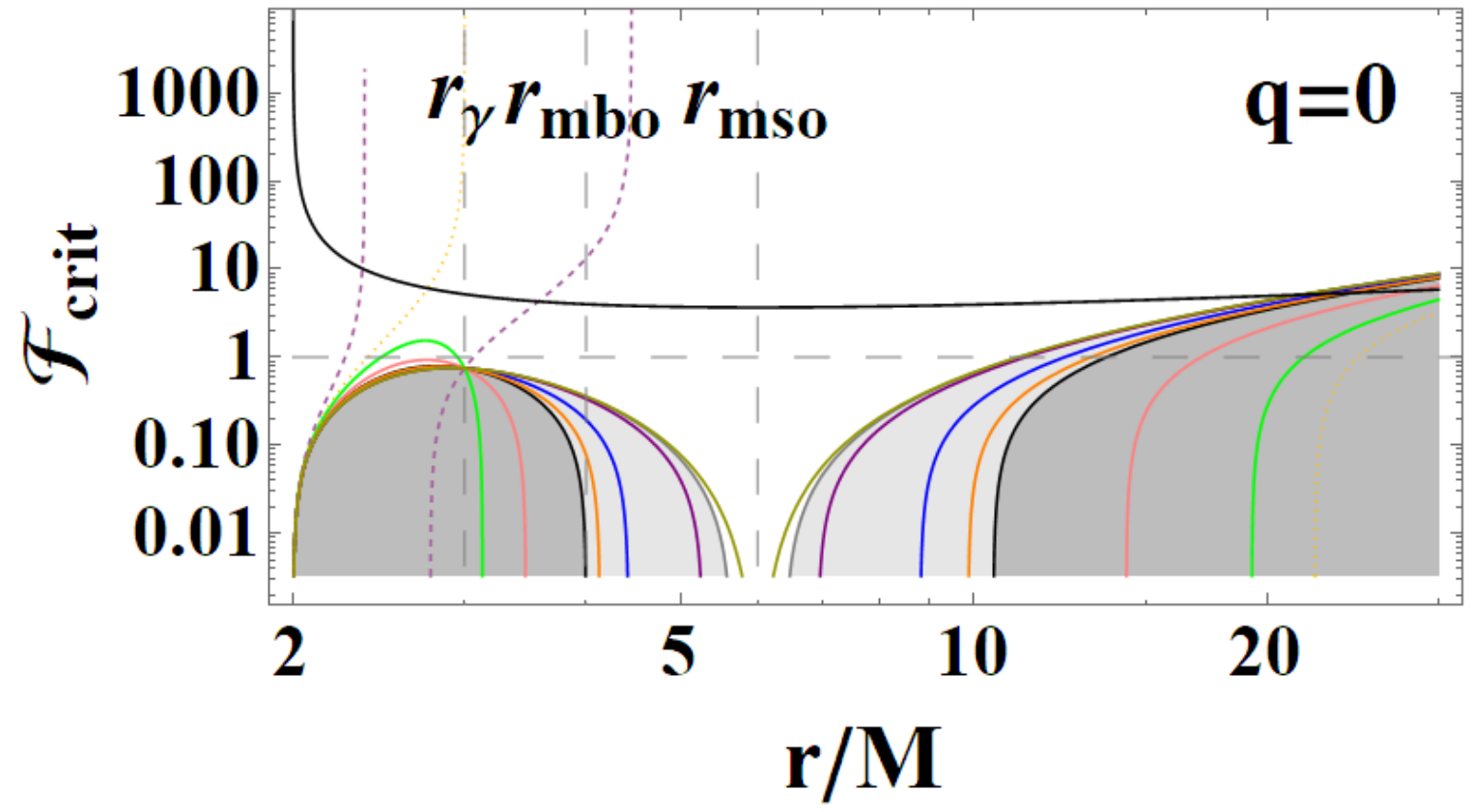}
  \includegraphics[width=8cm]{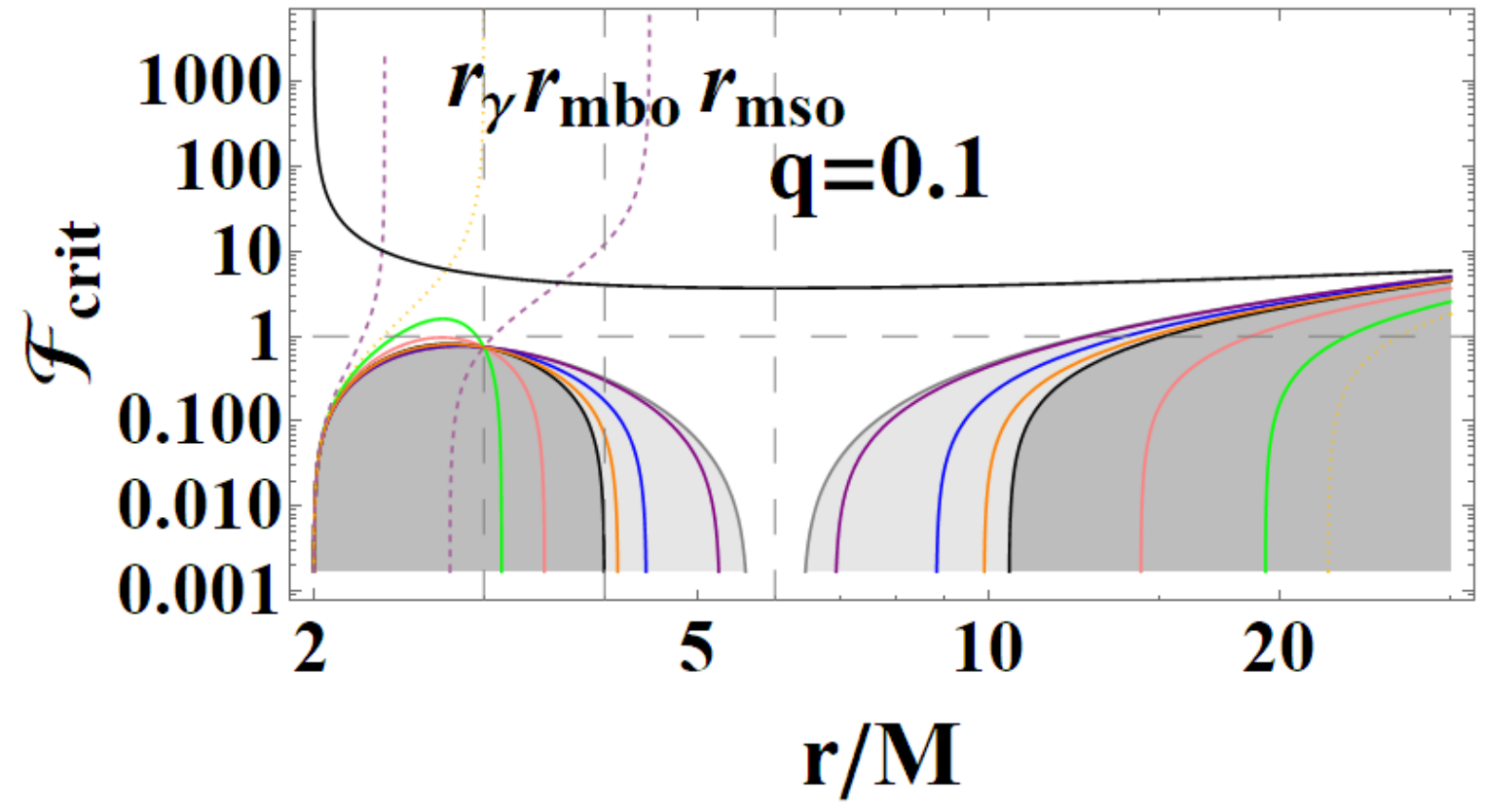}
     \includegraphics[width=8cm]{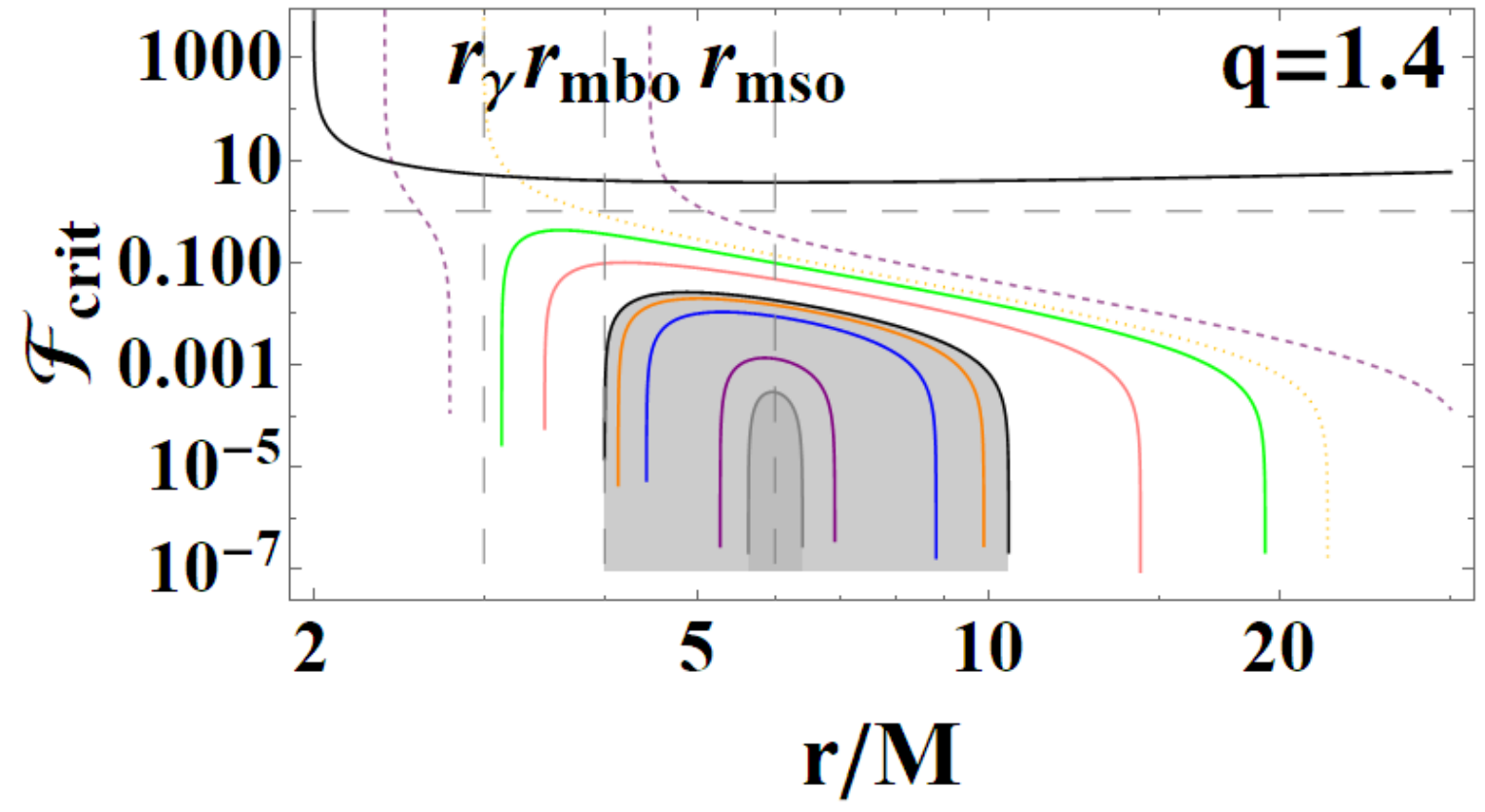}
    \includegraphics[width=8cm]{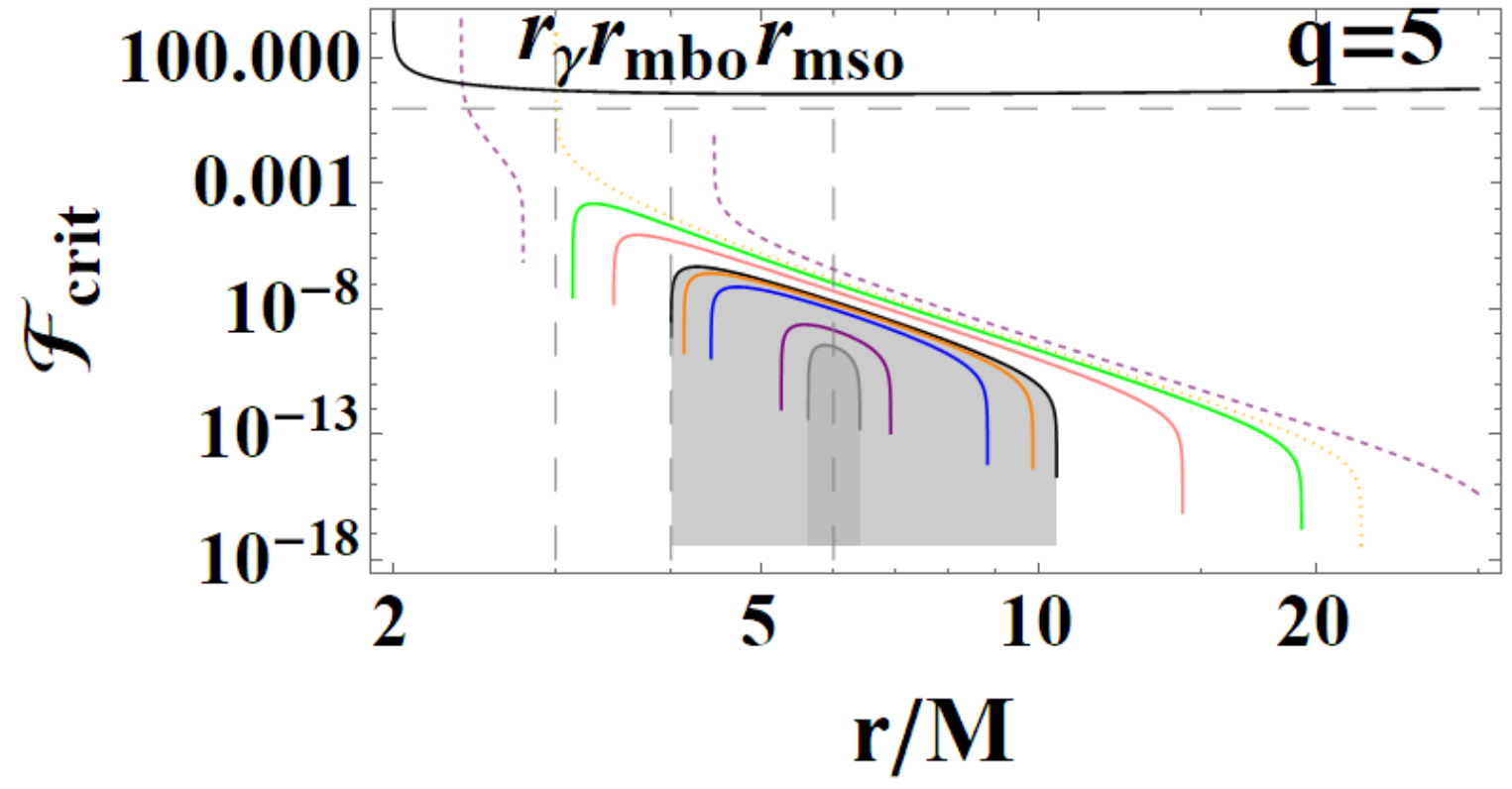}
  \caption{Magnetized tori. Function
$\Fa_{crit}$ defined in equation\il\ref{Eq:Sacrit} for fixed values of $q$ parameters. Note the different distributions in relation with the location of the marginally stable circular orbit $r_{mso}$, marginally bounded orbit $r_{mbo}$ and photon orbit $r_{\gamma}$. It is also clear that  $q<1$ represents a singular region in the space of the   $q$ parameter, , as discussed in \protect\cite{mnras}, in the case of central Kerr attractor.}\label{Fig:procePlot8}
\end{figure*}

\begin{figure*}
   \includegraphics[width=8.4cm]{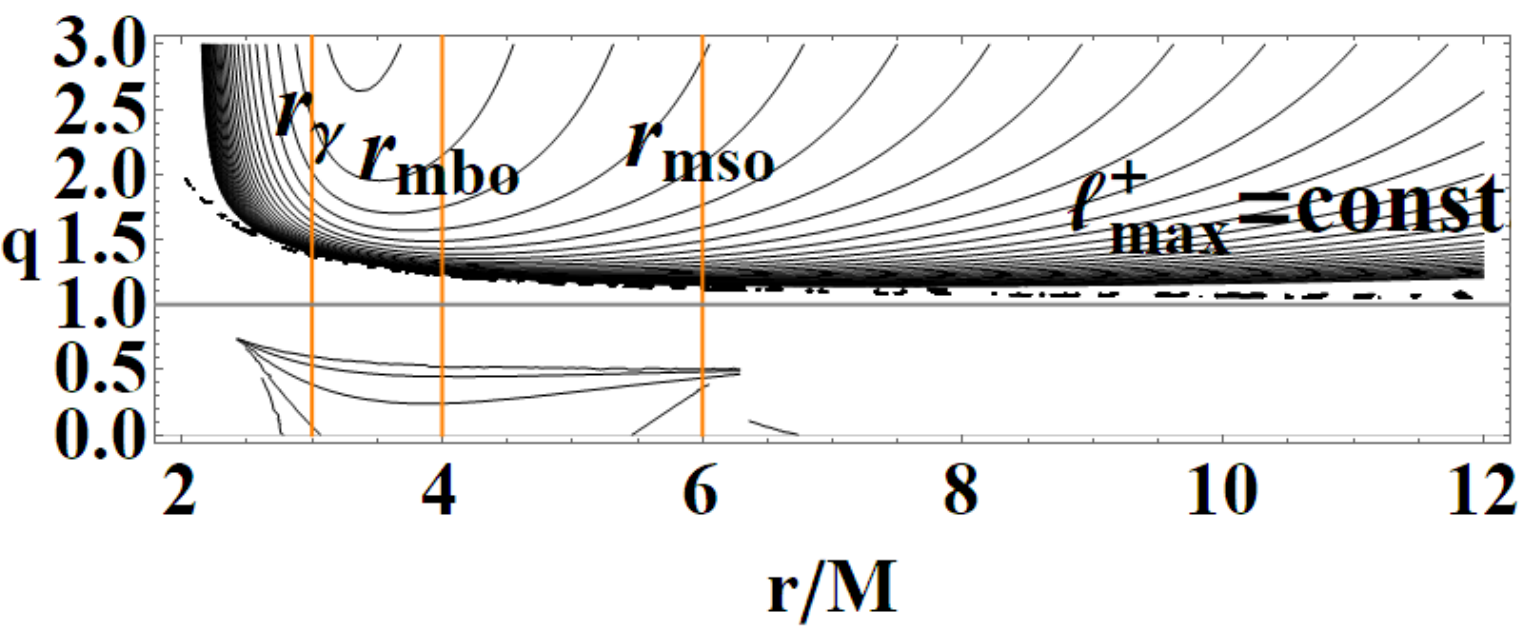}
     \includegraphics[width=8.4cm]{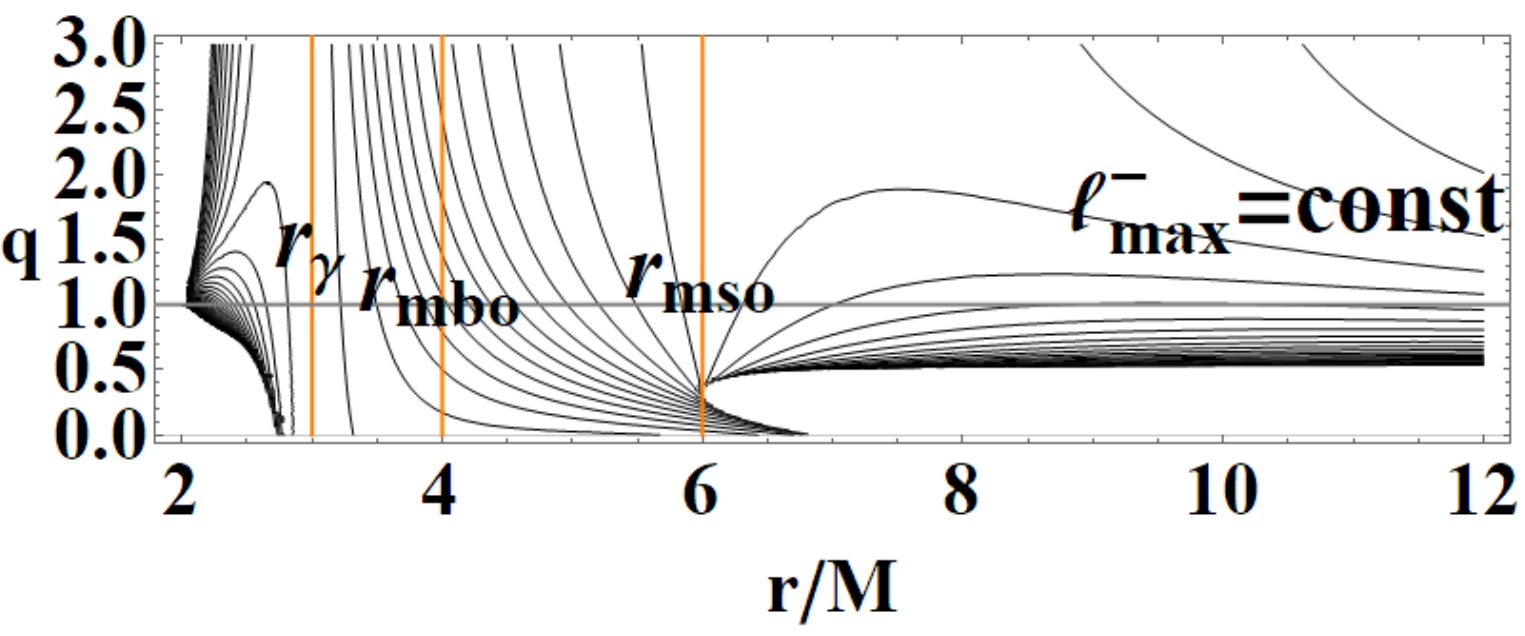}
     \includegraphics[width=8.4cm]{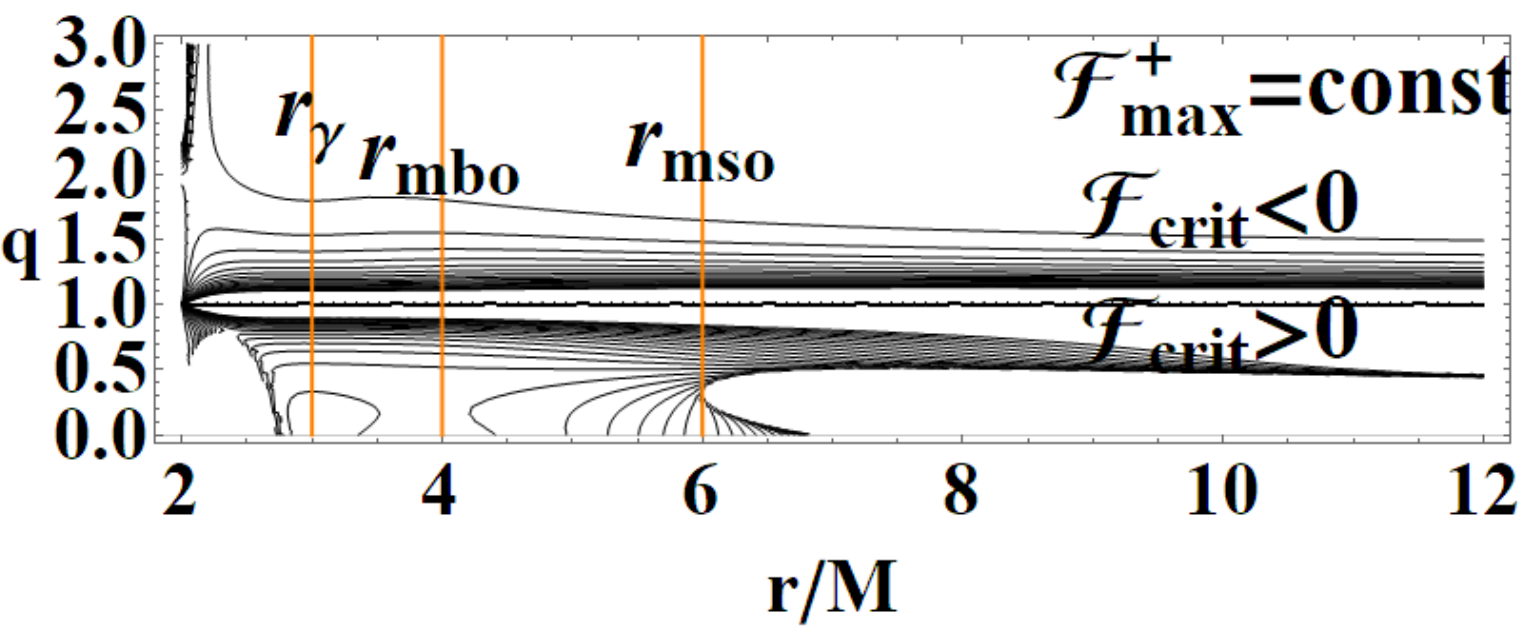}
    \includegraphics[width=8.4cm]{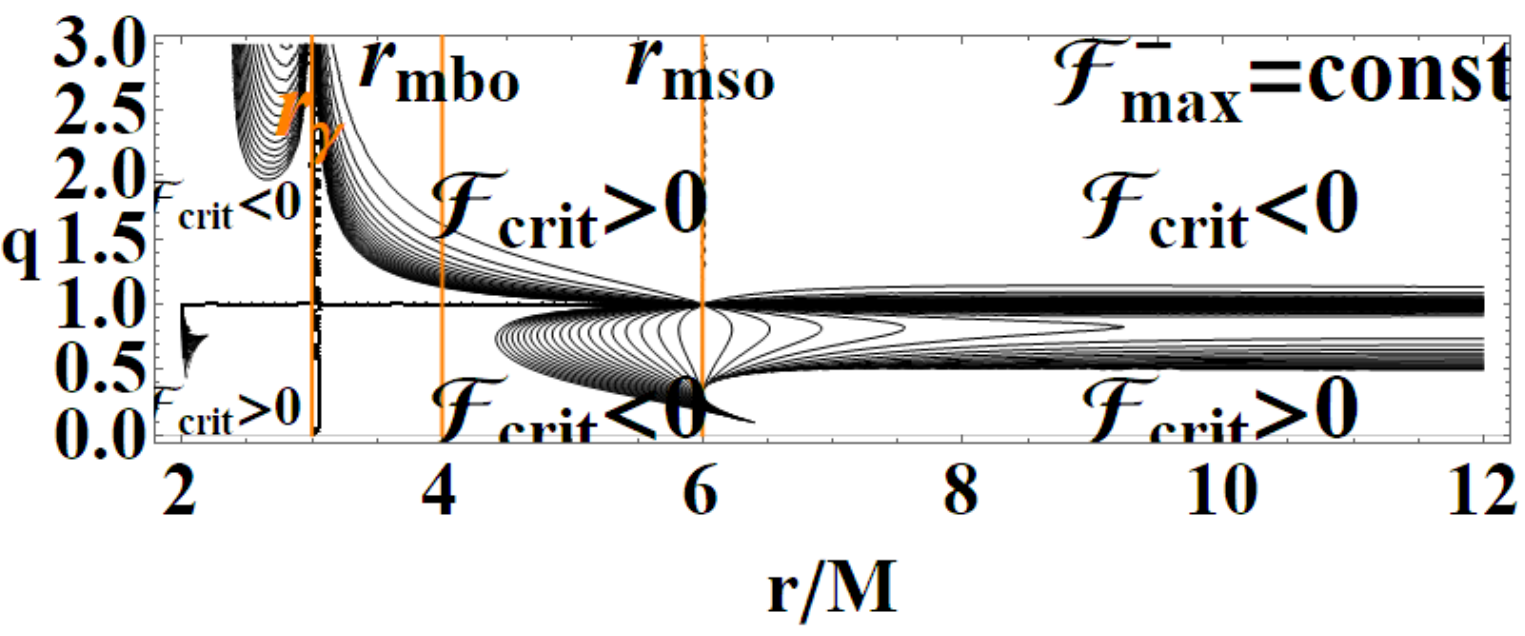}
     \includegraphics[width=8.4cm]{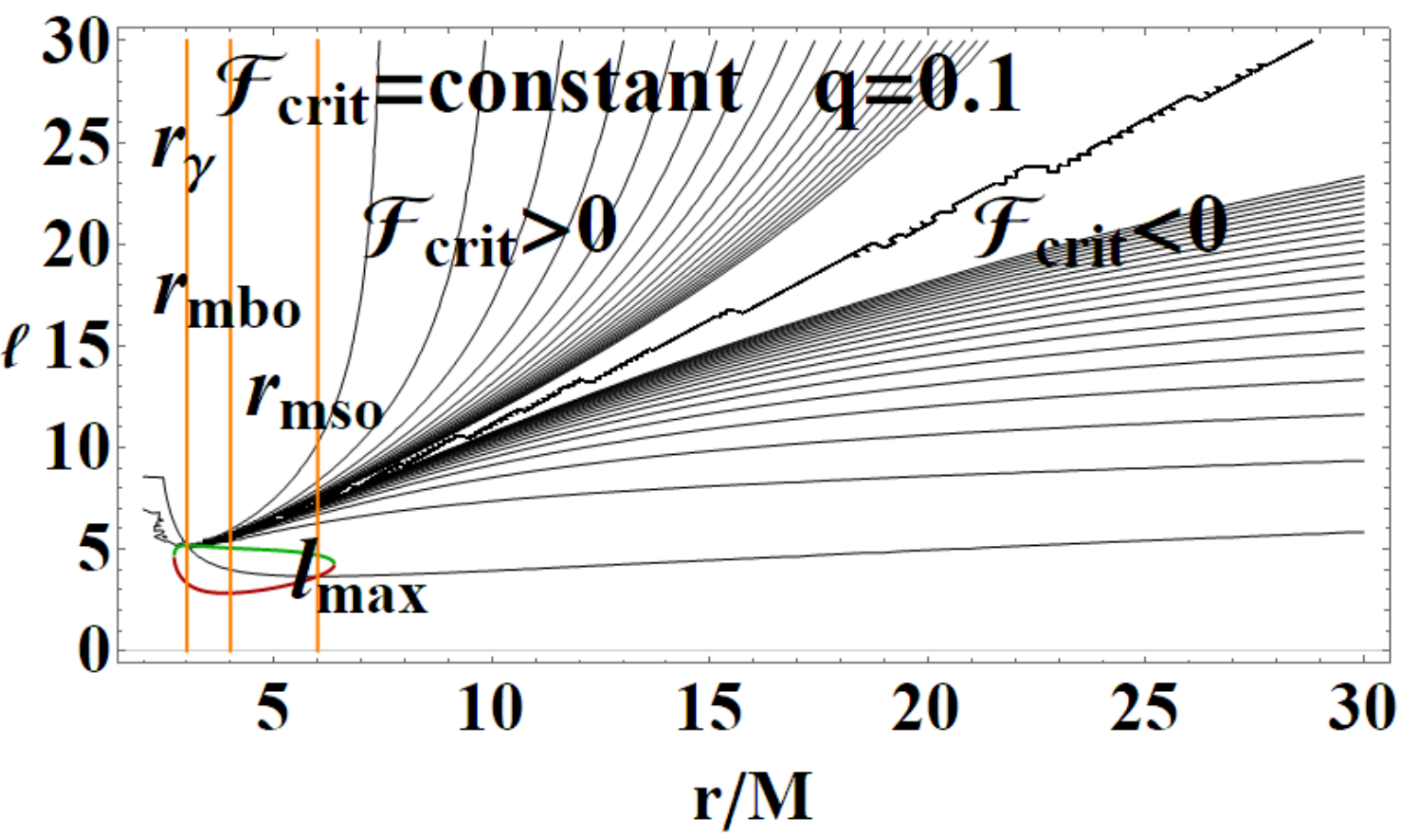}
   \includegraphics[width=8.4cm]{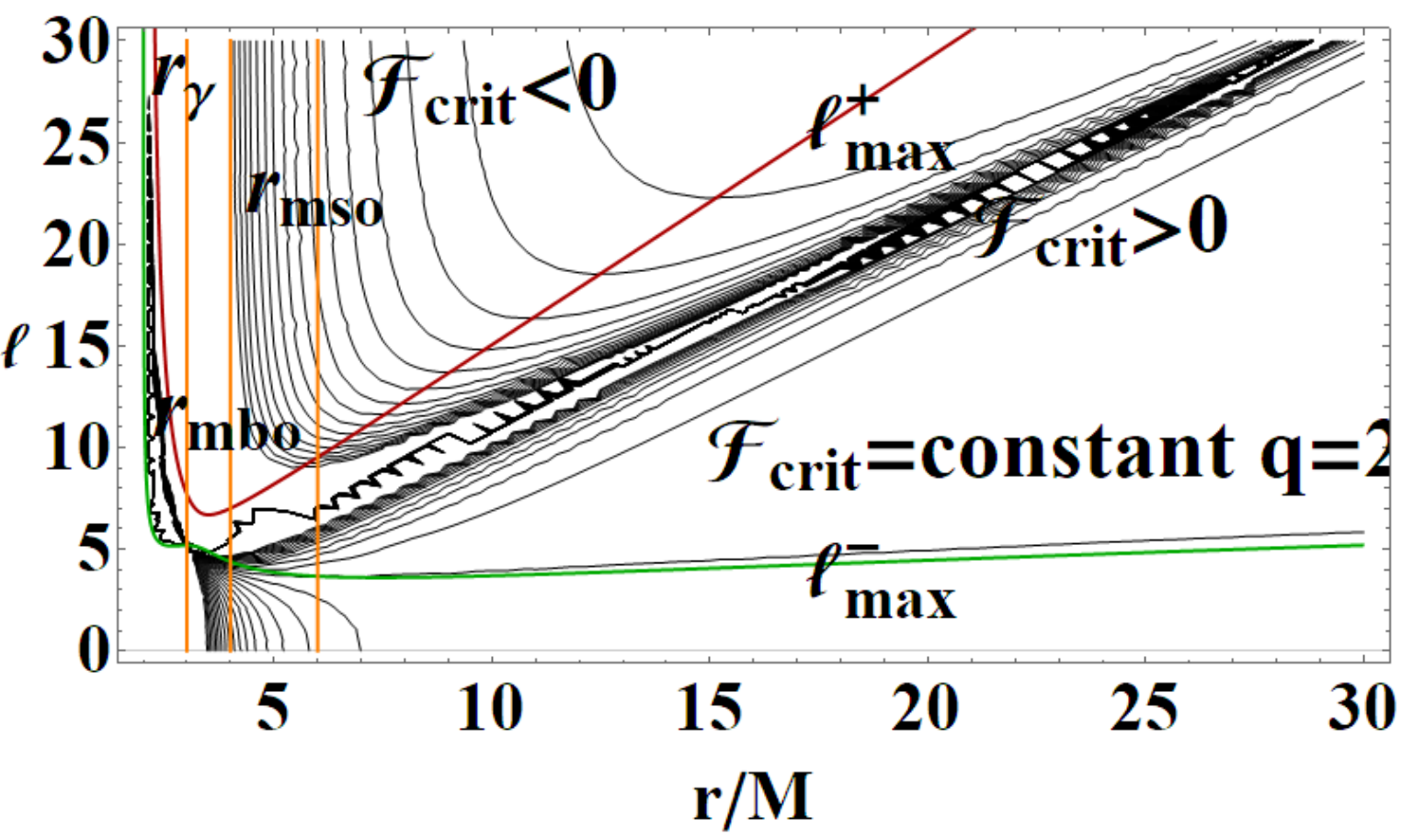}
    \includegraphics[width=8.4cm]{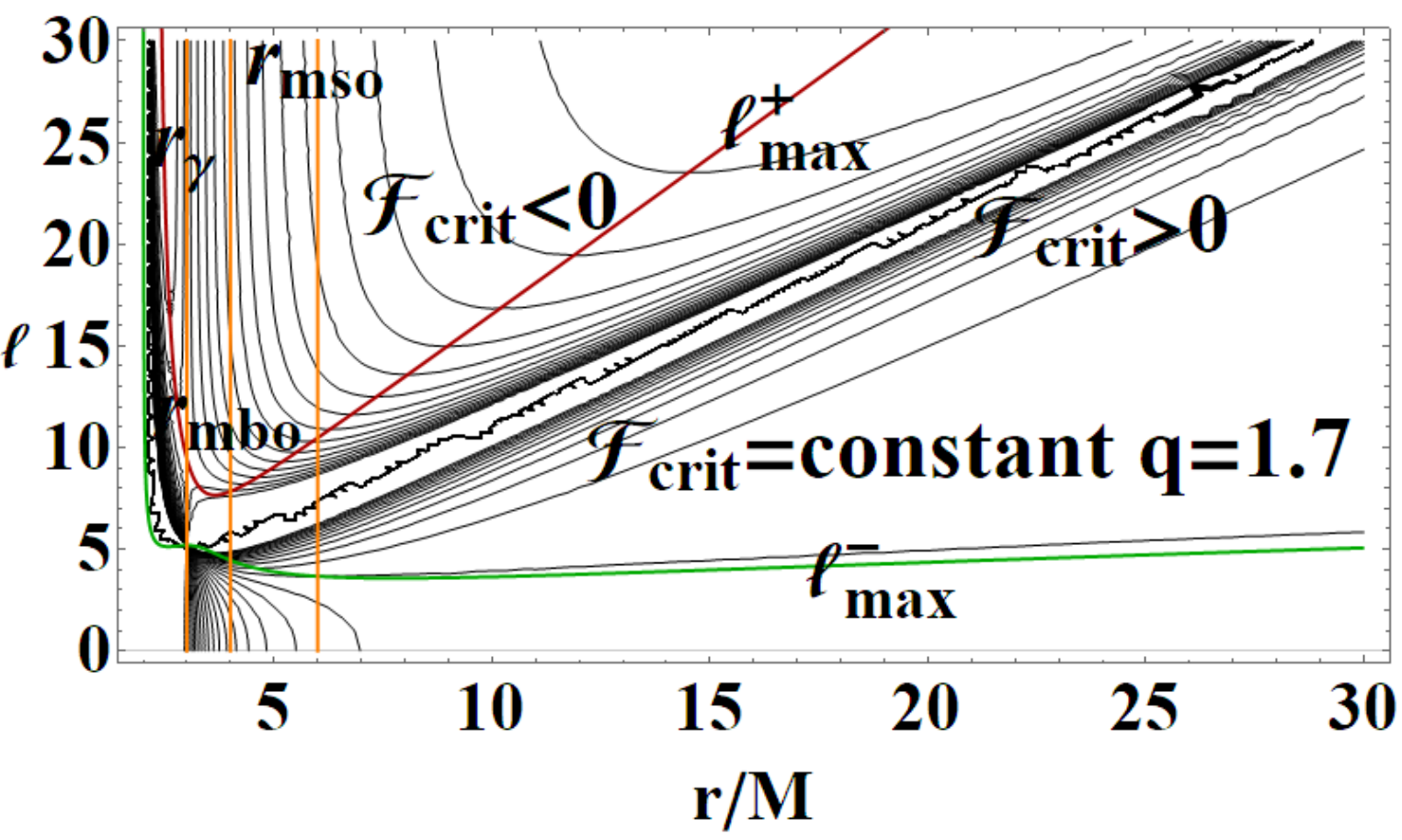}
     \includegraphics[width=8.4cm]{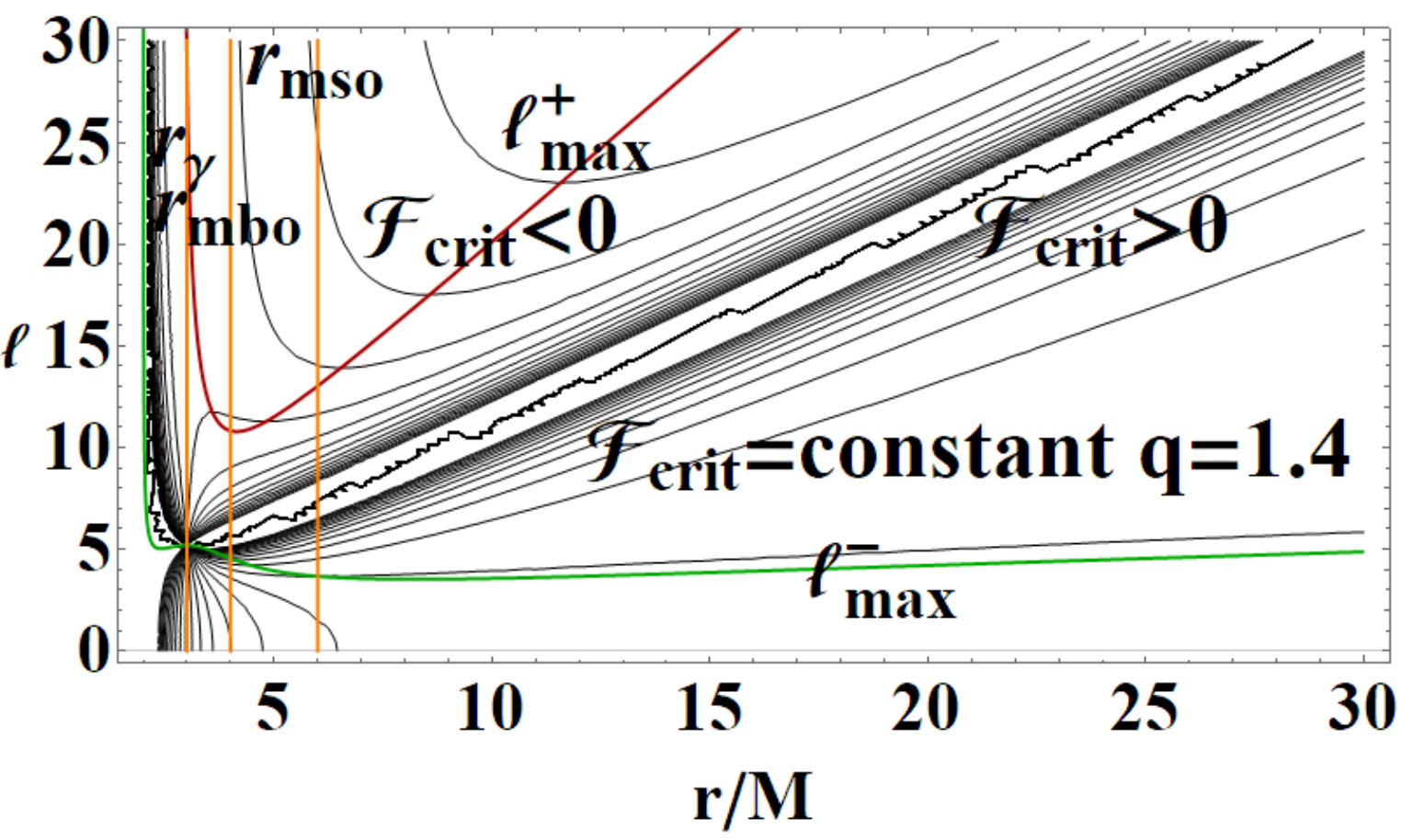}
  \caption{First panel line: curves $\ell_{\max}^{\pm}=$constant defined in equation\il\ref{Eq:lu-ul} as maximum of the $\Fa_{crit}$ in the plane  ($r/M$, $q$). It is clear the difference between the regions $q>1$ and $q<1$. Second panel line: curves $\Fa_{\max}^{\pm}\equiv \Fa_{crit}(\ell_{\max}^{\pm})=$constant in the plane  ($r/M$, $q$).
  Third and fourth  panel lines:
  Classes of tori with equal $\Fa_{crit}=$constant (see also \protect\cite{mnras}) defined in equation\il\ref{Eq:Sacrit} in the plane $(\ell,r)$ for different values of the magnetic $q$ parameter. Particularly there  are shown the regions $\Fa_{crit}>0$ and $\Fa_{crit}<0$, associated to the regions $q>1$ and $q<1$. Curve $\ell_{\max}^{\pm}$ as functions of $r/M$ are also shown. }\label{Fig:procePlot}
\end{figure*}

In this section we show a possible alternative choice for  the \textbf{RAD} leading function $\ell(r)$ in equation\il\ref{Eq:lqkp} (which was coincident with the \textbf{RAD} rotational law), considering a distribution function of alternative parameters that fix the toroids in the \textbf{RAD}, for the \textbf{eRAD} case this approach has been fully considered in \cite{mnras}--see also \cite{Agol:1999dn,Bugli,Das:2017zkl,Del-Zanna,Grasso:2000wj,Montero:2007tc,Parker:1955zz,
Parker:1970xv,Luci,Safarzadeh:2017mdy,Karas:2014rka,assum,thisc,lowcgq}. In order to do that we consider a properly adapted \textbf{RAD} model with toroids having toroidal magnetic field.
The tori considered in this model are regulated, according to equations\il\ref{Eq:Eulerif0}, by the balance of   the gravitational and centrifugal  effects with $p$ being the hydrostatic pressure.   In  \cite{mis(a)} we discussed some  characteristics related to the tori and \textbf{RAD} energetics dependent on the   polytropic index $\gamma$ different for  each toroidal component.
 \textbf{RAD} components  can be formed however  in different  periods of the  \textbf{BH} life,  having consequently different matter compositions. (On the other hand, the constraints on the inner edge variation range, as discussed in \cite{dsystem},  guarantee the viability of this analysis for large part of models where the curvature effects of the background are significant).
Here we consider the  \textbf{RADs} where some of their components are  magnetized tori with toroidal magnetic field introduced in \cite{Komissarov:2006nz} and also   discussed in \cite{abrafra,epl,adamek,Hamersky:2013cza,Montero:2007tc,Fragile:2017lbx,Gimeno-Soler:2017qmt,Zanotti:2014haa}, using  the approach considered in \cite{mnras,epl} where the magnetic pressure  is treated as a deformation of the potential term in the force balance equation.  For an extensive discussion and comparison on different solutions of the Euler equation for this problem see for example \cite{adamek,Hamersky:2013cza}, (it should be noted  that, as discussed in these references,  the total pressure  extremes are unchanged with respect to the hydrodynamic case).
We use the analysis presented  in \cite{mnras}, adapted to our case of a static attractor as done in \cite{epl}.  For such tori the entropy per particle is constant on the flow lines of each torus.
The magnetic pressure is $p_B=\mathcal{M} \omega ^q \left(-g_{tt} g_{\phi \phi }\right){}^{q-1}$, where $
(\mathcal{M}, \omega, q)$ are constants related to the magnetic field and the enthalpy at the center of the configurations.
Here we do not want to focus our investigation on the discussion of the properties of the magnetic field definition and role in the magnetized accretion disks models, but  rather use this case to verify the situation related to the \textbf{RAD} construction and stability  with misaligned disks when the \emph{leading \textbf{RAD} function} has changed to an alternative definition, and  a modification of the force balance equation has to  be considered at least for a \textbf{RAD} component. Thus  our purpose is to derive the new leading function, and show how tori distribution varies according to the new parametrization,  testing  the validity of the study of the hydrodynamic case.
As such we adopt here approach of \cite{epl}  allowing to explicit  these considerations.
%
The toroidal magnetic field $B^{\phi}$  component is:
\bea
B^{\phi}=\sqrt{\frac{2 p_B}{\ell ^2 g_{\text{tt}}+g_{\phi \phi }}}=\sqrt{2} \sqrt{\frac{r \mathcal{M} [(r-2) r]^{q-1} \omega ^q}{r^3-(r-2) \ell ^2}}.
\eea
We introduce the following parameter $\Fa\equiv{q \mathcal{M} \omega ^{q-1}}/({q-1})$.
Explicitly the deformed (modified) fluid-magnetic potential $V_B$ is
\bea&&
V_B=
\frac{(r-2) r^2\sigma  e^{2 \Fa \left[(r-2) r \sigma\right]^{q-1}}}{r^3 \sigma-(r-2) \ell^2},
\eea
($\sigma\equiv\sin^2\theta$),
where the deformed fluid magnetic-specific momentum $\ell_b$ is
%
\bea\label{Eq:ellb}
\ell_b=\frac{\sqrt{r^3 \left[4 (q-1)^2 (r-1)^2 r S^2 [(r-2) r]^{2 q-1}+2 (q-1) (r-1)^2 r \Fa [(r-2) r]^q+(r-2)^2 r^2\right]}}{2 (q-1) (r-1) \Fa [(r-2) r]^q+(r-2)^2 r}.
\eea
%
%
The radial profile of this distribution of  fluid specific momentum is shown Figure\il\ref{Fig:coxxlinas} for different values of the model parameters, and as also noted in \cite{mnras} it is clear  that this  distribution  is strongly dependent on the magnetic parameters $q$ and $\mathcal{F}$. It is more convenient to  introduce the alternative critical parameter $\Fa_{crit}$
\bea\label{Eq:Sacrit}
\Fa_{crit}\equiv-\frac{\left[r^3-\ell^2 (r-2)^2\right] [(r-2) r]^{1-q}}{2 (q-1) (r-1) \left(r^3-\ell^2 (r-2)\right)},
\eea
for the toroids distribution in the \textbf{RAD}--Figures\il(\ref{Fig:procePlot8},\ref{Fig:procePlot}
).
On the other hand, $\Fa=0$ implies $\ell_b=\ell_K(r)$ in equation\il\ref{Eq:lqkp}.
The $\Fa_{crit}$ has a maximum as function of $r/M$ with fixes the $\Fa$ upper boundary  for the toroidal  solution of the Euler equation, therefore we can express the maximum as
%

\bea\label{Eq:lu-ul}
&&
\ell_{max}^{\mp}\equiv\sqrt{\frac{s_i\mp\text{\u \i}}{\hat{u}_i}},\quad\mbox{where}
\quad
s_i\equiv(r-2) (r-1) r^3 \left[q (r-1)^2-4 r+8\right]
\quad\hat{u}_i\equiv(r-2)^3 \left[2 q (r-1)^2+r (3-2 r)-2\right],\quad\mbox{and}
\\&&
\nonumber
\text{{\u\i}}\equiv\sqrt{(r-2)^2 r^6 \left[q^2 (r-3)^2 (r-1)^4-2 q (r-3) (r-2) r (r-1)^2-(r-2) [r \{r [r (2 r-23)+76]-90\}+36]\right]}.
\eea
%
Analysis of these cases in show in Figures\il\ref{Fig:coxxlinas},\ref{Fig:procePlot8},\ref{Fig:procePlot}.

\bsp	
\label{lastpage}
\end{document}